\documentclass[twocolumn,aps,pra,10pt,floatfix,longbibliography]{revtex4-1}

	\usepackage[usenames,dvipsnames]{xcolor}
	\usepackage{amsmath}
	\usepackage{amsfonts}
	\usepackage{amssymb}
	\usepackage[colorlinks=true,citecolor=blue,linkcolor=red]{hyperref}
	\usepackage{graphicx}
	\usepackage{bbold}					
	\usepackage[makeroom]{cancel}		
	\usepackage{multirow}				
	\usepackage[normalem]{ulem}        
	\usepackage{array} 
	\usepackage{mathrsfs} 	
	\usepackage{bibunits}
    \usepackage[section]{placeins}
    \usepackage{setspace}
    \usepackage{titlesec}

	\newcolumntype{x}[1]{>{\centering\let\newline\\\arraybackslash\hspace{0pt}}p{#1}}
		
	\DeclareMathOperator{\Pf}{Pf}  		
	\DeclareMathOperator{\tr}{tr}  		
	\DeclareMathOperator{\diag}{diag}  	

	\DeclareMathAlphabet{\mathbbold}{U}{bbold}{m}{n}
	\DeclareMathAlphabet{\mathpzc}{OT1}{pzc}{m}{it}
	\def\abs#1{\left|{#1}\right|}      	
	\def\norm#1{\lVert{#1}\rVert}
	\def\bs#1{\boldsymbol{#1}}			
	\def\imi{\mathrm{i}}				
	\def\imj{\mathrm{j}}				
	\def\imk{\mathrm{k}}				
	\def\e#1{\mathrm{e}^{#1}}				
	\def\de{\mathrm{d}}
	
	\def\eps{\varepsilon}					
 
	\def\mcH{\mathcal{H}}					
	\def\mcA{\mathcal{A}}					
	\def\mcF{\mathcal{F}}					
	
	\def\mcT{\mathcal{T}}					
	\def\mcP{\mathcal{P}}					
	\def\mcK{\mathcal{K}}					

	\def\intg{\mathbbold{Z}}					
	\def\ztwo{\mathbbold{Z}_2}					
	\def\triv{\mathbbold{0}}					
	\def\unit{\mathbbold{1}}					
	\def\reals{\mathbbold{R}}					
	
	\def\bra#1{\left<{#1}\right|}				
	\def\ket#1{\left|{#1}\right>}				
	\def\braket#1#2{\left<{#1}|{#2}\right>}				

	\newcounter{subeqn} %
	\makeatletter
	\@addtoreset{subeqn}{equation}
	\makeatother

\def\sectitle#1{\medskip\smallskip {\noindent \fontsize{11pt}{12.5pt}\selectfont {\textsf{\textbf{#1}}} \\ \noindent }}
\def\partitle#1{\medskip \noindent {\textsf{\textbf{#1}}} }

\newcommand{\PRLsep}{\noindent\makebox[\linewidth]{\resizebox{0.3333\linewidth}{1pt}{$\bullet$}}\bigskip}
	
\definecolor{TB}{rgb}{1,0.5,0}

\defaultbibliography{bib}

\usepackage{setspace}

\usepackage{titlesec}
\titleformat*{\section}{\normalsize\bfseries\centering}
\renewcommand\thesection{\Alph{section}}
\renewcommand\thesubsection{\Alph{section}\arabic{subsection}}
\makeatletter
\def\l@subsection#1#2{}
\def\l@subsubsection#1#2{}
\makeatother
\usepackage{inputenc}

\begin{document}
\begin{bibunit}
\title{\textsf{Non-Abelian reciprocal braiding of Weyl points and its manifestation in $\textrm{ZrTe}$}}

\author{Adrien Bouhon$^{1,2}$}\thanks{Contributed equally. Correspondence to \href{mailto:adrien.bouhon@su.se}{adrien.bouhon@su.se}, \href{mailto:quansheng.wu@epfl.ch}{quansheng.wu@epfl.ch}, and  \href{mailto:rjs269@cam.ac.uk}{rjs269@cam.ac.uk}.}
\author{QuanSheng Wu$^{3,4}$}\thanks{Contributed equally. Correspondence to \href{mailto:adrien.bouhon@su.se}{adrien.bouhon@su.se}, \href{mailto:quansheng.wu@epfl.ch}{quansheng.wu@epfl.ch}, and  \href{mailto:rjs269@cam.ac.uk}{rjs269@cam.ac.uk}.}
\author{Robert-Jan Slager$^{5,6}$}\thanks{Contributed equally. Correspondence to \href{mailto:adrien.bouhon@su.se}{adrien.bouhon@su.se}, \href{mailto:quansheng.wu@epfl.ch}{quansheng.wu@epfl.ch}, and  \href{mailto:rjs269@cam.ac.uk}{rjs269@cam.ac.uk}.}
\author{Hongming Weng$^{7,8}$}
\author{Oleg V. Yazyev$^{3,4}$}
\author{Tom\'{a}\v{s} Bzdu\v{s}ek$^{9,10,11}$}

\affiliation{\vspace*{0.4cm} $^{1}$Nordic Institute for Theoretical Physics (NORDITA), Stockholm, Sweden}
\affiliation{$^{2}$Department of Physics and Astronomy, Uppsala University, Box 516, SE-751 21 Uppsala, Sweden}
\affiliation{$^{3}$Institute of Physics, \'{E}cole Polytechnique F\'{e}d\'{e}rale de Lausanne, CH-1015 Lausanne, Switzerland}
\affiliation{$^{4}$National Centre for Computational Design and Discovery of Novel Materials MARVEL, Ecole Polytechnique F\'{e}d\'{e}rale de Lausanne (EPFL), CH-1015 Lausanne, Switzerland}
\affiliation{$^{5}$TCM Group, Cavendish Laboratory, University of Cambridge, J.~J.~Thomson Avenue, Cambridge CB3 0HE, United Kingdom}
\affiliation{$^{6}$Department of Physics, Harvard University, Cambridge, MA 02138}
\affiliation{$^{7}$Beijing National Laboratory for Condensed Matter Physics and Institute of Physics,
Chinese Academy of Sciences, Beijing 100190, China}
\affiliation{$^{8}$Songshan Lake Materials Laboratory, Guangdong 523808, China}
\address{$^{9}$Condensed Matter Theory Group, Paul Scherrer Institute, CH-5232 Villigen PSI, Switzerland}
\address{$^{10}$Department of Physics, University of Z\"{u}rich, Winterthurerstrasse 190, 8057 Z\"{u}rich, Switzerland}
\affiliation{$^{11}$Department of Physics, McCullough Building, Stanford University, Stanford, CA 94305, USA \vspace*{0.2cm}}

\date{\today}

\begin{abstract}
{\bf Weyl semimetals in three-dimensional crystals provide the paradigm example of topologically protected band nodes. It is usually taken for granted that a pair of colliding Weyl points annihilate whenever they carry opposite chiral charge. In a stark contrast, here we report that Weyl points in systems symmetric under the composition of time-reversal with a $\pi$-rotation are characterized by a non-Abelian topological invariant. The topological charges of the Weyl points are transformed via braid phase factors which arise upon exchange inside symmetric planes of the reciprocal momentum space. We elucidate this process with an elementary two-dimensional tight-binding model implementable in cold-atoms setups and in photonic systems. In three dimensions, interplay of the non-Abelian topology with point-group symmetry is shown to enable topological phase transitions in which pairs of Weyl points may scatter or convert into nodal-line rings. By combining our theoretical arguments with first-principles calculations, we predict that Weyl points occurring near the Fermi level of zirconium telluride (ZrTe) carry non-trivial values of the non-Abelian charge, and that uniaxial compression strain drives a non-trivial conversion of the Weyl points into nodal lines.} 
\end{abstract}

\maketitle


The robust and illustrious properties of topological order, such as protected edge states and the possibility of excitations that exhibit non-trivial braiding statistics~\cite{Wen95_top}, open up routes to potentially translate mathematical understanding of the physical phenomena to a new generation of quantum technology. This has arguably also fuelled the discovery of topological band structures~\cite{HasanKane10_RMP,Qi11_RMP} that can effectively give rise to such physical features~\cite{HasanKane10_RMP,Qi11_RMP,Majorana}. The past decade has witnessed considerable progress in cataloguing topological insulators and semimetals~\cite{Kitaev09_AIP, Ryu10_NJP, Fu11_PRL, Slager12_NatPhys,  Bzduvsek16nodal, Fang:2016, Bzdusek:2017, Kruthoff17_PRX, Bouhon:global_top, Po17_NatCommun, Bradlyn17_Nat, Slager2019, Holler18_PRB,Zhang2019,topomat}, thereby providing an increasingly viable platform for bringing the potential of topological materials to experiment. Notably, Weyl semimetals were convincingly shown to exhibit topologically protected surface Fermi arcs and chiral-anomaly-induced negative magnetoresistance~\cite{Wan:2011,Weng:2015,Lv:2015,Xu:2015,Huang:2015}.

Here we report that Weyl points in three-dimensional (3D) systems with $C_2\mcT$ symmetry  (time reversal composed with a $\pi$-rotation) carry non-Abelian topological charges. These charges are transformed via non-trivial phase factors that arise upon braiding the nodes inside the reciprocal momentum space. This discovery extends the previous theoretical works on  non-Abelian disclination defects in nematic liquids~\cite{Poenaru:1977,Volovik:1977,Madsen:2004,Alexander:2012,Prx2016}, Dirac lines in space-time-inversion symmetric metals~\cite{Wu:2018b,Tiwari:2019}, and Dirac points of twisted bilayer graphene~\cite{Ahn:2019,Ahn:2018,Cao_2018,Ashvin_2019,Song2019,Bouhon2018_fragile}. We show that interplay of the non-Abelian topology with point-group symmetry greatly enriches the range of topological phase transitions for Weyl points in 3D crystals. 

Below, we illustrate the reciprocal braiding with an elementary 2D model, which is directly implementable in cold-atoms and photonic systems. We relate the descriptions of the non-Abelian topology via quaternion numbers~\cite{Wu:2018b} resp.~Euler class~\cite{Ahn:2019}, and we numerically implement the latter. It is shown that the interplay of the non-Abelian charges with the chiralities of Weyl points and with mirror symmetry results in non-trivial conversions between Weyl points and nodal lines. Finally, we combine our formalism with first-principles modeling of the existing material zirconium telluride (ZrTe) and related compounds, which provide simple examples for the predicted nodal conversions.

\sectitle{Elementary braiding protocol}
The ability of band nodes in $C_2\mcT$-symmetric systems to pairwise annihilate crucially depends on the presence of band nodes in other band gaps~\cite{Wu:2018b,Tiwari:2019,Ahn:2019,Ahn:2018}. This enables non-trivial ``reciprocal braiding'' inside the momentum ($\bs{k}$) space, illustrated in Fig.~\ref{fig:braiding}. For three-band models we introduce the following terminology. The main gap of interest is called ``principal gap''. Accordingly, band nodes in this gap are described as ``principal nodes''. The other band gap, as well as the corresponding band and nodes, are called  ``adjacent''. 

As an illustrative model for node braiding in two dimensions, we consider a three-band Hamiltonian $H(\bs{k};t)$ [see Eq.~(\ref{eqn:Ham-matrix}) of Methods], where $t\in [-10,10]$ is a tuning parameter (``time''). The practical implementation of the model requires tuning only three tight-binding parameters, namely the potential on one site, and the hopping amplitude between the other two sites along the horizontal resp.~the vertical direction. 

The model exhibits nodal points along the $(1\bar{1})$ and $(11)$ diagonals of the Brillouin zone (BZ). In  Fig.~\ref{fig:protocol}{\bf c}, we show snapshots of the band structure along the two diagonals during the braiding protocol with solid resp.~dashed curves. At the initial time $t\!=\!-10$, the bands are energetically separated. At $t\!=\!-8$, the adjacent gap exhibits a pair of nodal points moving from $\Gamma$ (where they were created) towards M along $(1\bar{1})$. At $t\!=\!-4$, there are additional two principal nodes moving from M (where they were created) towards $\Gamma$ along $(11)$. At $t\!=\!-2$, the principal nodes meet at $\Gamma$. Remarkably, instead of annihilating, we find that the principal nodes ``bounce'' in the $(1\bar{1})$ direction, where they follow their adjacent counterparts, as visible at $t\!=\!0$. Fig.~\ref{fig:protocol}{\bf b} shows the full 2D band structure at this very time. At $t\!=\!2$, the two adjacent nodes meet at M and also fail to annihilate, as can be seen at time $t\!=\!4$ where they progress towards $\Gamma$. At $t\!=\!8$, the adjacent nodes have been annihilated at $\Gamma$. Finally, at $t\!=\!10$, the principal nodes have been annihilated at M, and the bands have become fully separated again. Panel Fig.~\ref{fig:protocol}{\bf d} displays the configuration of the nodal points at a few times, keeping track of their past trajectory. 

\begin{figure}[t]
\centering
\includegraphics[width=0.96\linewidth]{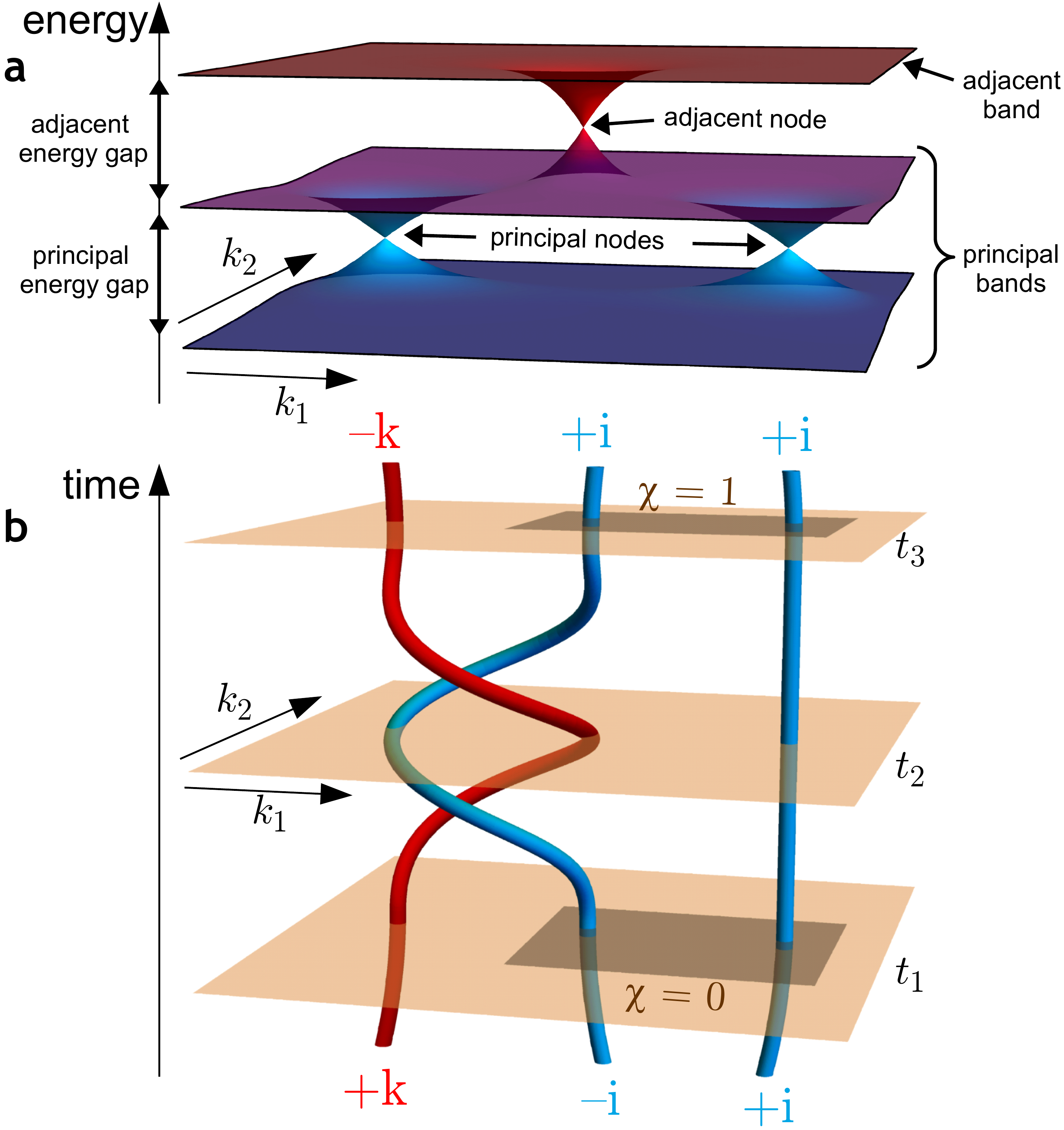}
\caption{\textsf{
\textbf{Reciprocal braiding of band nodes.} 
\textbf{a.} The terminology used in the text. 
We inspect the ability of the ``principal'' nodes, assumed to be near the Fermi level, to pairwise annihilate. We consider two ``principal'' bands that form nodes (blue), and a third ``adjacent'' band which enables additional species of band nodes (red) formed by the unoccupied bands. \textbf{b.} By adjusting the Hamiltonian parameters as a function of time (orange planes $t_{1,2,3}$), the node trajectories form braids in the momentum ($k_1,k_2$) space. The braid converts topological charges of the nodes (indicated by quaternion numbers $\pm \imi$ and $\pm \imk$~\cite{Wu:2018b}, and here dubbed ``frame-rotation charges'') and affects their ability to pairwise annihilate. Equivalently, the same property is encoded using Euler class $\chi$~\cite{Ahn:2019}, which changes value on the dark region during the braiding process. Note that the band structure in panel \textbf{a} corresponds to the situation in panel \textbf{b} at time $t_2$.}}
\label{fig:braiding}
\end{figure}

\begin{figure*}[t]
\centering
\begin{tabular}{lcl}
\begin{tabular}{l}
 \begin{tabular}{lcl}
     \textbf{a} & & \textbf{b} \\
     \includegraphics[width=0.23\linewidth]{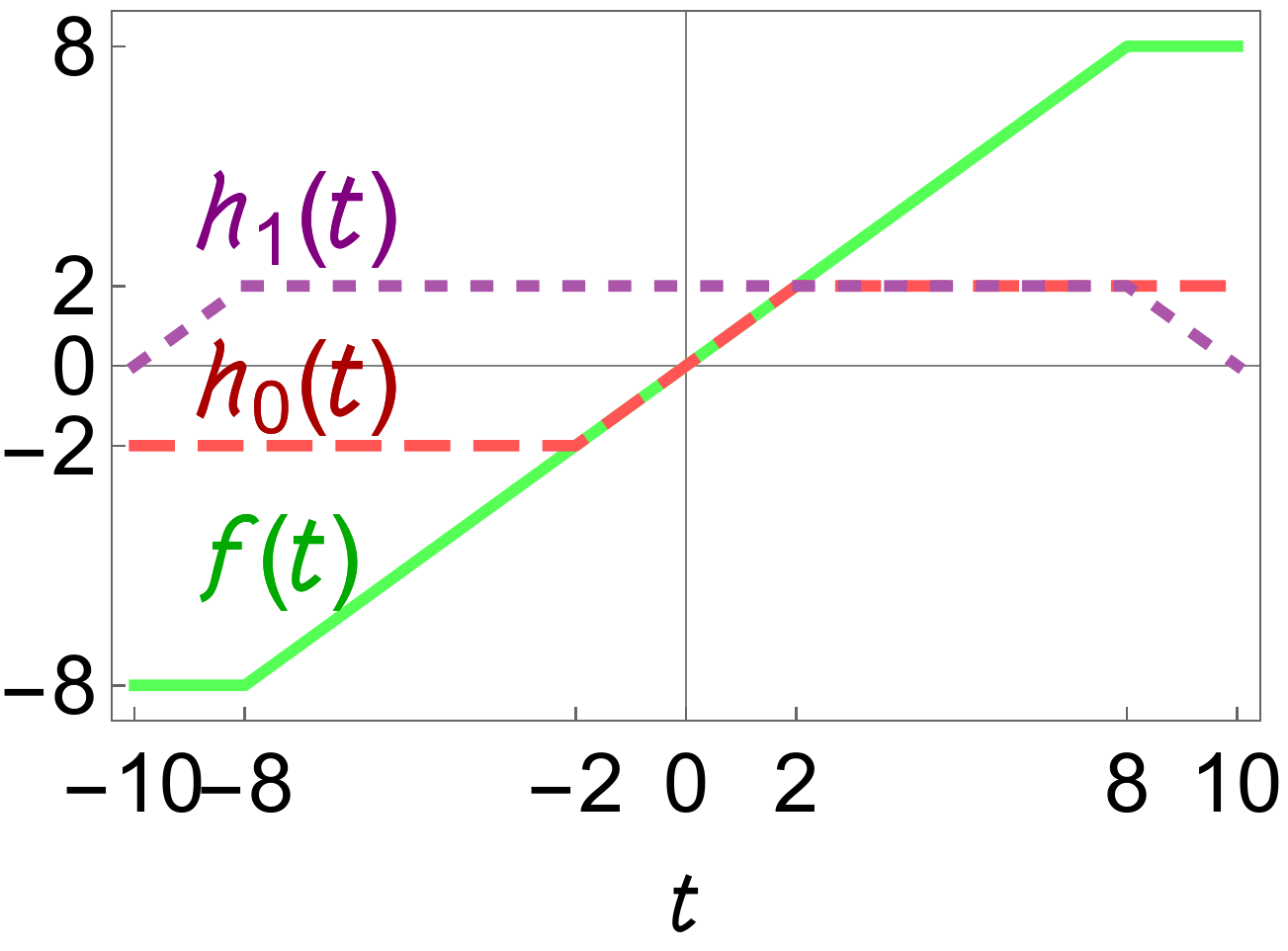}  & 
     \hspace{0.7cm}     & 
     \includegraphics[width=0.35\linewidth]{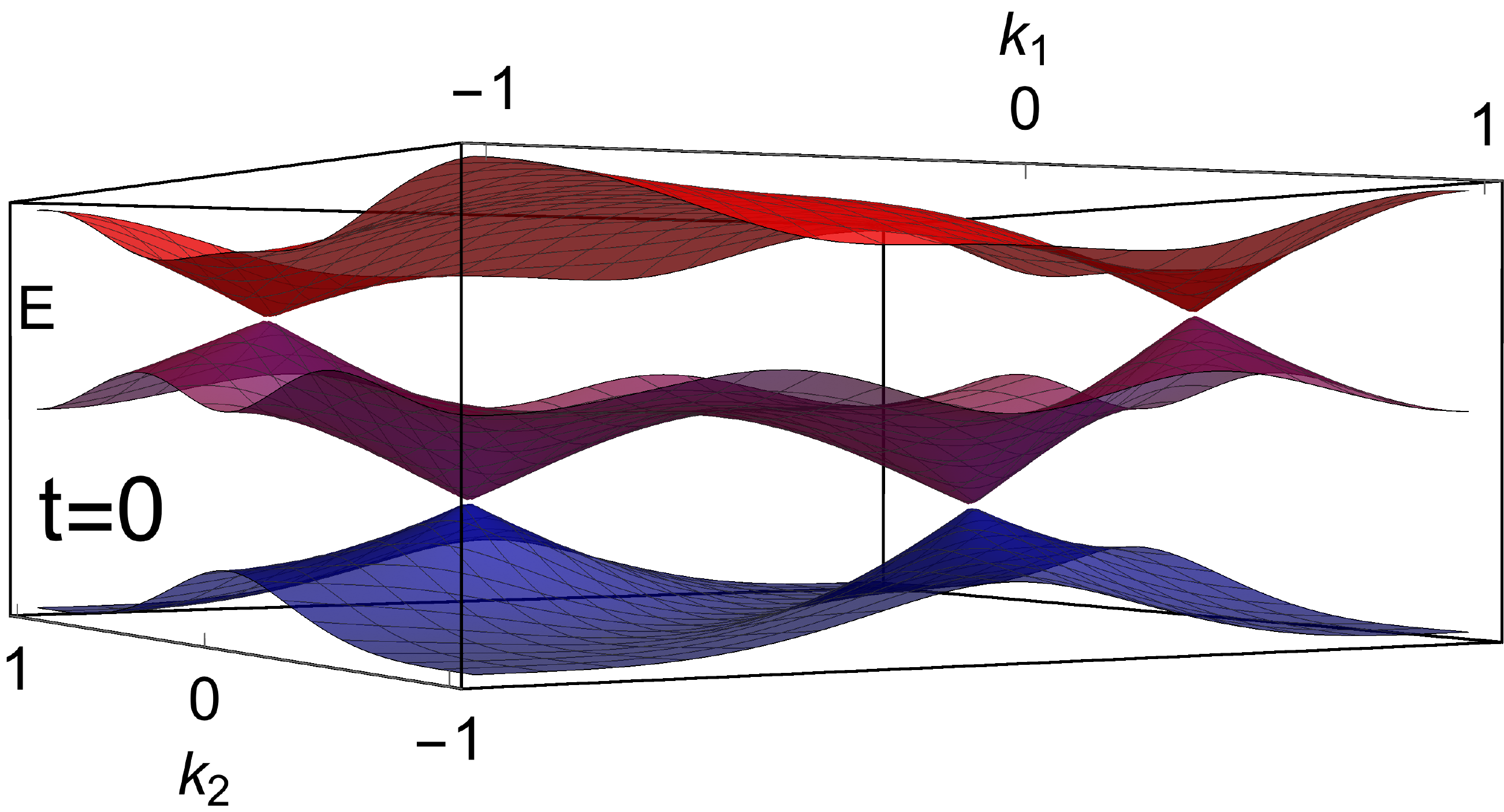} 
 \end{tabular}\\
 \textbf{c} \\
 \begin{tabular}{lll} 
	    $t=-10$ &
	    $t=-8$ &
	    $t=-4$ \\
	    \includegraphics[width=0.22\linewidth]{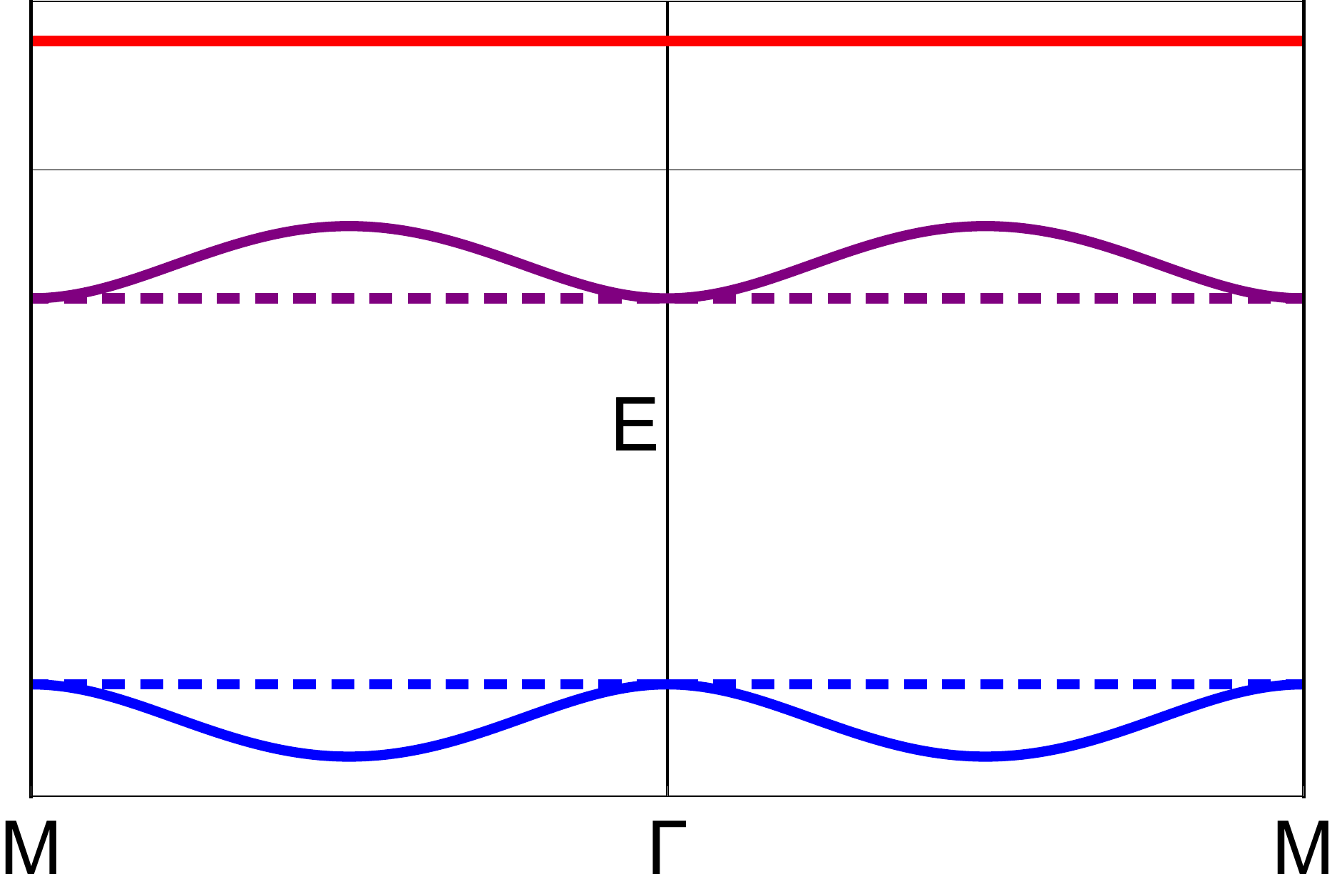} &
	    \includegraphics[width=0.22\linewidth]{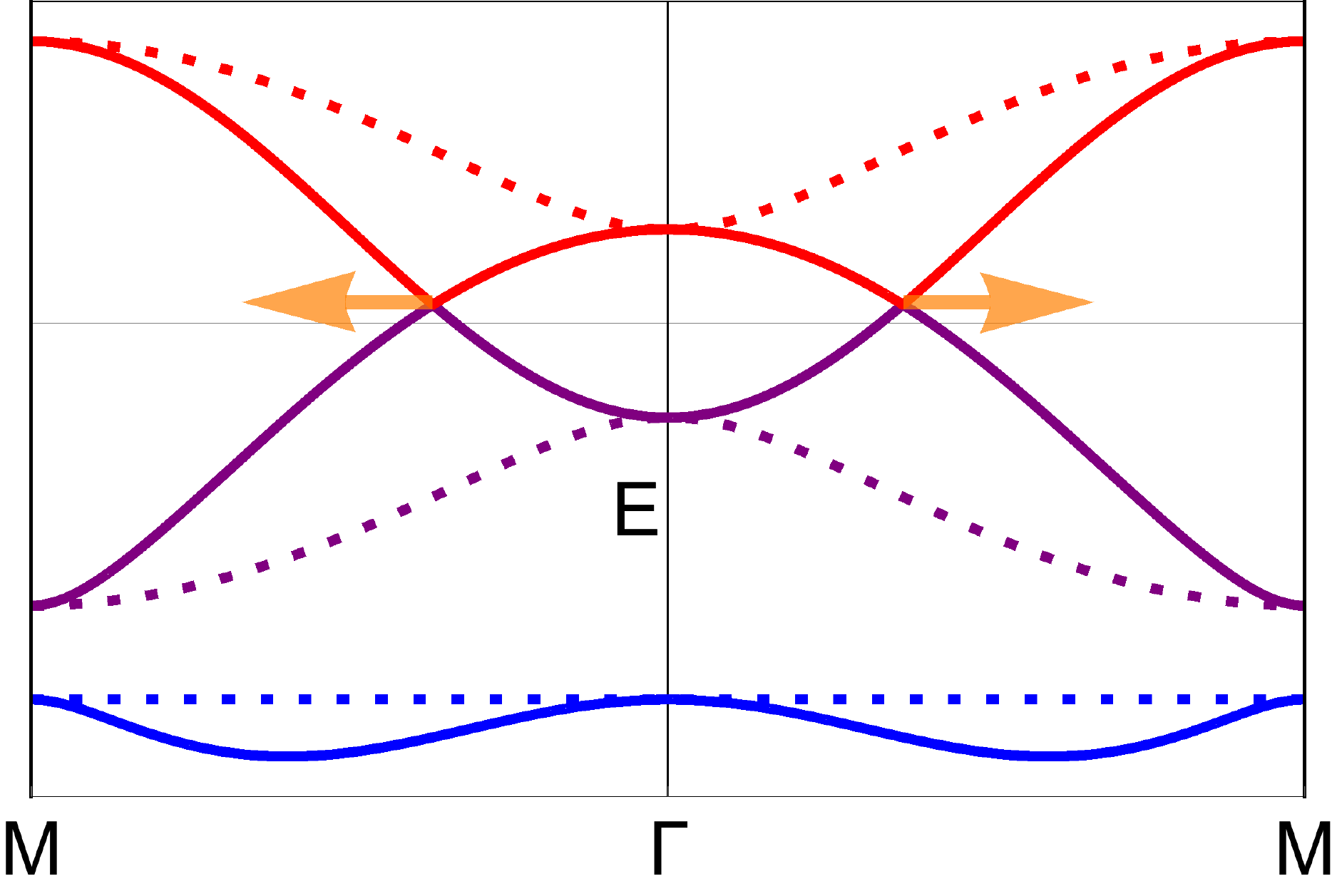} &
	    \includegraphics[width=0.22\linewidth]{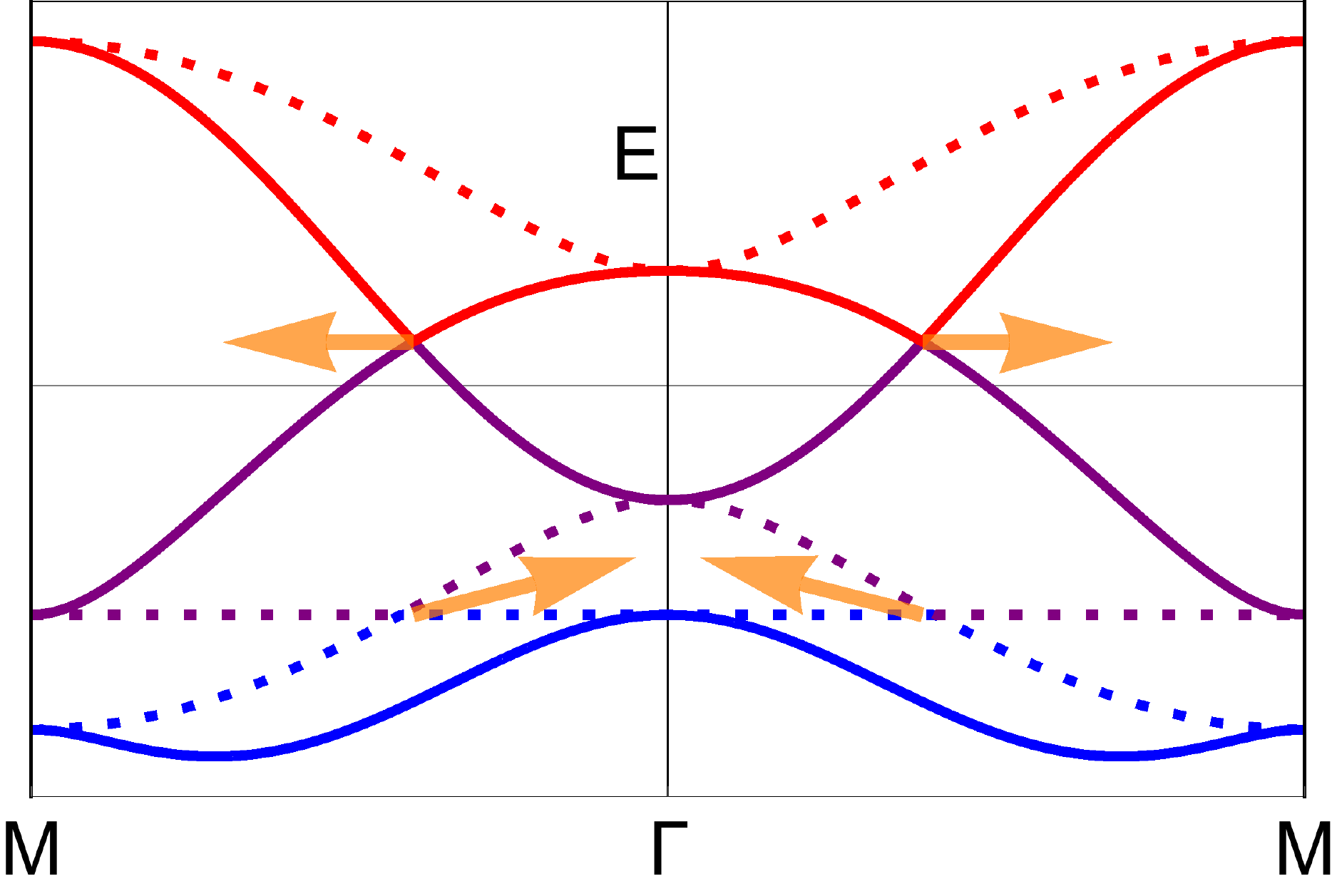} \\
	    $t=-2$ &
	    $t=0$ &
	    $t=2$  \\
	    \includegraphics[width=0.22\linewidth]{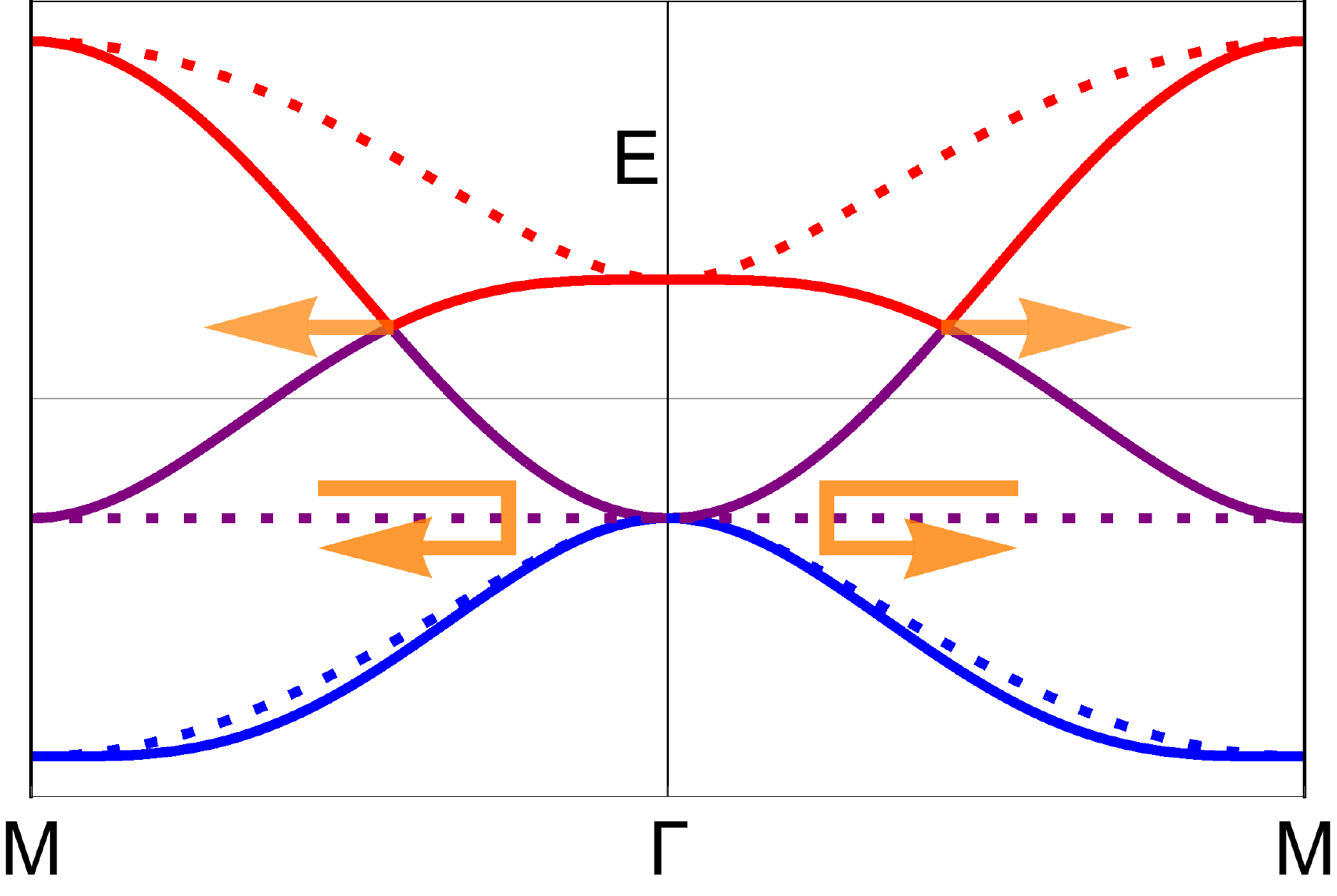} &
 	    \includegraphics[width=0.22\linewidth]{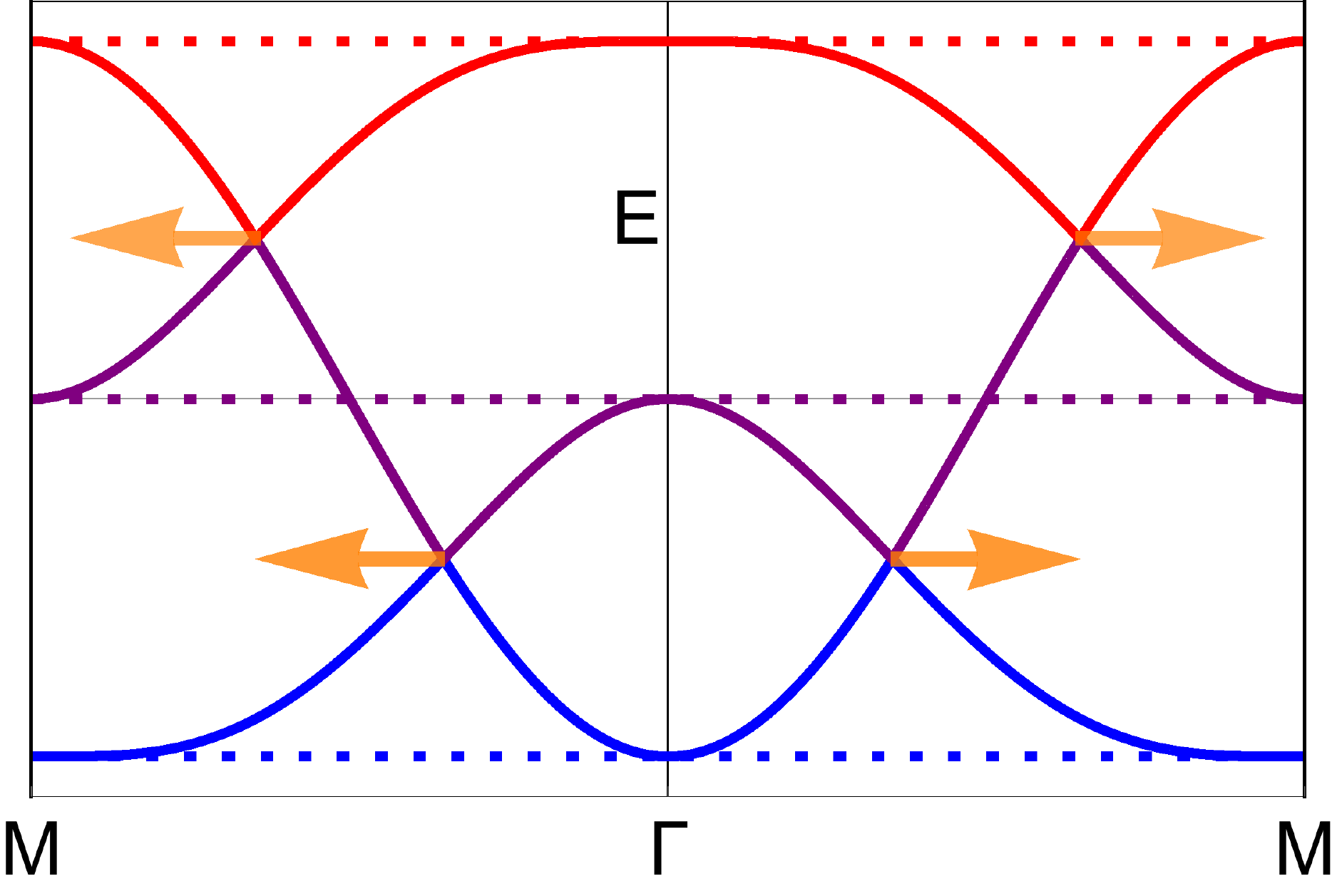} &
 	    \includegraphics[width=0.22\linewidth]{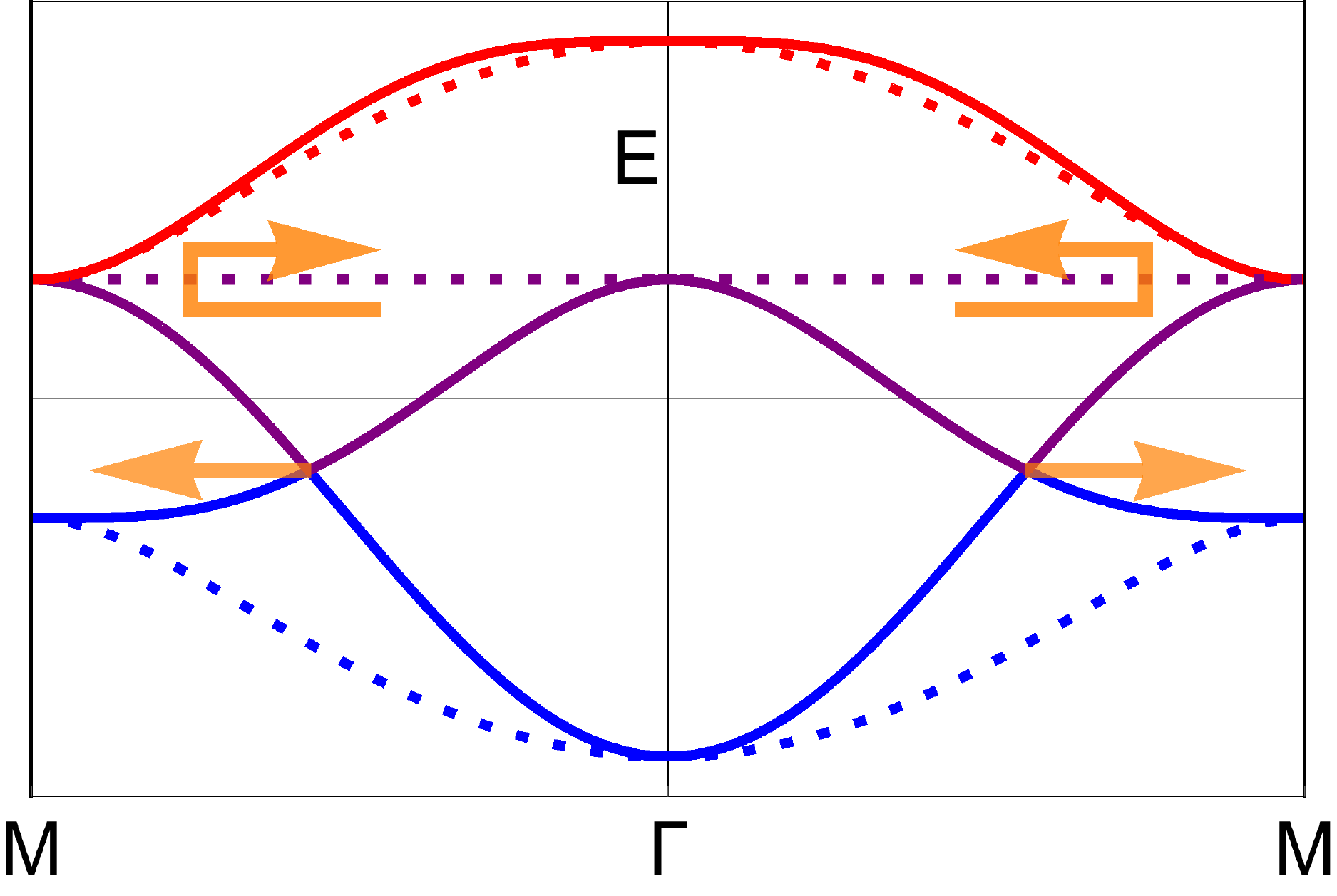} \\
 	    $t=4$ &
 	    $t=8$ &
 	    $t=10$ \\
 	    \includegraphics[width=0.22\linewidth]{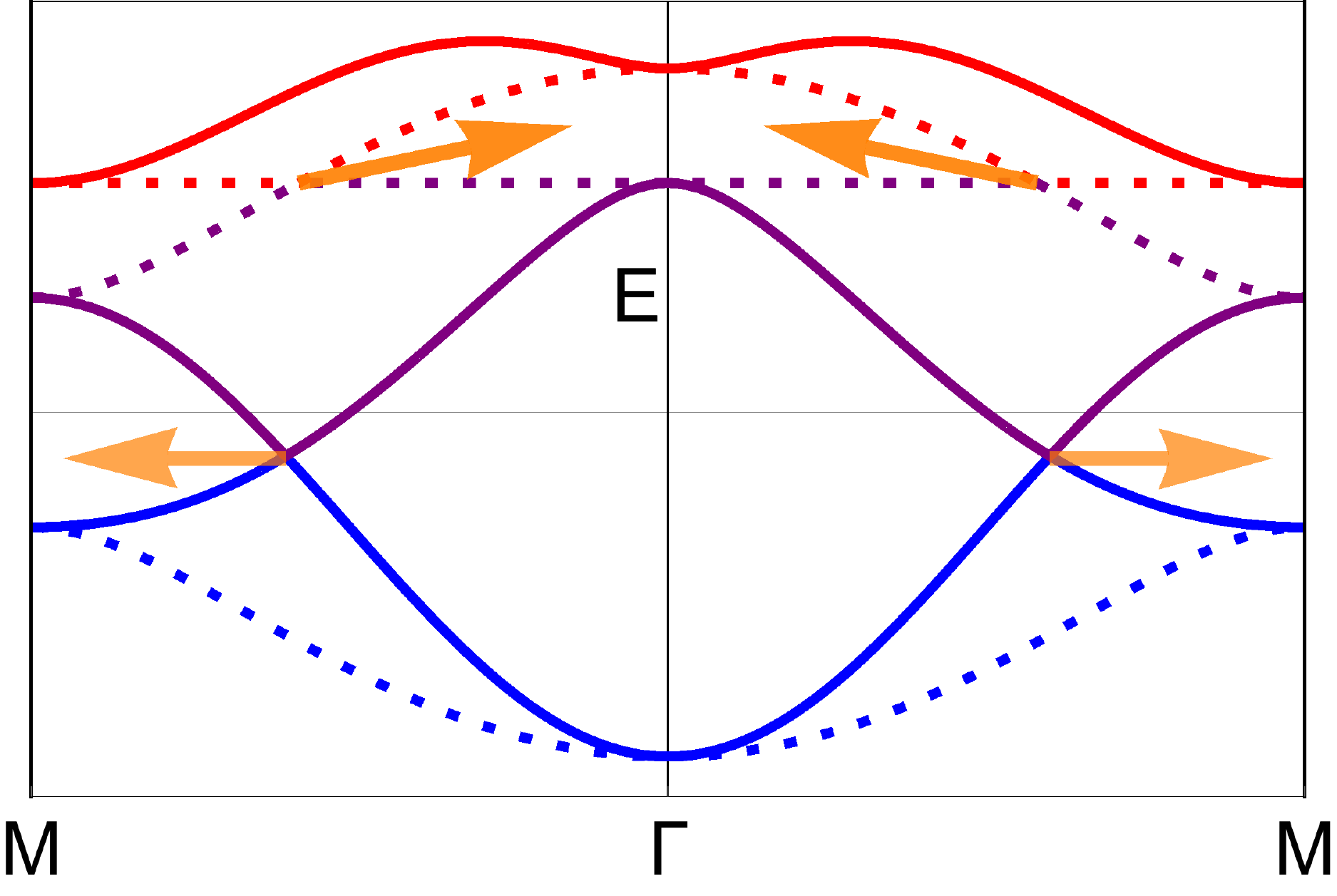} &
 	    \includegraphics[width=0.22\linewidth]{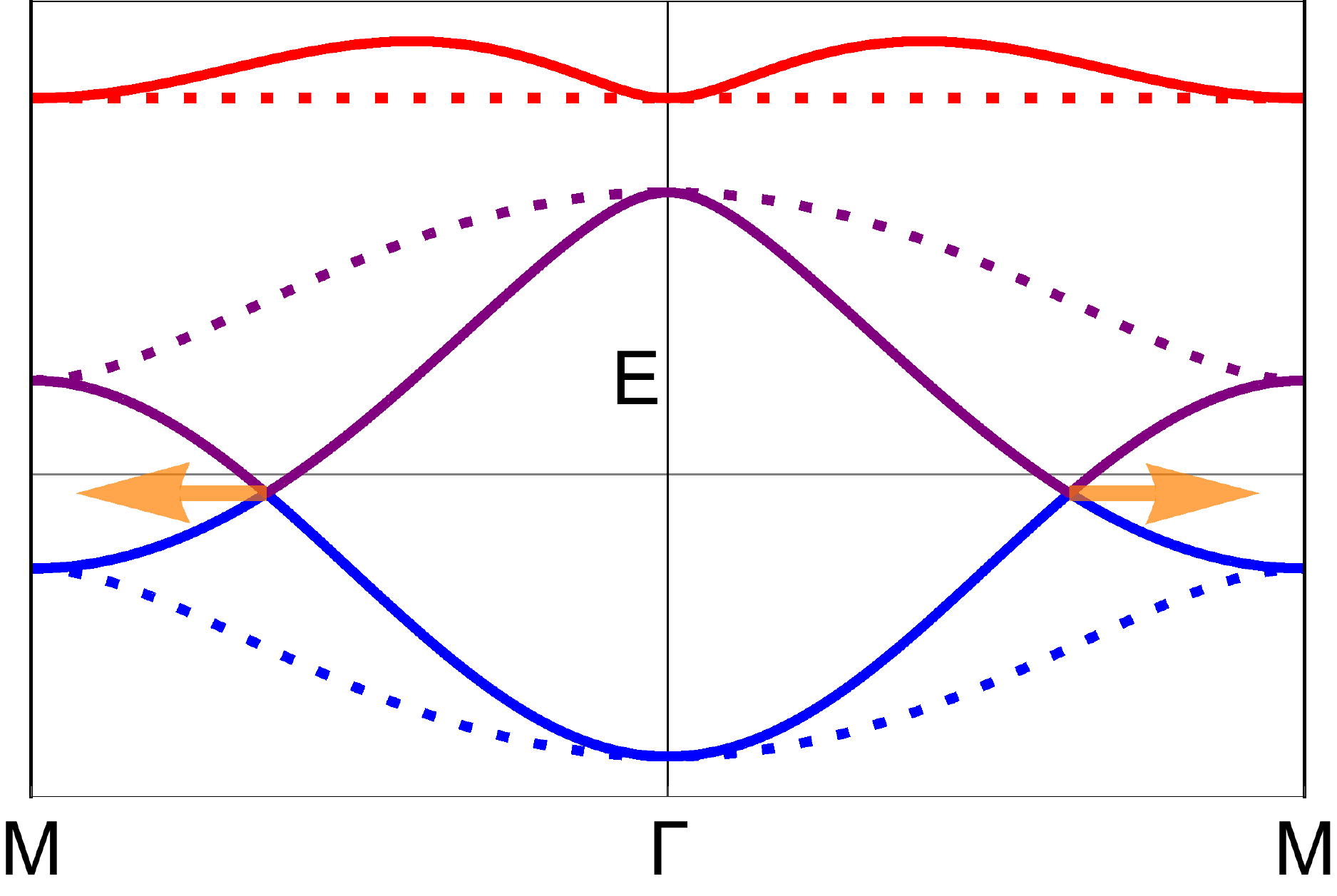} &	
 	    \includegraphics[width=0.22\linewidth]{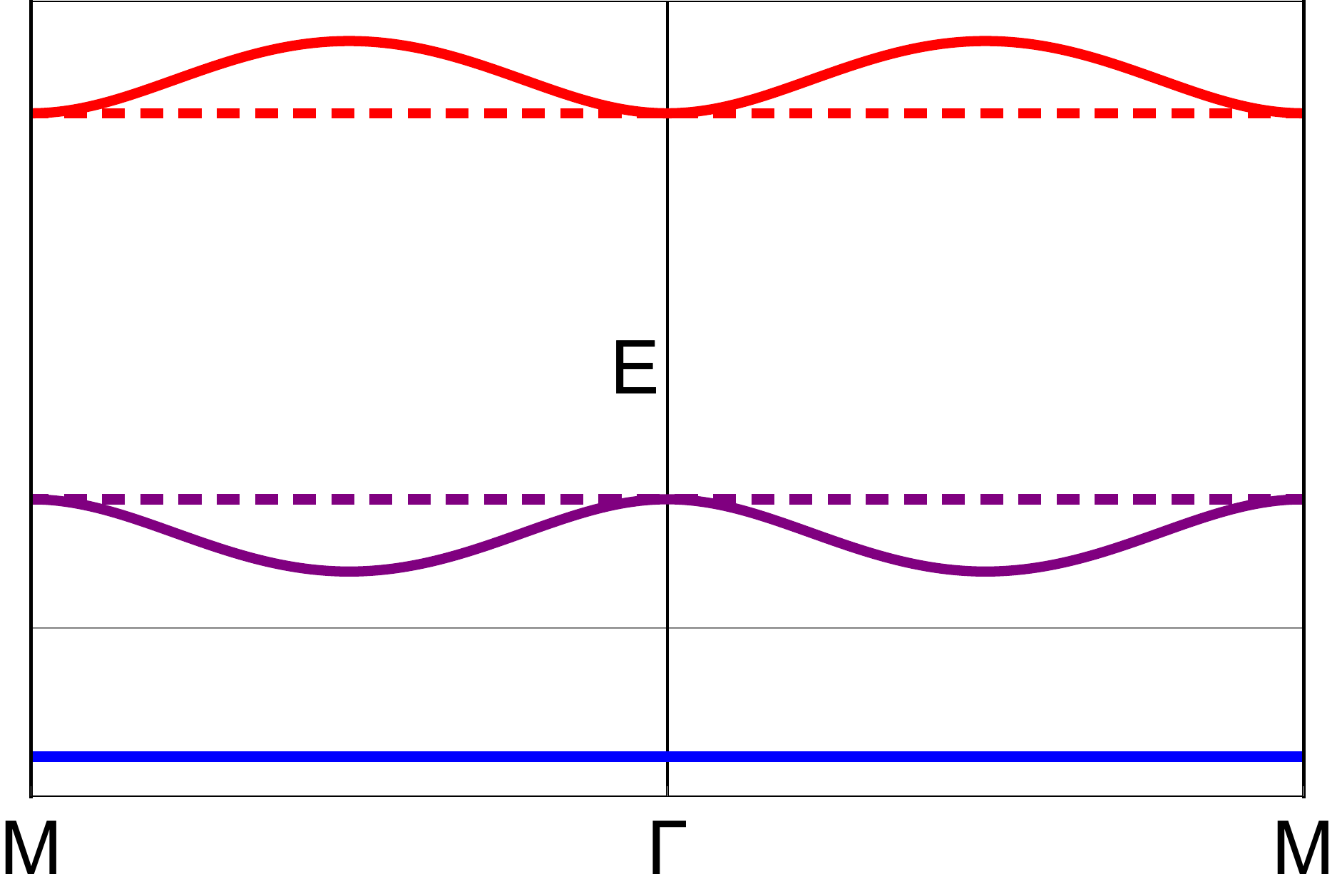} 
     \end{tabular} 
 \end{tabular} & \hspace{0.5cm} & 
     \begin{tabular}{l}
         \textbf{d} \\
         \includegraphics[width=0.18\linewidth]{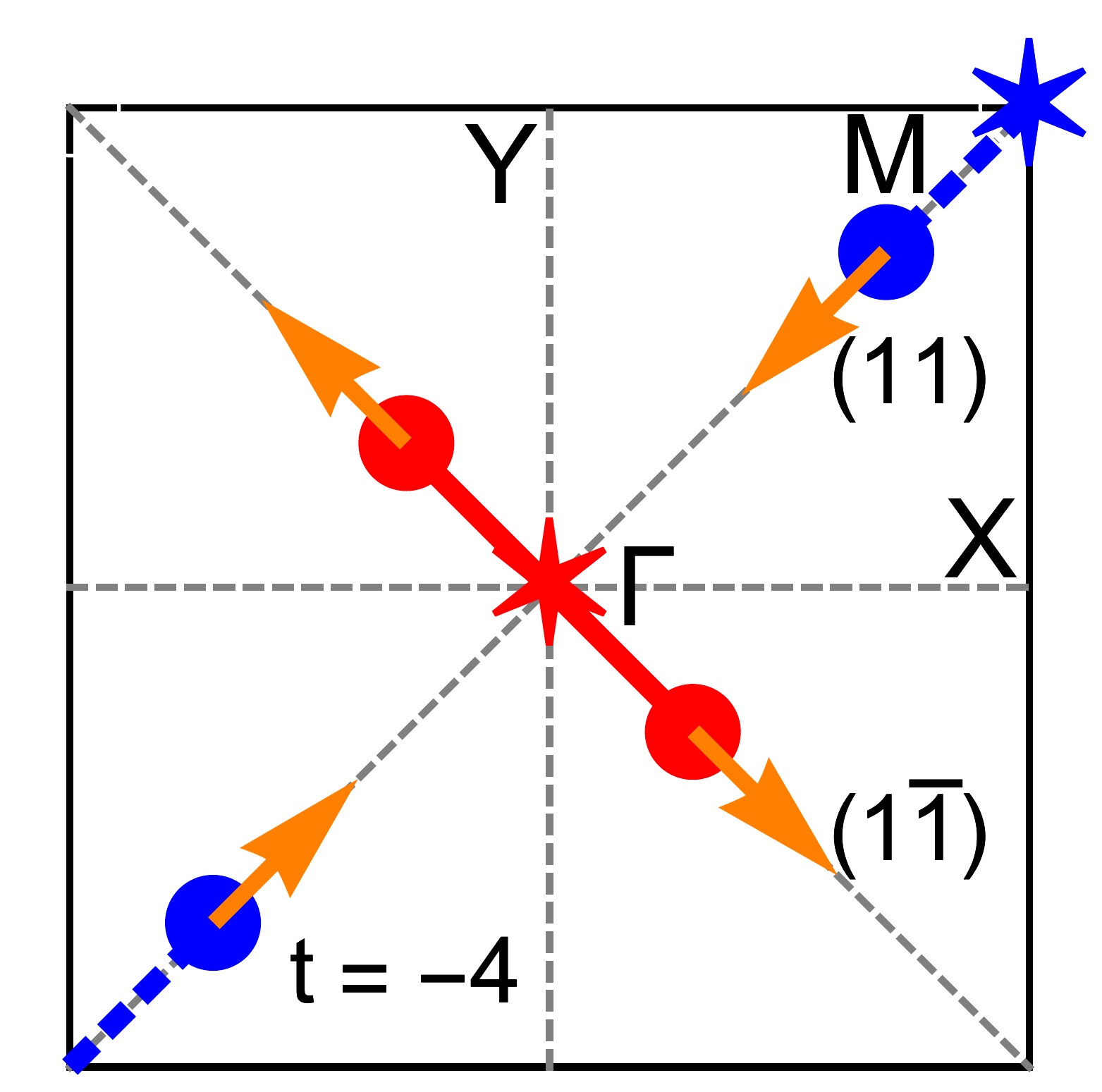} \\
         \includegraphics[width=0.18\linewidth]{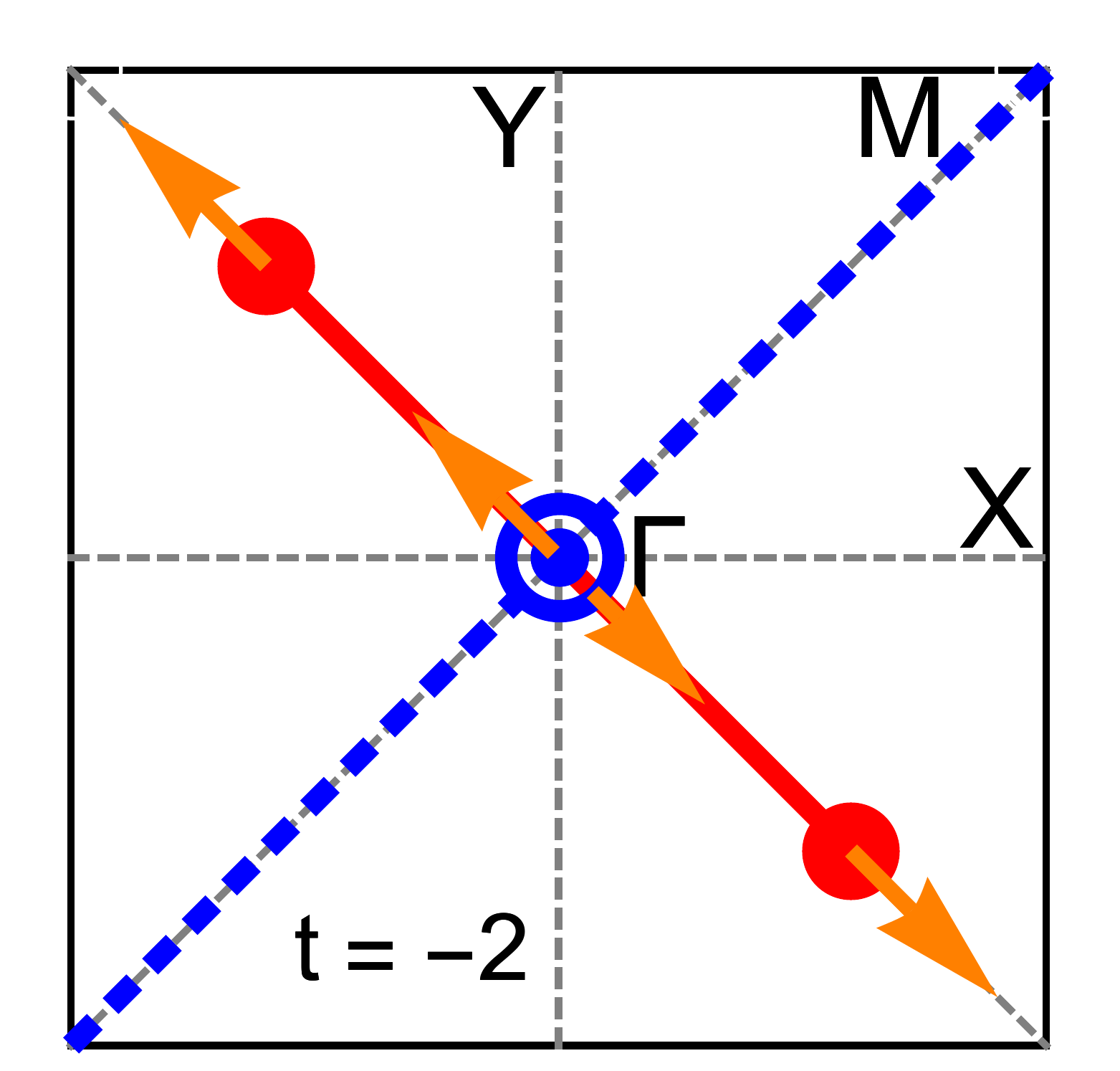} \\
         \includegraphics[width=0.18\linewidth]{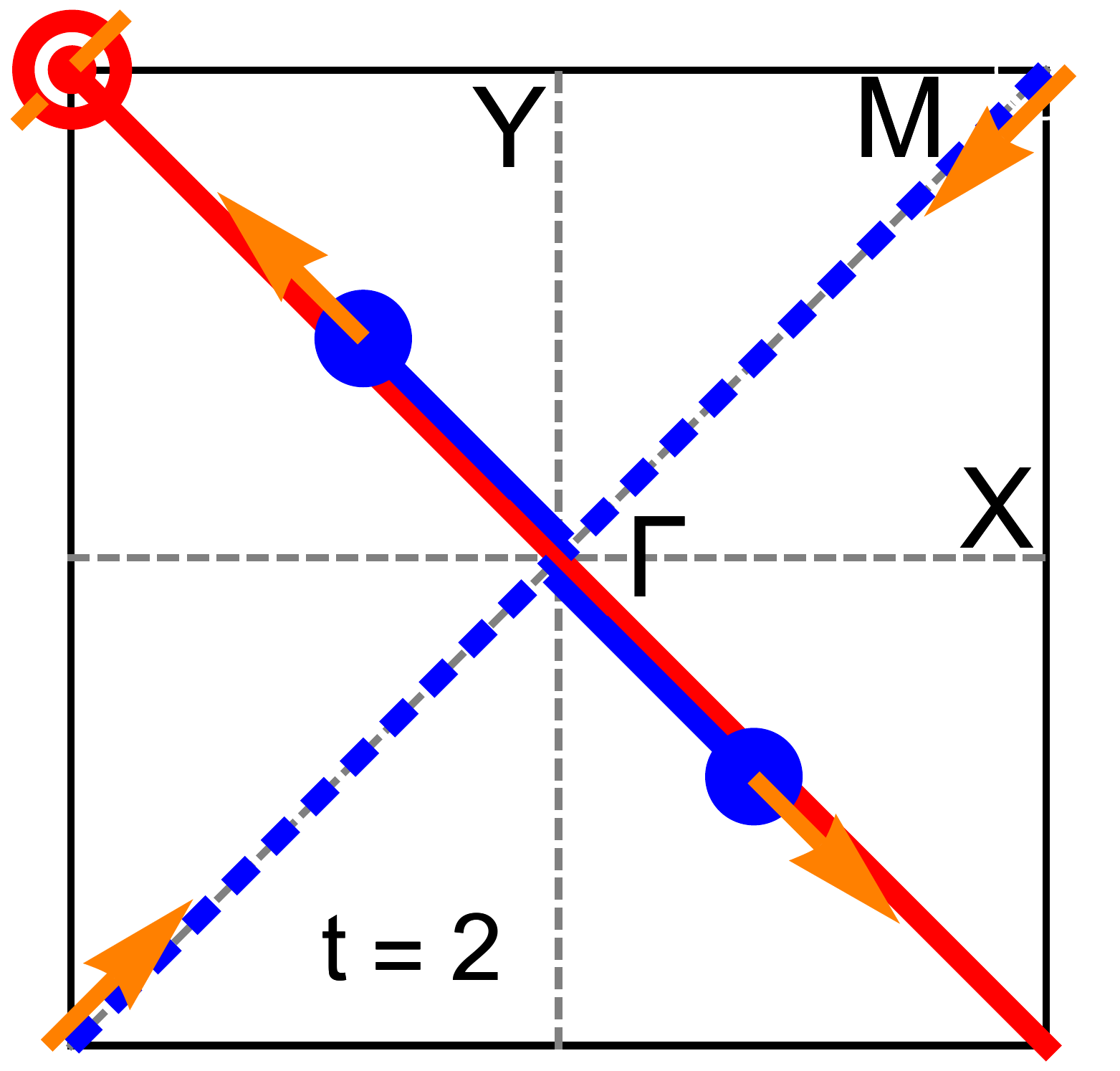} \\
         \includegraphics[width=0.18\linewidth]{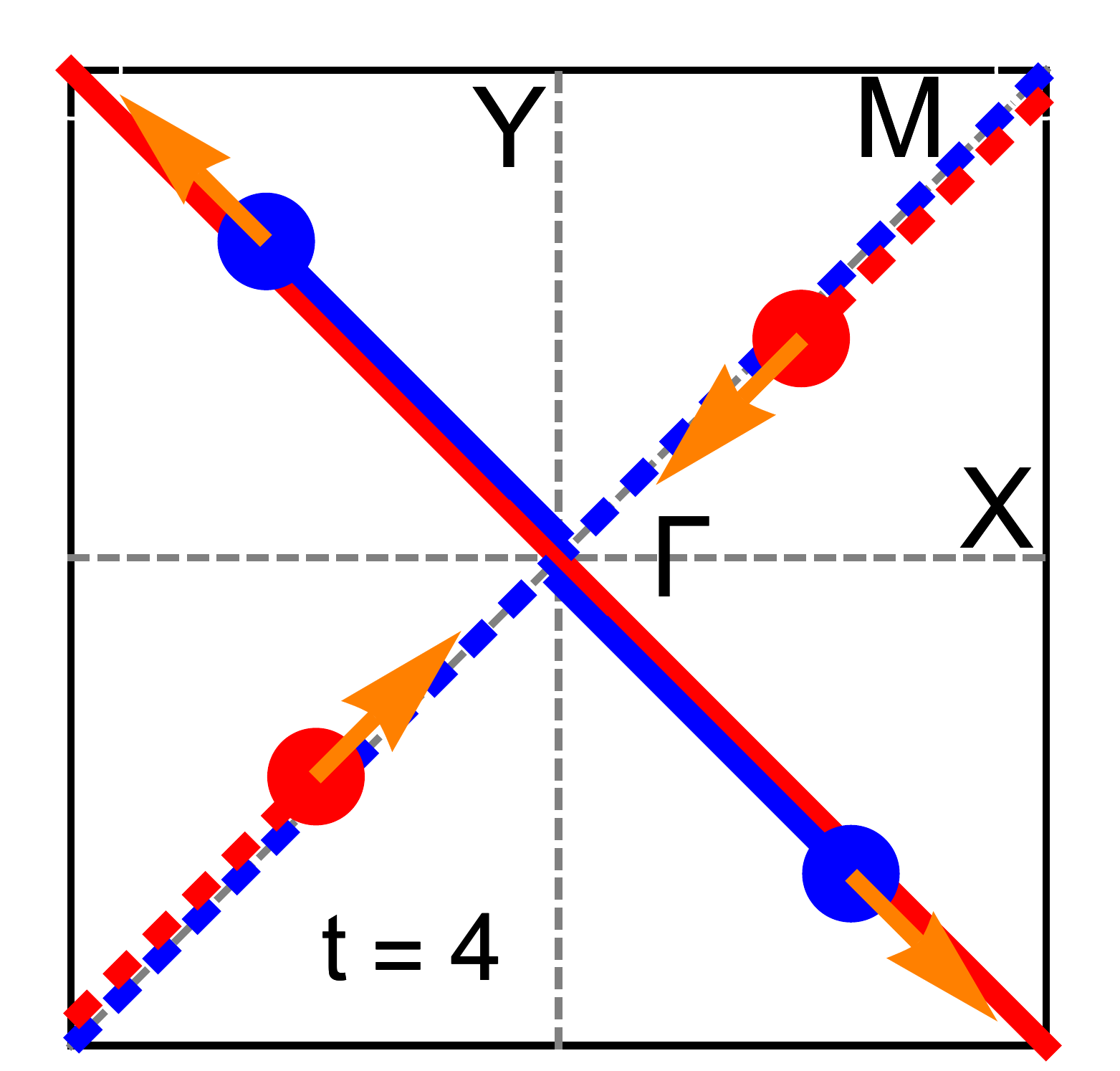}
     \end{tabular} 
 \end{tabular}
 \caption{\textsf{\textbf{Elementary protocol for braiding band nodes}. {\textbf a.} Control parameters of the braiding protocol as a function of adiabatic time $t$ [for details see Eq.~(\ref{eqn:Ham-matrix}) of Methods]. {\textbf b.} 2D band structure at $t=0$ where both principal and adjacent nodes coexist along the diagonal $(1\bar{1})$. {\textbf c.} Band structures along the two diagonals of the Brillouin zone (BZ), i.e.~$(1\bar{1})$ (full lines) and $(11)$ (dashed), at successive instants of the parameter $t\in[-10,10]$. The orange arrows indicate the motion of the nodes upon increasing time. {\textbf d.} Schematic configuration of the nodal points over the 2D BZ with their past trajectory at a few instants (the dashing matches with {\textbf c}). At $t=-2$, the principal nodes meet at $\Gamma$ but fail to annihilate. At $t=2$, the adjacent nodes meet at M without annihilating each other. (A supplementary model where the braided band nodes never cross the BZ boundary is presented in Sec.~B of SI).}}
\label{fig:protocol} 
 \end{figure*}

\sectitle{Non-Abelian topology}
The path-dependent capability of band nodes to annihilate is a consequence of underlying non-Abelian band topology~\cite{Wu:2018b,Tiwari:2019,Ahn:2019,Ahn:2018}. Before delving into its formal mathematical description, we attempt to visualize the non-Abelian obstruction in simple terms. Note that for two-dimensional systems $C_{2}\mcT$ symmetry implies the existence of a basis in which the Bloch Hamiltonian $H(\bs{k})$ is a real symmetric matrix [for proof see Sec.~C of Supplementary Information (SI)]. 

For momenta where the energy bands are non-degenerate, we form an energetically ordered set of $N$ \emph{real} Bloch states, $\{\ket{u^j(\bs{k})}\}_{j=1}^N =: {F}(\bs{k})$, which constitutes an \emph{orthonormal frame}~\cite{Wu:2018b}. The frame is well defined only up to the $\pm$ sign of each eigenstate, implying a gauge degree of freedom. This coincides with the gauge description of the order-parameter space of biaxial nematics~\cite{Prx2016}, which are known to exhibit disclination defects described by a non-Abelian group~\cite{Poenaru:1977,Volovik:1977,Madsen:2004,Alexander:2012}.

The constructed frames allow us to assign a \emph{frame-rotation charge} to each closed path that avoids band nodes. If one varies the momentum along a closed path based at $\bs{k}_0$, the Hamiltonian returns to its original form. Nonetheless, the initial and the final frame at $\bs{k}_0$ may differ by a gauge transformation, altering the vielbein spanning the frame. Notably, such a transformation occurs if one encircles a band node. As one moves along a tight loop around the node formed by a pair of bands, then the two Bloch states describing those bands perform a $\pi$-rotation, while the other states are essentially constant.   

Given two nodes inside the same band gap, one may wonder how their associated frame rotations compose. One possibility is that the second rotation undoes the first, e.g.~if we rotate by $\pi$ in the reverse direction. In that case the two nodes annihilate when brought together. Alternatively, the rotations could revolve in the same direction. Although the total $2\pi$-rotation looks like a do-nothing transformation, the Dirac's belt trick~\cite{Francis:1994} reveals that a $2\pi$-rotation cannot be trivially undone (while a $4\pi$ rotation can). Mathematically, this corresponds to the non-trivial fundamental group $\pi_1[\mathsf{SO}(N)] = \ztwo$ for $N>2$. Physically, this implies that a pair of nodes associated with a $2\pi$ frame rotation cannot annihilate~\cite{Sjoeqvist_2004}. 

For the elementary braiding Hamiltonian [Eq.~(\ref{eqn:Ham-matrix}) in Methods], we study in Fig.~\ref{fig:rotationcharge} the accumulated frame-rotation angle on two paths that enclose a pair of principal nodes. We decompose the 3D rotation matrix along the path using the rotation generators $L_{x,y,z}$ as $F(\bs{k})^\top \cdot F(\bs{k}_0) = \textrm{exp}[\alpha L_x + \beta L_y + \gamma L_z]$, and define the rotation angle $\varphi \!=\! \sqrt{\alpha^2 + \beta^2 + \gamma^2}$. As expected, we find that the total rotation angle equals $0$ ($2\pi$) if the nodes can (cannot) annihilate. The difference for the two paths originates from non-commutativity of rotations, namely the rotation angle $\alpha$ acquired as one circumnavigates a principal node is \emph{reversed} after conjugation with the overall $\pm \pi$ rotation associated with the adjacent node ($\e{\pi L_j}\e{\alpha L_i} \e{\pi L_j} \! =\!\e{-\alpha L_i}$ for $i\neq j$). As a result, the topological charge of principal nodes \emph{anticommutes} with the topological charge of adjacent nodes. This property has been modelled by the non-Abelian quaternion group $\mathsf{Q}=\{\pm 1,\pm\imi,\pm\imj,\pm\imk\}$~\cite{Wu:2018b} (indicated in Fig.~\ref{fig:braiding}{\bf b}) resp.~using Dirac strings~\cite{Ahn:2019}. We show the equivalence of the two approaches in Sec.~G of SI. 

\begin{figure}[t]
\centering
 \includegraphics[width=0.98\linewidth]{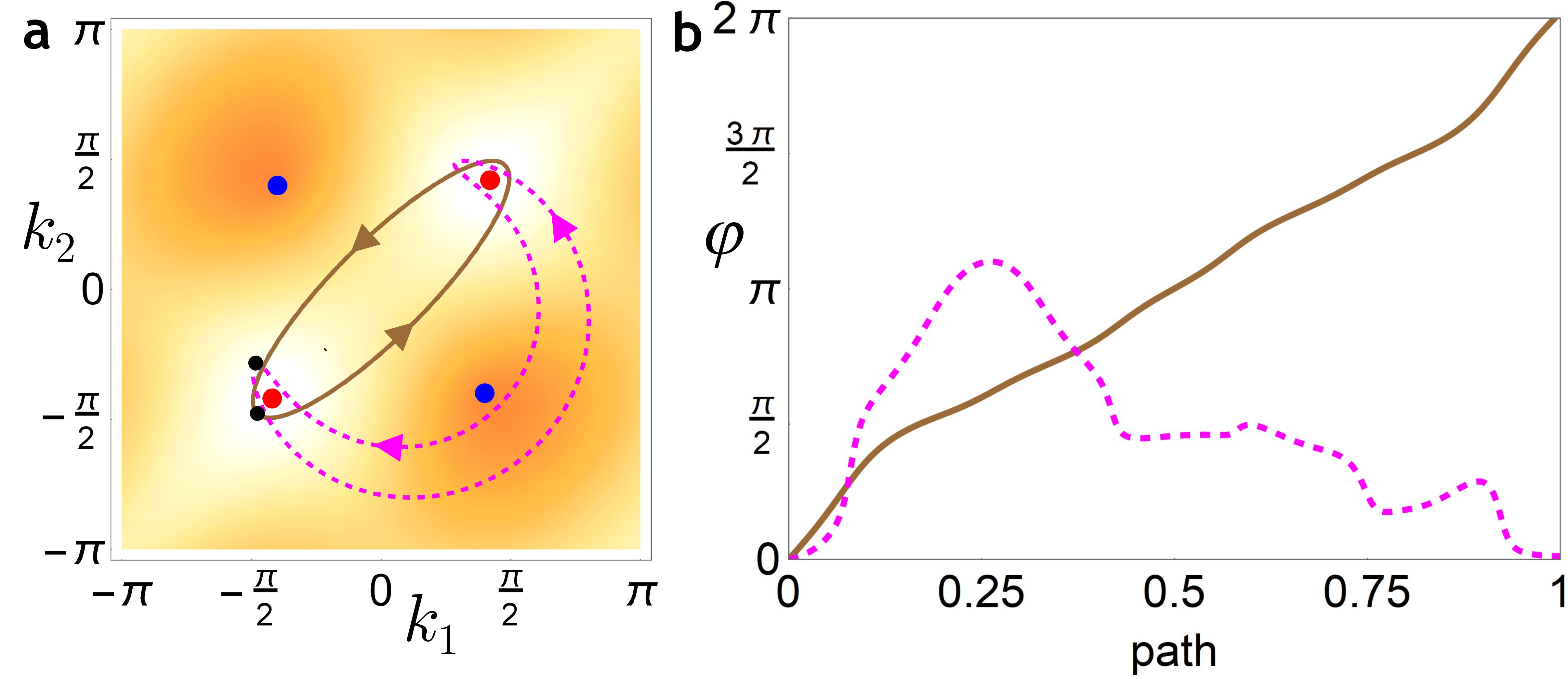}
\caption{\textsf{\textbf{Frame-rotation charge.} \textbf{a.} The blue (red) points indicate the location inside the Brillouin zone of the principal (adjacent) nodes of the model in Eq.~(\ref{eqn:Ham-matrix}) for $t=-4$, cf.~Fig.~\ref{fig:protocol}\textbf{d}. The shades of orange on the background indicate the magnitude of the principal gap (white denotes gapless points, darker shades mean a larger gap). The oriented solid brown resp.~dashed purple lines indicate two trajectories to enclose the principal nodes, with base points $\bs{k}_0$ marked with black dots. \textbf{b.} The accumulated frame-rotation angle along the two trajectories in panel \textbf{a}.
}}
\label{fig:rotationcharge}
 \end{figure}


\sectitle{Euler class} 
While the frame rotations computed in Fig.~\ref{fig:rotationcharge} faithfully predict the ability of band nodes to annihilate, the method is computationally too costly for many-band models. Instead, to study real materials, we utilize a tool introduced by Ref.~\cite{Ahn:2019} to describe a fragile-topological phase of twisted bilayer graphene near the magic angle~\cite{Cao_2018,Bouhon2018_fragile,Ashvin_2019,Song2019}. If $\ket{u^1(\bs{k})}$ and~$\ket{u^2(\bs{k})}$ are real Bloch states of a pair of bands, their Euler form is~\cite{Zhao:2017}
\begin{equation}
\textrm {Eu}(\bs{k}) = \bra{\bs{\nabla} u^1 (\bs{k})} \times \ket{\bs{\nabla} u^2 (\bs{k})},\label{eqn:Eu-form-def}
\end{equation}
and the integral of Euler form over a closed surface defines integer topological invariant called \emph{Euler class}~\cite{Ahn:2018}. Below, we recast this notion to describe \emph{stable} topology of band nodes in many-band models, while also assuming an unconventional partitioning of energy bands. We first examine the geometric meaning of this mathematical object, not clarified by the earlier works~\cite{Ahn:2019,Ahn:2018}, and apply it to band node analysis in two dimensions. In the next section we generalize to 3D systems, and discuss the interplay of Euler class with point-group symmetry.

The concepts of Euler form and Euler class can be understood as refinements of Berry curvature and of Chern numbers (a detailed exposition appears in Methods). To reveal their relation, consider a two-band complex Hamiltonian $H(\bs{k}) \!=\! \bs{h}(\bs{k})\!\cdot\!\bs{\sigma}$, where $\{\sigma_i\}_{i=1}^3$ are the Pauli matrices and $\bs{h}(\bs{k})$ is a three-component real vector. The integral of Berry curvature over an infinitesimal domain $\de k_1 \de k_2$ can be expressed~\cite{HasanKane10_RMP} as one half of the solid angle
\begin{equation}
\de \Omega = \bs{n}\cdot \left(\partial_{k_1} \bs{n} \times \partial_{k_2} \bs{n}\right)\de k_1 \de k_2\label{eqn:solid-angle}
\end{equation}
covered by unit vector $\bs{n}(\bs{k})\!=\!\bs{h}(\bs{k})/\norm{\bs{h}(\bs{k})}$ as $\bs{k}$ ranges over the domain, cf.~Fig.~\ref{fig:unit-spheres}{\bf a}. If momentum ranges over a \emph{closed} manifold, then $\bs{n}$ wraps around the unit sphere an integer number of times. Berry curvature thus integrates to integer multiples of $2\pi$, defining the Chern number. The theory of characteristic classes~\cite{Milnor:1975} predicts that the quantization persists in models with more than two band. 

 \begin{figure}[t]
 \centering
 \includegraphics[width=0.88\linewidth]{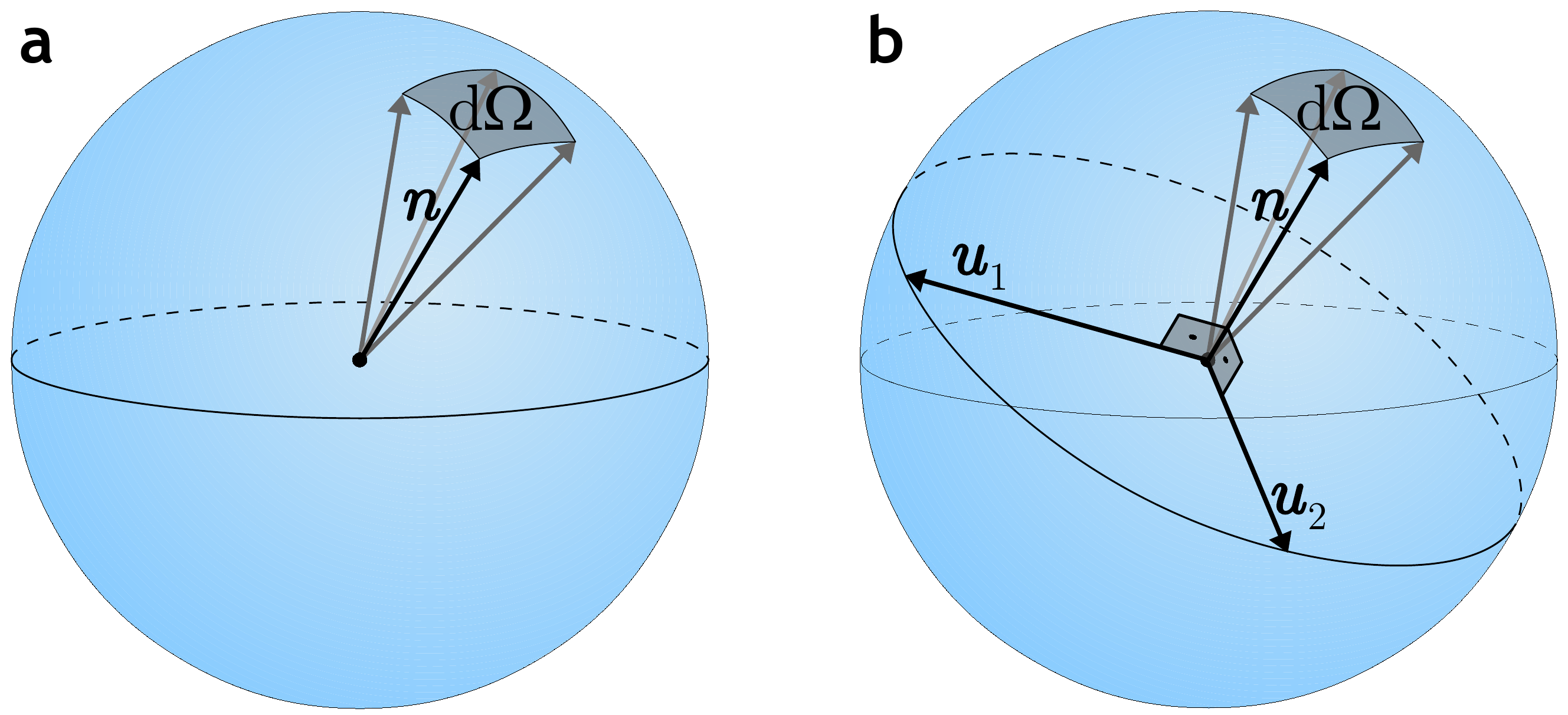}
\caption{\textsf{\textbf{Berry curvature vs.~Euler form.} \textbf{a.} For a complex two-band  Ha\-mil\-to\-nians, Berry curvature is equal to half of the solid angle $\de\Omega$ spanned by the unit vector $\bs{n}$ that encodes the Hamiltonian using the Pauli matrices. \textbf{b.} Similarly, for three-band real Hamiltonians, Euler form over two bands $\bs{u}_{1,2}$ corresponds to the solid angle spanned by the unit vector $\bs{n} = \bs{u}_1 \!\times\! \bs{u}_2$.}}
\label{fig:unit-spheres}
 \end{figure}

 \begin{figure*}[t]
 \centering
 \includegraphics[width=0.91\linewidth]{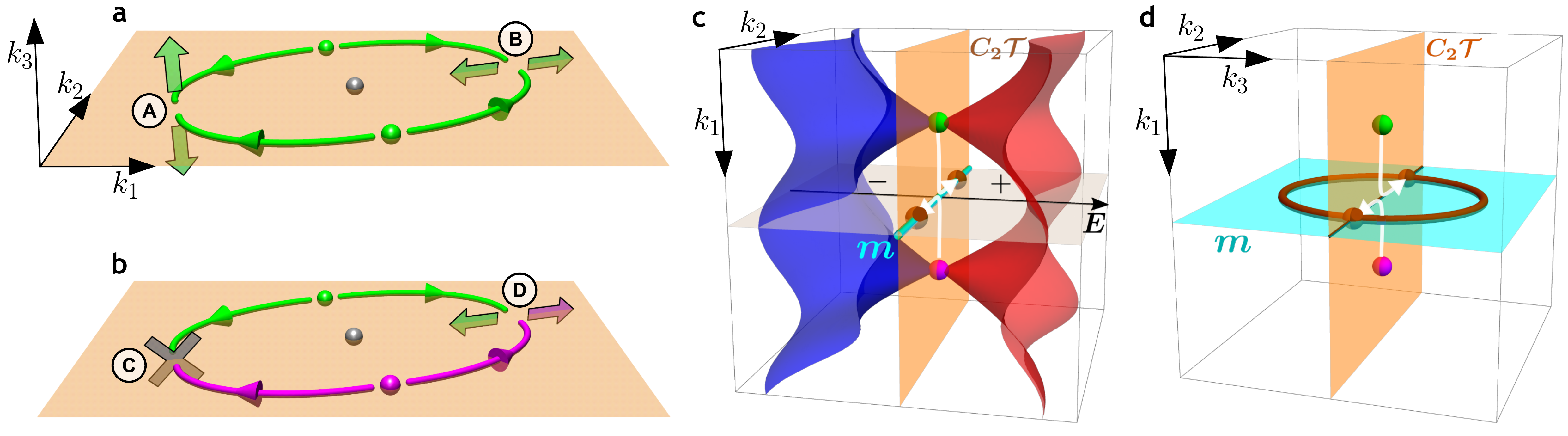}
 \caption{\textsf{\textbf{Conversion of Weyl points (WPs) in 3D momentum space.} \textbf{a.} Two principal WPs (spheres) with \emph{equal} chirality (green color), pinned to plane $k_3 \!=\! 0$ (orange sheet), can be brought together on two sides of an adjacent WP (grey sphere). After their collision, the two principal WPs can either symmetrically leave the plane (\textbf{A}), or remain pinned inside the plane (\textbf{B}), depending on their Euler class (i.e.~the frame-rotation charge). \textbf{b.} Analogous situation involving two principal WPs with \emph{opposite} chirality (green vs.~magenta color). Depending on their Euler class, the WPs either annihilate (\textbf{C}) or remain pinned inside the plane (\textbf{D}) after colliding. \textbf{c.} In the presence of an additional mirror ($m$) symmetry (cyan line), the scenario (\textbf{D}) corresponds to bouncing (white arrows) of two $m$-related WPs to band nodes (brown spheres) lying inside  $m$-invariant plane. This scenario occurs when the two principal bands (blue and red sheet) carry opposite eigenvalues of $m$ (indicated by $+$ and $-$). \textbf{d.} The nodal manifold after the collision extends into a nodal line (NL, brown line) inside the $m$-plane (cyan sheet), which is stabilized by the mirror $m$ symmetry. The reverse process, i.e.~producing a pair of WP by colliding two NLs inside $m\cap C_2\mcT$, is controlled by the Euler class, see~Fig.~\ref{fig:zrte}\textbf{e} for a material example.
}}
\label{fig:braiding3D}
 \end{figure*}

In analogy, the simplest scenario with a non-trivial Euler class of a pair of bands occurs in three-band models. We find (see Fig.~\ref{fig:unit-spheres}{\bf b} and Sec.~D of SI), that the integral of the Euler form over an infinitesimal domain $\de k_x \de k_y$ is equal to the solid angle $\de \Omega$ in Eq.~(\ref{eqn:solid-angle}) (without the $1/2$ prefactor), where $\bs{n} \!=\! \bs{u}_1 \!\times\! \bs{u}_2$ is the cross product of the two Bloch states. Using arguments as before, one concludes that for three-band models the Euler form on closed manifolds integrates to integer multiples of $4\pi$. Although this simple geometric interpretation of the Euler form becomes insufficient in many-band models, it again follows from the theory of characteristic classes~\cite{Milnor:1975} that the quantization to integer multiples of $2\pi$ persists, defining an integer topological invariant akin to the Chern number. We show in Methods that the Euler form of two real states $\ket{u^1}$ and $\ket{u^2}$ equals to the Berry curvature of complex state $\left(\ket{u^1}{+}\imi\ket{u^2}\right)/\sqrt{2}$, thus enabling an efficient computational implementation~\cite{Fukui:2005} (see Sec.~H of~SI).

\sectitle{Node diagnosis and band partitioning}
The gauge degree of freedom may alter the overall sign of the Euler form. Consequently (see Secs.~C and~D of SI), $\textrm{Eu}(\bs{k})$ can be defined around principal nodes, but not around adjacent nodes. Therefore, to predict whether a pair of principal nodes annihilate when brought together along a specified trajectory, we propose the following strategy, which generalizes the methods of Ref~\cite{Ahn:2019}. We choose a region (disc) $\mathcal{D}$ that (\emph{i}) contains the trajectory, and that (\emph{ii}) does not contain any additional principal nor adjacent nodes. On the boundary $\partial\mathcal{D}$ we construct the \emph{Euler connection} ${\textbf{a}}(\bs{k}) \!=\! \braket{u^1(\bs{k})}{\bs{\nabla}u^2(\bs{k})}$. We define \emph{Euler class over $\mathcal{D}$},
\begin{equation}
\chi(\mathcal{D}) = \frac{1}{2\pi}\left[\int_\mathcal{D} \textrm{Eu}(\bs{k})\, \de k_1 \de k_2 - \oint_{\partial\mathcal{D}} \textbf{a}(\bs{k})\cdot\de\bs{k}\right],\label{eqn:Euler-Stokes}
\end{equation}
which is an integer topological invariant whenever the disc $\mathcal{D}$ contains an \emph{even} number of principal nodes and no adjacent nodes (for further details see Sec.~F of SI). Crucially (as elaborated in Sec.~E of SI), the real Bloch states pertaining to the principal bands exhibit a singularity at principal nodes, which invalidates the Stokes' theorem and prevents the cancellation of the two integrals in Eq.~(\ref{eqn:Euler-Stokes}).

If the principal nodes inside $\mathcal{D}$ are able to collectively annihilate, then $\chi(\mathcal{D})$ must be zero. This is because annihilating all the nodes makes the Euler form exact in terms of the Euler connection, i.e. $\textrm{Eu} = \boldsymbol{\nabla}\times\textbf{a}$ in $\mathcal{D}$, in which case the Stokes' theorem guarantees cancellation of the two integrals. Conversely, non-vanishing $\chi(\mathcal{D})$ indicates an obstruction for annihilating the principal nodes. We confirm this for the model in Eq.~(\ref{eqn:Ham-matrix}) in Sec.~H of SI using a computational algorithm detailed therein. 

We have also developed a complementary algorithm that allows us to find the Euler class of a collection of principal nodes using the increasingly appreciated paradigm of Wilson-loop flows, which we discuss in Methods and illustrate in Extended Data Fig.~\ref{two_flows}.

For many-band models, such as when modelling real materials, we require the two principal bands to be separated inside $\mathcal{D}$ by an energy gap on \emph{both} sides, i.e.~from above and from below. Such partitioning of energy bands into three groups is in contrast with the contemporary paradigm in topological band theory~\cite{Kitaev09_AIP, Ryu10_NJP}, which is to partition the bands as occupied vs.~unoccupied via a single energy gap. This distinction explains why the stable integer invariant in Eq.~(\ref{eqn:Euler-Stokes}) has been overlooked by previous works classifying the band node topology~\cite{Bzdusek:2017,Fang:2016}.

 \begin{figure*}[t]
 \centering
 \includegraphics[width=0.9\linewidth]{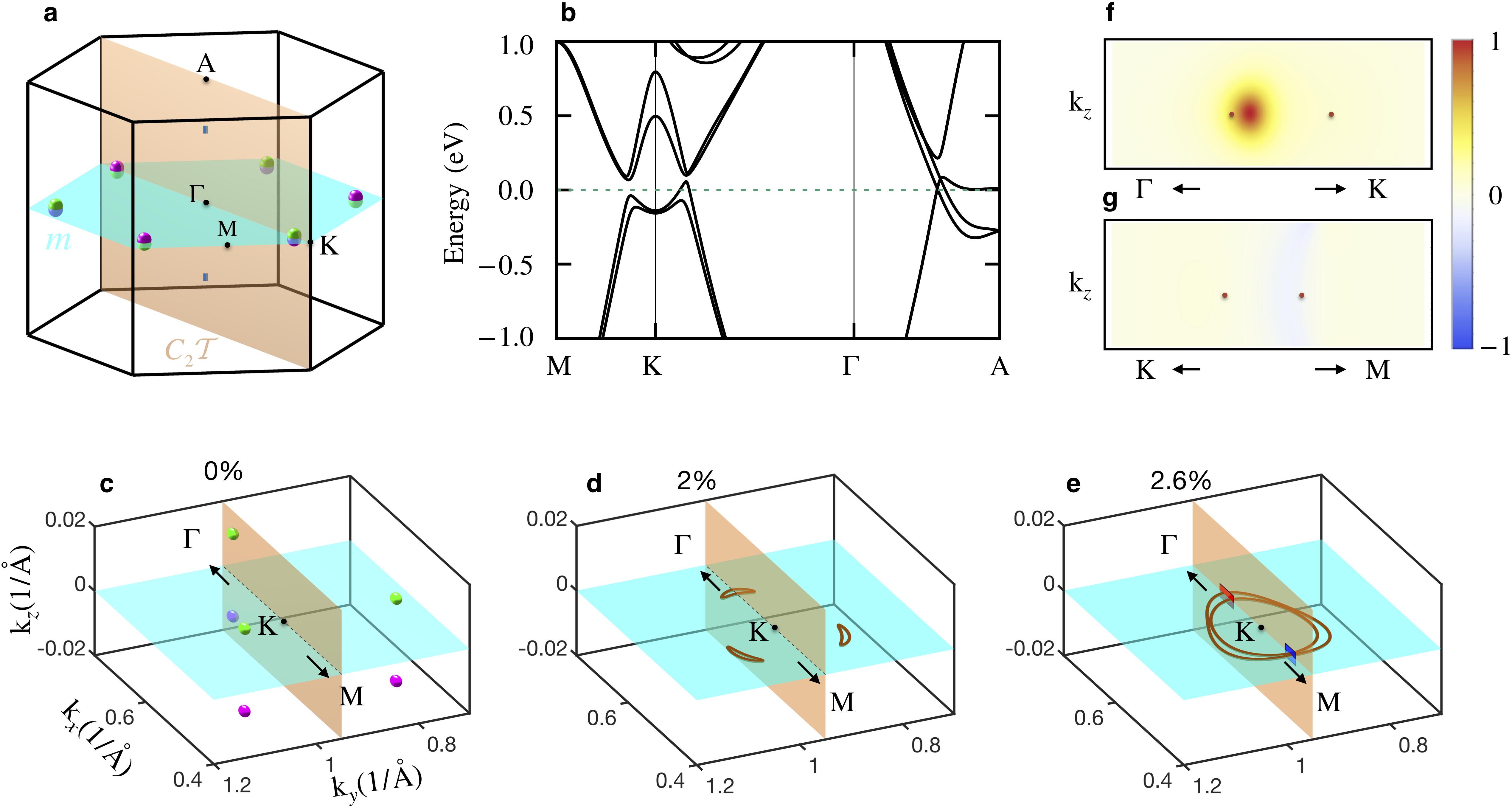}
\caption{\textsf{\textbf{Weyl points and nodal lines in ZrTe.} \textbf{a.} The Brillouin zone (BZ) and the principal band nodes of ZrTe under ambient conditions. Weyl points (colored green resp.~magenta according to their chirality) appear in pairs related by horizontal mirror $m$-plane (cyan), and lie within three vertical $C_2\mcT$-planes (one shown in orange). Four triple points, located along the $\Gamma$-$\rm A$ direction, constitute endpoints of two narrow nodal lines (blue). \textbf{b.} Band structure of ZrTe under ambient conditions. \textbf{c.} A zoom-in view of panel $\textbf{a}$ around $\rm K$ point. \textbf{d} and \textbf{e.} Nodal lines (brown) of ZrTe under 2\% and 2.6\% [001]-uniaxial compression strain inside the same region of BZ. \textbf{f} and \textbf{g.} Numerically computed Euler form, in arbitrary units as indicated by the bar diagram, inside the red resp.~the blue rectangular region within the $C_2\mcT$-plane as displayed in panel \textbf{e}. The Euler class [combined surface and boundary integrals in Eq.~(\ref{eqn:Euler-Stokes})] equals 1 (non-trivial) resp.~0 (trivial) for the two regions. (For additional first-principles data see Methods and Extended Data.)}}
 \label{fig:zrte} 
 \end{figure*}

\sectitle{Non-Abelian conversions in 3D}
We now shift our attention to three spatial dimensions. It is understood that $C_{2}\mcT$ symmetry can stabilize Weyl points (WPs) \emph{inside} high-symmetry ``$C_2\mcT$-planes'', as observed e.g.~in the $k_z \!=\! 0$ plane of $\textrm{WTe}_2$~\cite{Soluyanov:2015}, $\textrm{MoP}$~\cite{Lv2017} and $\textrm{TaAs}$~\cite{Weng:2015,Lv:2015,Xu:2015}. While such WPs are characterized by their chiral charge~\cite{Wan:2011}, the $C_2\mcT$ symmetry assigns them an extra frame-rotation charge defined by the Hamiltonian inside the symmetric plane. Importantly, these charges carry complementary pieces of topological information. While the chiral charge discloses whether a pair of WPs can annihilate, the frame-rotation charge conveys whether the two WPs can disappear from the 
$C_2\mcT$-plane.

Four scenarios are possible, as illustrated in Fig.~\ref{fig:braiding3D}{\bf a--b} which displays the braiding of two principal WPs (marked in green vs.~magenta to distinguish their chirality) around an adjacent node (gray). Starting with two WPs of equal chirality within the $C_2\mcT$-plane, they can either leave the plane ({\bf A}) or bounce within the plane ({\bf B}) when collided. Considering instead two principal WPs of opposite chirality, they can either annihilate ({\bf C}) or bounce ({\bf D}). In cases ({\bf B}) and ({\bf D}), the colliding WPs carry a nontrivial frame-rotation charge (i.e.~Euler class), which obstructs their disappearance from the $C_2\mcT$-plane, irrespective of their chirality. In cases ({\bf A}) and ({\bf C}), the frame-rotation charge is trivial, allowing the pair of principal WPs to disappear from the plane.

\sectitle{Non-Abelian Weyl points with mirror symmetry}
Additional crystalline symmetries may interact non-trivially with the non-Abelian band topology. Here, we consider the presence of a mirror symmetry $m$, which facilitates a mirror-invariant $m$-plane (cyan in Fig.~\ref{fig:braiding3D}\textbf{d}) perpendicular to the $C_2\mcT$-plane (orange in Fig.~\ref{fig:braiding3D}\textbf{d}). Recall that mirror symmetry can stabilize nodal-line degeneracies of bands with opposite mirror eigenvalues~\cite{Fang:2016, Bzduvsek16nodal}. 

Let us consider a WP (green in Fig.~\ref{fig:braiding3D}\textbf{c}) stabilized inside $C_2\mcT$-plane on one side of the mirror. Such a WP has a mirror-related partner of opposite chirality (magenta in Fig.~\ref{fig:braiding3D}\textbf{c}) on opposite side of the $m$-plane. Moving the two WPs together locally inverts the two bands along the intersection ${m}\!\cap \! {C_2\mcT}$ of the two planes. We distinguish two possibilities. (1) If the two bands have the same mirror eigenvalue, they hybridize inside the $m$-plane, resulting in an avoided crossing, and the two WPs annihilate upon collision. (2) If the two bands have opposite mirror eigenvalue, they cannot hybridize. Their crossing is stable, implying that the two WPs convert to a NL~\cite{Sun:2018}. Note that the resulting NL still crosses the $C_2\mcT$-plane at two points (brown dots in Fig.~\ref{fig:braiding3D}\textbf{c}), which corresponds to the ``bouncing'' of two nodes inside the ${C_2\mcT}$-plane (white arrows in Fig.~\ref{fig:braiding3D}\textbf{d}) as described above as ({\bf D}). Curiously, the fate of the reverse process, i.e.~two NLs colliding inside $m \cap C_2\mcT$, does not uniquelly follow from the relative mirror eigenvalue of the principal bands. Nevertheless, we find that the outcome of such collisions can be predicted by computing the Euler class of the two NLs passing through the $C_2\mcT$-plane.

\sectitle{Node conversion in \textbf{ZrTe}}
Zirconium telluride (ZrTe, space group \#187) material class includes triple-point metals with a non-saturating quadratic magnetoresistance~\cite{Zhu:2016,Weng:2016,He:2017,Ma:2018}. Under ambient conditions, the band structure of ZrTe further 
exhibits 6 pairs of WPs (green and magenta in Fig.~\ref{fig:zrte}\textbf{a}) related by mirror ($m$) symmetry (cyan sheet in  Fig.~\ref{fig:zrte}) near the $\textrm{K}$ point of BZ, all of them lying inside vertical $C_2\mcT$-planes~\cite{Weng:2016} (orange sheet in Fig.~\ref{fig:zrte}\textbf{a}). The WPs are located only $50\,\textrm{meV}$ above the Fermi level, which could possibly be further lowered by doping. These properties make ZrTe an ideal platform to study the interplay of the non-Abelian band topology with crystalline symmetry. 

By performing high-precision first-principles computation including hybrid functional HSE06~\cite{vasp,wannier90,wanniertools} (see Methods and Extended Data Figs.~\ref{fig:mirror-bs} and~\ref{fig:zrte-pbe-hse06}), we find that a $2\%$ compressive uniaxial strain in $z$-direction brings the pairs of WPs of ZrTe together at the $m$-plane. Although experimentally challenging, the relatively low Young modulus ($122\,\textrm{GPa}$ in the $z$-direction~\cite{Guo:2017}) may allow for angle-resolved photoemission spectroscopy (ARPES) of ZrTe under large strains. The relative mirror eigenvalue of the two bands forming the WPs are opposite (see Extended Data Fig.~\ref{fig:mirror-bs}), hence the colliding WPs convert into NLs (brown lines in Fig.~\ref{fig:zrte}\textbf{d}). Further increasing the strain to $2.6\%$ fuses three elongated NLs into two concentric NL rings (Fig.~\ref{fig:zrte}\textbf{e}).

The reverse process, i.e.~relaxing the strain of compressed $\textrm{ZrTe}$, exhibits collisions of two NLs within the $C_2\mcT$-plane along both the $\Gamma\textrm{K}$ and the $\textrm{KM}$ high-symmetry lines. However, only the collisions along $\Gamma\textrm{K}$ eject pairs of WPs, while the collisions along $\textrm{KM}$ do not. We compare the two situations by computing the Euler class~(\ref{eqn:Euler-Stokes}) inside the red resp.~the blue rectangular regions shown in Fig.~\ref{fig:zrte}\textbf{e}, each traversed by two nodal lines, using the algorithm outlined in Sec.~H of SI. We find that the ejection of WPs from colliding NLs corresponds to a non-trivial Euler class on the rectangular region, cf.~Fig.~\ref{fig:zrte}\textbf{f} and~\textbf{g}, consistent with our theoretical predictions.

\sectitle{Further material candidates}
We find that ZrTe is not the only WC-type material that supports WPs with a non-trivial Euler class. Through first-principles calculations, we confirm that WC-type MoP and NbS also exhibit WPs under ambient conditions. Among them, the WPs of MoP (plotted in Extended Data Fig.~\ref{fig:zrte-mop}) were observed with ARPES~\cite{Lv2017}. We further find that there are NLs in ambient MoC and MoN. Those nodal lines can be converted into WPs by appropriate pressure. However, these four materials are not ideal candidates to validate our theory because (\emph{i}) their Young modulus is larger than that of ZrTe, and (\emph{ii}) the WPs and NLs are located about $1\,\textrm{eV}$ above or below the Fermi level. Notably, the pairs of WPs in the $C_2\mcT$-invariant plane of TaAs, which are experimentally well established~\cite{Lv:2015,Xu:2015}, also carry nontrivial Euler class (see Extended Data Figs.~\ref{fig:taas-nodalconversion} and~\ref{fig:taas-eulercurvature}). Additional first-principles data for all the mentioned materials appear in Extended Data Fig.~\ref{fig:9materials} and are discussed in Methods.

\sectitle{Other experimental realizations}
While the presented data for the ZrTe material class underpin the predictive power of our analysis, we also mention other possible general directions. In particular, we anticipate that recent progress in the analysis of van-der-Waals heterostructures should entail an interesting research avenue. Studies into twisted bilayer graphene~\cite{Cao_2018} have identified the existence of non-trivial band insulators in terms of the above characterization under $C_2\mcT$ symmetry~\cite{Ashvin_2019, Ahn:2019,Bouhon2018_fragile,Song2019}. The flexibility of the stacking direction (also under stress, strain and voltage/potential differences for e.g~the $p_z$ band~\cite{Ashvin_2019}) could potentially be utilized to realize the reciprocal braiding phenomenon in such systems. Regarding explicitly three-dimensional Weyl semimetals, we remark the pioneering experimental efforts to move the WPs of $\textrm{WTe}_2$ through $\bs{k}$-space by driving an optical ``shear'' phonon mode~\cite{Sie:2019}. 

Apart from the solid-state route, given the concrete nature of the braiding protocol presented in Fig.~\ref{fig:protocol}, another promising direction to access this physics entails implementation in cold-atom or photonic systems. Indeed, recent experimental studies have reported techniques to measure geometric concepts related to Euler class, such as the Wilson phases~\cite{Li_2016}. Furthermore, upon the influence of an external force, elements of the Berry-Wilczek-Zee connection~\cite{Wilczek:1984} can be obtained, and the reconstruction of Berry curvature using tomography has been experimentally achieved~\cite{Flaschner_2016}. Similarly, photonic systems offer a platform to implement our scheme. Models with three or even four momentum dimensions are at present routinely implemented in topological photonics~\cite{Lu:2013,Lu:2015,Zilberberg_2018}, opening a possibility to emulate the interplay of the non-Abelian band topology with 3D point-group symmetry analogous to the ZrTe material class.

\sectitle{Conclusions}
We presented an elementary protocol that elucidates a non-Abelian braiding effect for momentum-space band nodes, and we discussed three mathematically rigorous ways to characterize these phenomena based on frame rotations~\cite{Wu:2018b}, Eules class~\cite{Ahn:2019}, resp.~Wilson loops~\cite{Bouhon2018_fragile}. We emphasize that the novel vantage point reached by considering different partitionings of bands via two energy gaps into three groups, namely (\emph{i}) many occupied, (\emph{ii}) two principal, and (\emph{iii}) many unoccupied bands, opens up an avenue to new theoretical as well as experimental studies, as it allows for more intricate topological structures. Importantly, we considered the interplay of the non-Abelian topology with point-group symmetry in 3D, uncovering novel phase transitions of Weyl points in $C_2\mcT$-symmetric systems. The discussed models and phenomena are within current experimental reach, and the predictive power of our framework is reflected in specific material predictions.

\sectitle{Online content} 
The \textsf{\textbf{Methods}} section contains additional information on (\emph{i}) the elementary braiding Hamiltonian, (\emph{ii}) topology of complex and real vector bundles, (\emph{iii}) the Wilson-flow algorithm, and (\emph{iv}) the first-principles calculations. The \textsf{\textbf{Supplementary Information}} (SI) file contains information about (\emph{a}) $C_2\mcT$-symmetric tight-binding models, (\emph{b}) $\bs{k}$-local tight-binding braiding protocol,  (\emph{c}) the reality condition in $C_2\mcT$-symmetric models, (\emph{d}) geometric interpretation of Euler form in three-band models, (\emph{e}) behavior of Euler form at principal nodes, (\emph{f}) Eq.~(\ref{eqn:Euler-Stokes}) for manifolds with a boundary, (\emph{g}) relation between Euler class and frame-rotation charge, and (\emph{h}) the Euler-form integration algorithm. The \textsf{\textbf{Extended data}} contain one figure illustrating the Wilson-flow algorithm, and six figures with additional first-principles data for ZrTe, for other WC-type materials, and for TaAs. 


\sectitle{References}
\vspace{-0.4cm}
%


\onecolumngrid
\pagebreak
\phantom{
\color{white}
\pagebreak}
\twocolumngrid

\fontsize{8}{10}\selectfont

\sectitle{Methods}
\vspace{-0.5cm}

\partitle{Elementary braiding Hamiltonian.} As an illustrative model for node braiding in two dimensions, we consider in the main text the Hamiltonian [for more details on the model construction see Sec.~A of the Supplementary Information (SI)]
\begin{equation}
H(\bs{k};t) = \left(\begin{array}{ccc}
f(t)       &   g(\bs{k})      &   g^*(\bs{k})    \\
g^*(\bs{k})    &   0              &   h(\bs{k};t)      \\
g(\bs{k})      &   h^*(\bs{k};t)  &   0   
\end{array}\right)\label{eqn:Ham-matrix} 
\end{equation}
with on-site energy $f(t) = F_{8}^-(t) $, couplings $g(\bs{k}) = (\e{-\imi k_1 \pi}-\e{-\imi k_2\pi})$ and $h(\bs{k};t) = h_0(t) + h_1(t) (\e{\imi k_1 \pi}+\e{\imi k_2\pi})$. Here $h_0(t) = -F_{2}^-(t)$ and $ h_1(t) = [F_{8}^+(t)-10]$. The dependence on the tuning parameter $t\!\in\! [-10,10]$ is defined through $F_{\nu}^{\pm}(t) \!=\! \tfrac{1}{2}\left(\abs{t+\nu}\!\pm\!\abs{t-\nu}\right)$, which is a piecewise-linear function with shoulders at $\pm\nu$, see Fig.~\ref{fig:protocol}{\bf a}. 

The physical degree of freedom $\phi_A$ is an $s$-wave orbital symmetric under $C_{2z}$, and $\phi_{B,C}$ are two $p_z$-wave orbitals related to each other by $C_{2z}$ (all orbital wave functions are assumed to be real). The model is symmetric under time reversal $\mathcal{T}=\mathcal{K}$ (complex conjugation), under $\pi$-rotation $C_{2z} = 1 \!\oplus\! \sigma_x$ [permutation matrix $(123)\!\leftrightarrow\!(132)$], and under their composition $C_{2z} \mcT = (1 \!\oplus\! \sigma_x) \mcK$. The model is brought to a real-symmetric form by a unitary rotation $V\!\cdot\! H(\bs{k})\cdot V^\dagger$ where $V \!=\! \sqrt{1 \!\oplus\! \sigma_x}$. The model is also symmetric under $\pi$-rotations around in-plane axes, namely $C_{2,(11)} = 1\!\oplus\! \sigma_z$ and $C_{2,(1\bar{1})} = 1\!\oplus\! (-\sigma_x)$. These additional symmetry relates the two $p_z$-orbitals and imposes the motion of the nodal points along the BZ diagonals as observed in Fig.~\ref{fig:protocol}{\bf c}.

\partitle{Topology of general vector bundles.} Below, we outline the rigorous mathematical definition of Euler form and of Euler class for real vector bundles. To make the analogy with Berry curvature and Chern number manifest, we first review here the definitions and the properties of these more familiar objects. We adopt the language of differential forms~\cite{Nakahara:2003} as it allows for concise expressions of the studied objects and of the relations between them. 

We first recall the basic terminology. A collection of $n$ bands over a base space $B$ defines a rank-$n$ vector bundle $E\to B$, which is generically \emph{complex}. Here we assume that the vector bundle is \emph{smooth} (this assumption is lifted in Sec.~E of SI when discussing the analytic properties near band nodes). We order the states $\{\ket{u^a(\bs{k})}\}_{a=1}^n$ 
as columns into a rectangular matrix $\mathfrak{U}(\bs{k})$, and we construct the Berry-Wilczek-Zee (BWZ) connection~\cite{Wilczek:1984}
\begin{equation}
\mcA(\bs{k})=\mathfrak{U}^\dagger(\bs{k}) d \mathfrak{U}(\bs{k}),\label{eqn:BWZ-A-def}
\end{equation}
where ``$d$'' is the exterior derivative (i.e.~the differentiation $d k^i \partial_{k_i}$ followed by antisymmetrization in covariant indices~\cite{Nakahara:2003}). Mathematically, $\mcA(\bs{k})$ is a 1-form with values in Lie algebra $\mathfrak{u}(n)$, and it can be expressed componentwise as
\begin{equation}
\mcA^{ab}_i(\bs{k})=\braket{u^a(\bs{k})}{\partial_{k_i} u^b(\bs{k})},
\end{equation}
where $i$ is a momentum component (the 1-form part), and $a,b$ are band indices (the Lie algebra part). The BWZ connection is skew-Hermitian in band indices because the Lie algebra $\mathfrak{u}(n)$ corresponds to skew-Hermitian matrices. Mixing the $n$ states with a unitary matrix $X(\bs{k}) \in \mathsf{U}(n)$ as $\tilde{\mathfrak{U}}(\bs{k}) = \mathfrak{U}(\bs{k}) X(\bs{k})$ (i.e.~performing a \emph{gauge transformation}), transforms the connection as
\begin{equation}
\tilde{\mcA} = X^{\dagger} \mcA X + X^\dagger d X, \label{eqn:BWZ-conn-GT}
\end{equation}
where we dropped the momentum arguments for brevity. 

The BWZ \emph{curvature} is defined as
\begin{equation}
\mcF = d \mcA + \mcA\wedge \mcA, \label{eqn:BWZ-F-def}
\end{equation}
which is a 2-form with values in $\mathfrak{u}(n)$. Componentwise,
\begin{eqnarray}
\mcF^{ab}_{ij}&=&\braket{\partial_{k_i} u^a}{\partial_{k_j} u^b} - \braket{\partial_{k_j} u^a}{\partial_{k_i} u^b},\label{eqn:BWZ-F-componentwise} \\
&+&\!\braket{u^a}{\partial_{k_i} u^c}\braket{u^c}{\partial_{k_j}u^b} \!-\! \braket{u^a}{\partial_{k_j} u^c}\braket{u^c}{\partial_{k_i}u^b} \nonumber
\end{eqnarray}
which is skew-symmetric in momentum coordinates (the 2-form part), and skew-Hermitian in band indices (the Lie algebra part). The curvature transforms covariantly under gauge transformations, 
\begin{equation}
\tilde{\mcF} = X^{\dagger} \mcF X ,\label{eqn:BWZ-GT}
\end{equation}
which allows us to define a gauge-invariant object, $\mathsf{F} = -\imi \tr (\mcF)$, called \emph{Berry curvature}. The trace in this definition runs over band indices, i.e. we perform a projection $\mathfrak{u}(n)\to\mathfrak{u}(1)$. Assuming the Einstein summation convention, this amounts to 
\begin{equation}
\mathsf{F}_{ij} \!=\! -\imi \mcF^{aa}_{ij} \!=\! -\imi\left(\braket{\partial_{k_i} u^a}{\partial_{k_j} u^a} \!-\! \braket{\partial_{k_j} u^a}{\partial_{k_i} u^a}\right)\label{eqn:Bery-curv}
\end{equation}
where the two terms in the second line of Eq.~(\ref{eqn:BWZ-F-componentwise}) have cancelled each other. One can similarly define \emph{Berry connection} $A = \tr \mathcal{A}$. Since the expression $[\mcA\wedge\mcA]_{ij} = A_i A_j - A_j A_i$ in Eq.~(\ref{eqn:BWZ-F-def}) has zero trace, it follows that $\mathsf{F} = d A$.

\partitle{Topology of real vector bundles.} We now discuss Euler connection and Euler form of $C_2\mcT$-symmetric systems. We adopt the reality condition justified in Sec.~C of SI, and assume the real gauge for Bloch states $\ket{u^a(\bs{k})}$. Then $E\to B$ becomes a \emph{real} vector bundle. To preserve the reality, we consider only $\mathsf{O}(n)$ gauge transformations. As a consequence, the connection 1-form and the curvature 2-form take values in the orthogonal Lie group, $\mathfrak{so}(n)$. As these correspond to skew-symmetric matrices, the components of both of these objects are skew-symmetric in band indices. In particular, it follows that $\mathsf{F} = -\imi \tr \mcF = 0$, i.e.~Berry curvature of a real Hamiltonian vanishes whenever well-defined. The ``well-defined'' condition fails only when the matrix of states $\mathfrak{U}$ is not a differentiable function of $\bs{k}$, i.e.~at band nodes. Indeed [as we elaborate in Sec.~E of SI] band nodes of real Hamiltonians are known to carry a singular $\pi$-flux of the curvature.

While the change of Lie algebra $\mathfrak{u}(n) \to \mathfrak{so}(n)$ trivializes Berry curvature, it also enables \emph{new} gauge-invariant and topological objects. Decomposing into the basis 1-forms, $\mcA = \mcA_i d k^i$, the prefactors $\mcA_i$ are just skew-symmetric matrices. If we limit our attention to the case of an \emph{even} number $n$ of bands, then we can define \emph{Euler connection}
\begin{equation}
\textrm{a} := \Pf(\mcA_i) d k^i \label{eqn:Eu-a-def}
\end{equation}
where $\Pf$ denotes Pfaffian. Below, we express the construction in Eq.~(\ref{eqn:Eu-a-def}) simply as ``$\mathrm{a} = \Pf(\mcA)$''. Gauge transformation of Euler connection [i.e.~combining  Eq.~(\ref{eqn:Eu-a-def}) and Eq.~(\ref{eqn:BWZ-conn-GT}) with $X\in\mathsf{O}(n)$] for a general number of bands is a non-trivial task. However, for $n=2$ bands, the Pfaffian is a linear function of the matrix entries, therefore we can split
\begin{eqnarray}
\tilde{\textrm{a}} &=& \Pf(X^\top \mcA X) + \Pf(X^\top d X) \nonumber \\
&=&\det(X) \textrm{a} + \Pf(X^\top d X)\quad \textrm{(for $n=2$ bands),}\label{eqn:Euler-conn-GT}
\end{eqnarray}
where we used the well-known identity that for an antisymmetric matrix $\mathpzc{M}$ and an arbitrary matrix $X$, 
\begin{equation}
\Pf(X^\top \mathpzc{M} X) = \det(X)\Pf(\mathpzc{M}). \label{eqn:PF-GT-identity}
\end{equation}
The sign $\det(X) = \pm 1$ in Eq.~(\ref{eqn:Euler-conn-GT}) conveys whether we perform a proper (orientation-preserving) or an improper (orientation-reversing) gauge transformation.

Note that Euler connection is \emph{not} a matrix anymore, i.e.~it can be treated as an ordinary differential 1-form. In particular, one can study the exterior derivative $d \textrm{a}$. For a general number of bands, commuting the exterior derivative with the Pfaffian operator in Eq.~(\ref{eqn:Eu-a-def}) is a difficult task. However, for $n=2$ bands, the linearity of the Pfaffian implies that $[d,\textrm{Pf}]=0$. We thus define 
\begin{equation}
\textrm{Eu} := d \mathrm{a} \qquad \textrm{(for $n=2$ bands),}\label{eqn:Eu-form-def-2}
\end{equation}
which is called the \emph{Euler curvature} or \emph{Euler form}. With a bit of manipulation, we obtain
\begin{equation}
\textrm{Eu} 
= \Pf (d\mcA) = \Pf(\mcF - \mcA\wedge\mcA) = \Pf(\mcF) \;\; \textrm{(for $n=2$ bands),}\label{eqn:Euler-curv-2-band}
\end{equation}
where we used that $(\mcA\wedge\mcA)_{ij} = A_i A_j - A_j A_i = 0$ for $\mathfrak{so}(2)$ matrices due to commutativity of their product. However, owing to the skew-symmetry of $\mcF$, the last line of Eq.~(\ref{eqn:Euler-curv-2-band}) makes sense for \emph{arbitrary} even $n$. We therefore lift this equation to be the \emph{definition} of Euler form for an arbitrarily large collection of real bands,
\begin{equation}
\textrm{Eu} := \Pf(\mcF_{ij}) d k^i \wedge d k^j. \qquad \textrm{(for arbitrary $n$),}
\end{equation}
Combining Eqs.~(\ref{eqn:BWZ-GT}) and~(\ref{eqn:PF-GT-identity}), we find 
\begin{equation}
\tilde{\mathrm{Eu}} = \det(X) \mathrm{Eu} ,\label{eqn:Gaug-invar-Eu}
\end{equation}
meaning that Euler form is invariant under orientation-preserving $\mathsf{SO}(n)$ transformations of the $n$ bands, while it flips sign under orientation-reversing transformations.

\partitle{Quantization of Euler class.} If the rank of the bundle is $n=2$, and the base space is two-dimensional and parameterized by momenta $k_x$ and $k_y$, then the Euler curvature 
\begin{equation}
\textrm{Eu} = \left(\braket{\partial_{k_x} u^1}{\partial_{k_y}  u^2} - \braket{\partial_{k_y} u^1}{\partial_{k_x} u^2}\right) d k^x\wedge d k^y,\label{eqn:Euler-form-two-bands}
\end{equation}
is a single-component object. If further both the base space and the vector bundle are \emph{orientable}, Euler curvature can be treated as a volume form, and integrated over the base space $B$. Note that the nilpotence $d^2 = 0$ implies that the exterior derivative of Euler curvature is zero. However, the potential may not be globally defined, therefore Euler class of an oriented real bundle on a base space $B$ defines an element of the de Rham cohomology $H^2_\textrm{dR}(B)$. In fact, it can be shown~\cite{Milnor:1975} that $\frac{1}{2\pi}\textrm{Eu}(\bs{k})$ integrates to an \emph{integer} if the base space does not have a boundary, therefore the Euler-form integral defines an element of the singular cohomology with \emph{integer} coefficients, $H^2(B;\intg)$~\cite{Hatcher:bundles}. The integer
\begin{equation}
\chi(E) = \frac{1}{2\pi}\oint_B \textrm{Eu} \qquad \textrm{(for $n=2$ bands)}\label{eqn:Euler-class-defi}
\end{equation}
is called \emph{Euler class} of vector bundle $E$. The name is motivated by the observation that for a tangent bundle $TM$ of a two-dimensional manifold $M$ without a boundary, the integer $\chi(TM)$ reproduces the \emph{Euler characteristic} of $M$~\cite{Nakahara:2003}. This observation for $n=2$ is a special case of the more general Chern-Gauss-Bonnet theorem~\cite{Chern:1945}, which applies to manifolds without boundary of higher even dimensions. Note also that the Euler characteristic of any odd-dimensional closed manifold is zero~\cite{Hatcher:2002}, justifying our motivation to consider even-rank bundles and the Pfaffian operator. 

Observe that Eq.~(\ref{eqn:Euler-class-defi}) is analogous to the definition of the first Chern number
\begin{equation}
c_1(E) = \frac{1}{2\pi}\oint_B \mathsf{F} \label{eqn:1-chern-def}
\end{equation}
which is an element of $H^2(B;\intg)$ for \emph{complex} vector bundles. The mathematical arguments guaranteeing the quantization of $\chi(E)$ resp.~$c_1(E)$ are essentially identical~\cite{Hatcher:bundles}, and based on considering a covering of the base space $B$ with open discs $\{\mathcal{D}_\alpha\}_{\alpha=1}^N$. To outline the argument, let us explicitly consider the case of $B$ being a 2-sphere ($S^2$). The sphere is covered by $N=2$ discs, e.g.~the northern hemisphere $\mathcal{D}_{\textrm{north}}$ and the southern hemisphere $\mathcal{D}_{\textrm{south}}$, which meet at the equator $\gamma_{\textrm{eq.}}$. Since disc is a contractible manifold, we can use Stokes' theorem to relate $\int_{\mathcal{D}_\alpha} \!\!\textrm{Eu}$ to $\oint_{\partial \mathcal{D}_\alpha} \!\!\mathrm{a}$ (and analogously $\int_{\mathcal{D}_\alpha} \!\!\mathsf{F}$ to $\oint_{\partial \mathcal{D}_\alpha} \!\!A$ for the complex case) on each hemisphere. The resulting two integrals run around the equator in opposite directions, therefore
\begin{eqnarray}
2\pi\chi(E) &=&\oint_{S^2} \textrm{Eu} = \int_{\mathcal{D}_{\textrm{north}}} \!\!\!\!\!\!\textrm{Eu} \;\;+ \int_{\mathcal{D}_{\textrm{south}}} \!\!\!\!\!\!\textrm{Eu} \nonumber \\
&=& \oint_{\gamma_\textrm{eq.}} \!\!\left(\mathrm{a}_\textrm{north} - \mathrm{a}_\textrm{south}\right) \nonumber \\
&=& \oint_{\gamma_\textrm{eq.}}\!\! \Pf[X^\top dX]\quad  \textrm{(for $n=2$ bands)}\label{eqn:Euler-quantization-proof}
\end{eqnarray}
where we used that connections $\mathrm{a}_\textrm{north}$ and $\mathrm{a}_\textrm{south}$ on for rank-2 bundles are related by a gauge transformation in Eq.~(\ref{eqn:Euler-conn-GT}), and we assumed that the orientation of the vector bundle is fixed on the whole 2-sphere [thus $\det (X) = +1$]. Note that if we write the $\mathsf{SO}(2)$ matrix $X$ using the algebra element $\alpha \in \mathfrak{so}(2)$ as $X = \e{+\imi\alpha \sigma_y}$, then the last expression in Eq.~(\ref{eqn:Euler-quantization-proof}) reduces to integration of $\Pf[X^\top dX] = d \alpha$. Since the gauge transformation $X$ must return to its original form after traversing the equator, the value of $\alpha$ must increase by an integer multiple of $2\pi$ on $\gamma_{\textrm{eq.}}$. Therefore
\begin{eqnarray}
2\pi\chi(E) &=& \oint_{\gamma_\textrm{eq.}} \!\!d \alpha = 2\pi m\quad \\
    &\phantom{=}& \textrm{with $m\in\intg$}\quad \textrm{(for $n=2$ bands)}
\end{eqnarray}
which completes the proof of the quantization of the Euler class for real orientable rank-2 vector bundles on $S^2$. 

\partitle{Complexification of rank-2 real vector bundle.} We find a correspondence between the Euler form of two real states $\ket{u^{1,2}}$ and the Berry curvature of the complex state $\ket{\psi}=\tfrac{1}{\sqrt{2}}\left(\ket{u^1}+\imi\ket{u^2}\right)$. A simple calculation for the single-band Berry curvature according to Eq.~(\ref{eqn:Bery-curv}) reveals that
\begin{eqnarray}
\mathsf{F}_{ij} 
&=& -\imi\braket{\partial_{k_i}\psi}{\partial_{k_j}\psi} + \textrm{c.c.} \\
&=& -\frac{\imi}{2}\Big(\braket{\partial_{k_i} u^1}{\partial_{k_j} u^1} + \braket{\partial_{k_i}u^2}{\partial_{k_j}u^2} \label{eqn:first-line-of}\\
&\phantom{=}& \phantom{-\frac{\imi}{2}\Big(} +\imi\braket{\partial_{k_i} u^1}{\partial_{k_j} u^2} - \imi \braket{\partial_{k_i}u^2}{\partial_{k_j}u^1} \Big) + \textrm{c.c.} \nonumber 
\end{eqnarray}
($\textrm{c.c.}$ stands for ``complex conjugate''). Due to the reality condition on the states $\ket{u^{1,2}}$, the expressions $-\imi \braket{\partial_{k_i} u^a}{\partial_{k_j} u^a}$ for $a=1,2$ are purely imaginary, meaning that they drop under the combination with the $\textrm{c.c.}$ part. In contrast, the two terms in the second line of Eq.~(\ref{eqn:first-line-of}) are real, therefore combination with $\textrm{c.c.}$ leads to doubling. Therefore
\begin{equation}
\mathsf{F}_{ij} 
= \braket{\partial_{k_i} u^1}{\partial_{k_j} u^2} - \braket{\partial_{k_i}u^2}{\partial_{k_j}u^1} \Big) + \textrm{c.c.}  = \textrm{Eu}_{ij}
\end{equation}
where we recognized the componentwise version of Eq.~(\ref{eqn:Euler-form-two-bands}). The identification 
\begin{equation}
\textrm{Eu}\bigg[\ket{u^1},\ket{u^2}\bigg] = \mathsf{F}\bigg[\tfrac{1}{\sqrt{2}}\left(\ket{u^1}+\imi\ket{u^2}\right)\bigg]\label{eqn:Berry-Euler-rel}
\end{equation}
allows us to numerically compute Euler form with the help of various tricks known from the numerical computation of Berry curvature, such as the projection onto the complex state along an infinitesimal square path, cf.~Ref.~\cite{Fukui:2005}. We further comment on the relation in Eq.~(\ref{eqn:Berry-Euler-rel}) in Sec.~H of SI when presenting our numerical integration method~\cite{Bzdusek:Mathematica-Euler}. We emphasize that Eq.~(\ref{eqn:Berry-Euler-rel}) is true for a pair of bands in models of arbitrarily high rank.

\partitle{Wilson-loop formulation of Euler class in the presence of adjacent nodes.} We introduce a way of computing the frame-rotation charge (i.e.~Euler class) of two principal nodal points by utilizing Wilson loops in the case when the principal bands are connected to the rest of the band structure through adjacent nodes. On the one hand, this approach has the advantage that it does not require the explicit construction of a smooth gauge for the eigenstates, as compared to Eq.~(\ref{eqn:Euler-Stokes}). On the other hand, a specific flow of base loops must be designed in order to get around the adjacent nodes. 

Adopting the real gauge for the eigenstates, the Wilson loop computed over the two-principal bands on a closed path $l$,
\begin{equation}
\mathcal{W}_l = \displaystyle\mathrm{exp}\left\{ \oint_{l} \mathcal{A}(\boldsymbol{k}) \cdot d\boldsymbol{l}(\boldsymbol{k}) \right\},
\end{equation}
is an element of Lie group $\mathsf{SO}(2)$. Therefore, we can associate a Wilson-loop Hamiltonian $\mathcal{H}_{\mathcal{W}_l}$ to each Wilson-loop matrix through $\mathcal{W}_{l} = e^{\imi\mathcal{H}_{\mathcal{W}_l} }$. $\imi\mathcal{H}_{\mathcal{W}_l}$ is an element of Lie algebra $\mathfrak{so}(2)$ of real $2\times 2$ skew-symmetric matrices. We thus parametrize the Wilson loop through $\mathcal{W}_{l} = e^{\zeta(l) \imi \sigma_y }$, with $\zeta(l) \in \mathbb{R}$ defined by the Pfaffian 
\begin{equation}
	\zeta(l) = \mathrm{Pf}\left[ \imi\mathcal{H}_{\mathcal{W}_l} \right]  = \mathrm{Pf}\left[\log \mathcal{W}_{l}\right] \;.
\end{equation} 
The Wilson loop is periodic in  $\zeta(l)$ modulo $2\pi$ and it changes continuously under smooth deformations of the base loop $l$, as long as no adjacent nodes are crossed. A winding number of Wilson loop~\cite{Bouhon2018_fragile} is thus obtained, i.e.~the winding number of $\zeta(l)$, as a function of the flow of the base loop over a closed two-dimensional manifold avoiding the adjacent nodes. For this we devise a flow of Wilson loop (resp.~of the Pfaffian) over the \emph{punctured Brillouin zone} $\mathrm{BZ} {-} \mathcal{D}^{\epsilon}$, i.e.~we exclude the infinitesimal islands ($\mathcal{D}^{\epsilon}$) surrounding the adjacent nodes. Fixing a base point, $x_0$, we form oriented base loops, $l_{\nu}$, within the punctured BZ. Then the flow of Pfaffian is obtained by deforming the base loop smoothly over the punctured Brillouin zone from the base point ($l_0 {=} x_0$) to the boundary $l_1 {=} \partial \mathrm{BZ} {-} \partial \mathcal{D}^{\epsilon}$ (we label the deformation of $l_{\nu}$ by $\nu {\in} [0,1]$). 

As an example, we consider the case of a pair of principal and of adjacent nodes. We define two distinct flows, as illustrated in Extended Data Fig.~\ref{two_flows}{\bf a} and {\bf b}. The dashed lines mark the origin of the creation of the pair of principal nodes (black) and adjacent nodes (gray). Assuming that the principal nodes were created first, we know that the adjacent nodes can be annihilated along the dashed line between the two. Then Eq.~(\ref{eqn:Euler-Stokes}) implies that the Euler class $\chi[\mathrm{BZ} {-} \mathcal{D}^{\epsilon}] {=} 0$ over the region $\mathrm{BZ} {-} \mathcal{D}^{\epsilon}$ of Extended Data Fig.~\ref{two_flows}{\bf a}, while $\chi[\mathrm{BZ} {-} \mathcal{D}^{\epsilon}] {=} 1$ is finite (nontrivial frame-rotation charge) over the region $\mathrm{BZ} {-} \mathcal{D}^{\epsilon}$ of Extended Data Fig.~\ref{two_flows}{\bf b}.

We compare the predictions based on Eq.~(\ref{eqn:Euler-Stokes}) to the flow of the Wilson-loop Pfaffian in Extended Data Fig.~\ref{two_flows}{\bf c}--{\bf f} for the braiding model in Eq.~(\ref{eqn:Ham-matrix}). First, at $t=6$ (when the two adjacent nodes are located on top of each other at $\Gamma$, cf.~Fig.~\ref{fig:protocol}{\bf c}) we find that the winding number is indeed trivial, consistent with $\chi[\mathrm{BZ} - \mathcal{D}^{\epsilon}] = 0$ (see Extended Data Fig.~\ref{two_flows}{\bf c} and~{\bf e}). In contrast, for $t=2$ (when the two adjacent nodes are located on top of each other at $\textrm{M}$) the Wilson loops exhibits a non-trivial winding number, consistent with $\chi[\mathrm{BZ} - \mathcal{D}^{\epsilon}] = 1$ (see Extended Data Fig.~\ref{two_flows}{\bf d} and~{\bf f}). We observe that the predictions based on Eq.~(\ref{eqn:Euler-Stokes}) are consistent with calculating the winding of the Wilson-loop Pfaffian.

\partitle{First-principles calculations} Our first-principles calculations are performed using VASP (Vienna \emph{Ab initio} Simulation Package)~\cite{vasp, PhysRevB.59.1758} which relies on all-electron projector augmented wave (PAW) basis sets~\cite{PhysRevB.50.17953} combined with the generalized gradient approximation (GGA) with exchange-correlation functional of Perdew, Burke and Ernzerhof (PBE)~\cite{PhysRevLett.77.3865}. In order to better capture the correlation effects, Heyd-Scuseria-Ernzerhof (HSE06) screened hybrid functional~\cite{hse06} was used in first-principle calculations for ZrTe. Nevertheless, we still compare the band structure from PBE functional and PBE+HSE06 functional.

To obtain the band structure of WC-type ZrTe, we proceeded as follows. The cutoff energy for the plane wave expansion was set to 360~eV and a $\bs{k}$-point mesh of $12\times12\times12$ was adopted. The lattice constants are fully relaxed to $a=3.7807 \textrm{\AA}$ and $c= 3.8618 \textrm{\AA}$ which are comparable with the experimental values $a=3.7706 \textrm{\AA}$ and $c= 3.8605 \textrm{\AA}$~\cite{Orlygsson1999}. The WannierTools code~\cite{wanniertools} was used to search for nodal lines and Weyl points and to calculate chiralities of Weyl points in the Brillouin zone based on Wannier tight-binding model that was constructed by using the Wannier90 package~\cite{wannier90} with Zr $s,p,d$ and Te $p$ atomic orbitals as projectors. Spin-orbit coupling (SOC) effects were considered. The Fermi surface and the band structure of ZrTe along chosen high-symmetry lines is shown in Extended Data Fig.~\ref{fig:mirror-bs}. Weyl points and nodal lines are formed between two bands with different mirror eigenvalues. We show in Extended Data Fig.~\ref{fig:zrte-pbe-hse06} that HSE06 only enlarges the energy gap at $\textrm{K}$ point, but it does not change the band inversion character.

We further report that ZrTe is not the only candidate that has Weyl points with a non-trivial Euler class. We studied several other materials with the WC-type crystal structure, and find that at ambient conditions: (1) MoP and NbS have Weyl points, (2) MoC, WC and WN have nodal lines, and (3) TaN, NbN and MoN exhibit a gap near the $\mathrm{K}$ point. The nodal lines of MoC, WC and WN can be changed to Weyl points under appropriate pressure. The band structure obtained from first-principles calculations with PBE functional are shown in Extended Data Fig.~\ref{fig:9materials}. The discussed Weyl points, nodal lines, and band gap correspond to the two energy bands colored red and green in Extended Data Fig.~\ref{fig:9materials}. 

A suitable material candidate to study the nodal conversions, besides ZrTe, is MoP. We compare the band structure and the locations of nodal points of MoP vs.~ZrTe in  Extended Data Fig.~\ref{fig:zrte-mop}. We show that ZrTe and MoP have a very similar Weyl point distribution near the mirror invariant plane $k_z=0$. We remark that, as opposed to ZrTe, the Weyl points of MoP have already been confirmed in ARPES experiments~\cite{Lv2017}. 

Furthermore, TaAs is a theoretically predicted~\cite{Weng:2015} and experimentally confirmed~\cite{Xu:2015} Weyl-semimetal material. There are 24 Weyl points in the first Brillouin zone. Among them, 8 Weyl points are located inside a $C_2\mcT$-invariant plane. We plot the band structure and the distribution of the Weyl points of TaAs in Extended Data Fig.~\ref{fig:taas-nodalconversion}{\bf a}--{\bf d}. Using the numerical method detailed in Sec.~H of SI, we find that the Euler number of each pair of $C_2\mcT$-invariant Weyl points is non-trivial, implying that they cannot annihilate after collision. The computed Euler curvature near one such pair of Weyl points is shown in Extended Data Fig.~\ref{fig:taas-eulercurvature}. 

After applying a 5\% [001]-uniaxial strain, we observe pairs of Weyl points collide inside the $C_2\mcT$-invariant plane, and convert into nodal lines located inside the vertical mirror planes. The details are shown in Extended Data Fig.~\ref{fig:taas-nodalconversion}(e--h). While applying a static $5\%$ strain is clearly not experimentally viable, large values of strain have been achieved in another Weyl-semimetal compound $\textrm{WTe}_2$ by driving an optical ``shear'' phonon mode~\cite{Sie:2019}. Similar experiments might provide a way to test our ideas in a solid state setting.

\sectitle{Data availability} 
Source data are available for this paper~\cite{Wu:conversion-data}. All other data that support the plots within this paper and other findings of this study are available from the corresponding authors upon reasonable request. 

\sectitle{Code availability} 
\texttt{Mathematica} notebook for computing Euler class of a collection of band nodes by implementing the method presented in Sec.~H of SI is made available online~\cite{Bzdusek:Mathematica-Euler}.

\sectitle{References} 
\vspace{-0.4cm}
%

\sectitle{Acknowledgements} 
We acknowledge valuable discussions with C.~C.~Wojcik, A.~Vishwanath, and B.~A.~Bernevig. R.-J.~S acknowledges funding via A.~Vishwanath from the Center for Advancement of Topological Semimetals, an Energy Frontier Research Center funded by the U.S. Department of Energy Office of Science, Office of Basic Energy Sciences, through the Ames Laboratory under its Contract No. DE-AC02-07CH11358 and Trinity college as well as the Winton program at the University of Cambridge. T.~B. was supported by the Gordon and Berry Moore Foundation's EPiQS Initiative, Grant GBMF4302, and by the Ambizione Program of the Swiss National Science Foundation, Grant No.~185806. Q.-S.~W and O.~V.~Y acknowledge support from NCCR Marvel. H.M.~W acknowledges support from the Ministry of Science and Technology of China under grant numbers 2018YFA0305700, 2016YFA0300600. First-principles calculations have been performed at the Swiss National Supercomputing Centre (CSCS) under Project No.~s832 and the facilities of Scientific IT and Application Support Center of EPFL.

\sectitle{Author Contributions} 
A.~B., R.-J.~S., and T.~B.~contributed equally to the theoretical analysis in this work, and wrote the manuscript. Q.-S.~W.~discovered the nodal conversion in ZrTe, and obtained the presented first-principles data. H.M.~W.~and O.~V.~Y.~were involved in the discussion and analysis of the first-principles data. All authors discussed and commented on the manuscript.

\sectitle{Competing interests} 
The authors declare no competing interests.

\vfill

\FloatBarrier
\fontsize{10}{12}\selectfont
\pagebreak
 
\setcounter{figure}{0}
\renewcommand{\theHfigure}{E\arabic{figure}}
\renewcommand{\figurename}{\textsf{\textbf{Extended Data Fig.}}}

\begin{figure*}[h]
\centering
\begin{tabular}{ll} 
    {\bf a} & {\bf b} \\
	 \includegraphics[width=0.22\linewidth]{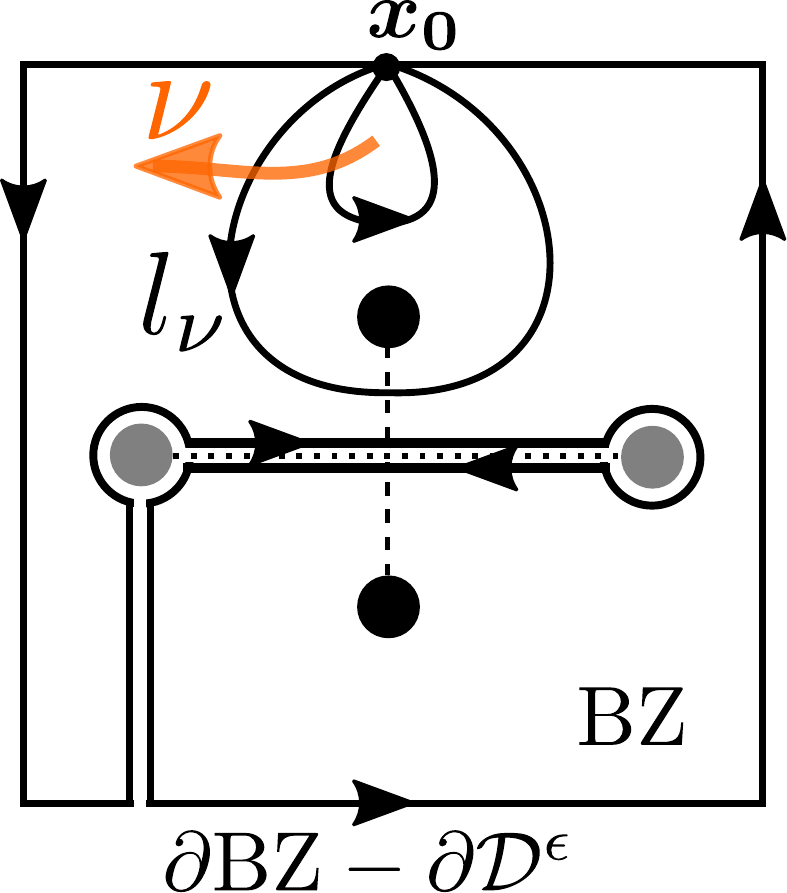} &
	\includegraphics[width=0.22\linewidth]{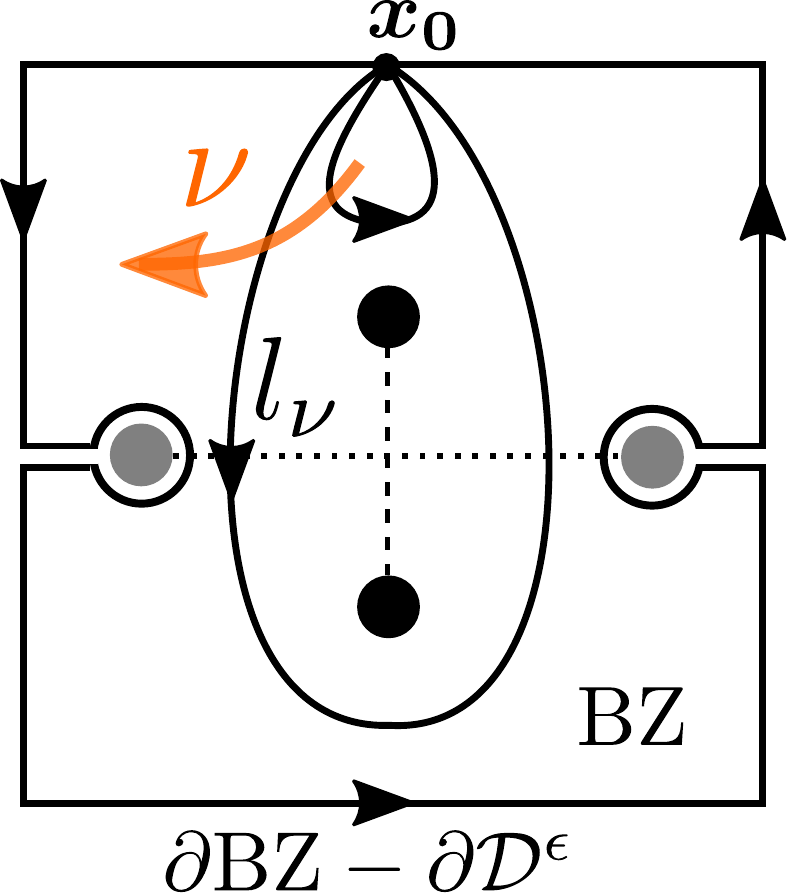} 
\end{tabular}\\
\caption{\textsf{\label{two_flows}
Wilson loop computation of Euler class of two principal nodes in the presence of two adjacent nodes. {\textbf a-b} Two possible flows of a base loop ($l_{\nu}$, $\nu\in[0,1]$) over the punctured Brillouin zone containing two principal nodes (black) and excluding two adjacent nodes (gray). The flow starts at a base point ($l_0=x_0$) and ends at the boundary $l_1=\partial \mathrm{BZ} - \partial \mathcal{D}^{\epsilon}$, with the orange arrow indicating the direction of the flow. Assuming that the pair of principal nodes were created first (along the dashed line between them), the Euler class is $\chi[\mathrm{BZ} -  \mathcal{D}^{\epsilon}] = 0$ in case {\textbf a}, and $\chi[\mathrm{BZ} -  \mathcal{D}^{\epsilon}] = 1$ in case {\textbf b}. {\textbf c-f} Wilson loop ({\textbf c},{\textbf d}) and Pfaffian ({\textbf e},{\textbf f}) as a function of the flow of base loop over the punctured Brillouin zone, i.e.~corresponding to {\textbf a} for ({\textbf c},{\textbf e}), and to {\textbf b} for ({\textbf d},{\textbf f}). {\textbf e} The zero winding of the Pfaffian, $\Delta \zeta = 0$, indicates a trivial frame-rotation charge of the two principal nodes (i.e.~zero Euler class on $\mathrm{BZ} - \mathcal{D}^{\epsilon}$). {\textbf f} The non-zero winding of the Pfaffian, $\Delta \zeta  / (2\pi)= 1$, indicates a nontrivial frame-rotation charge of the two principal nodes (i.e.~non-vanishing Euler class on $\mathrm{BZ} - \mathcal{D}^{\epsilon}$).
}}
\end{figure*}

\begin{figure*}[p!]
\centering
\includegraphics[width=0.98\linewidth]{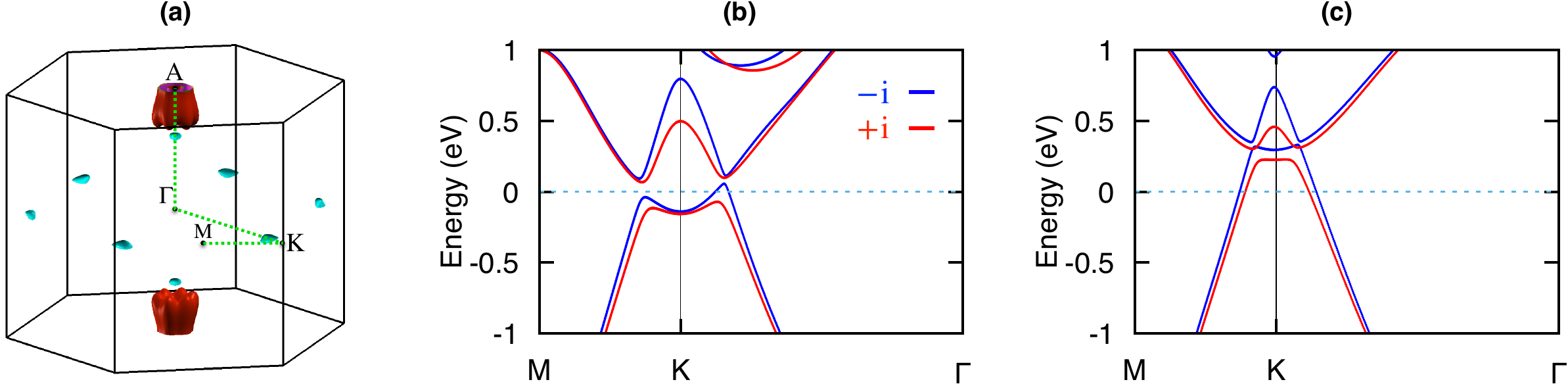}
\caption{\textsf{\textbf{Fermi surface and band structures of ZrTe with and without strain.} \textbf{a.} Fermi surface of ZrTe under ambient conditions, with pairs of Weyl nodes of opposite chirality located inside the cyan pockets close to $k_z=0$ plane. \textbf{b} and \textbf{c.} Band structures of ZrTe along high-symmetry lines of the Brillouin under 0\% resp.~under 2.6\% uniaxial compression strain. Path M-K-$\Gamma$ lies within the $k_z=0$ mirror-invariant plane. The mirror eigenvalues $+\textrm{i}$ and $-\textrm{i}$ are indicated by the red vs.~blue color of the corresponding energy band. The plotted data were obtained with PBE+HSE06 functional. }}\label{fig:mirror-bs}
\end{figure*}

\begin{figure*}[p!]
\centering
\includegraphics[width=0.72\linewidth]{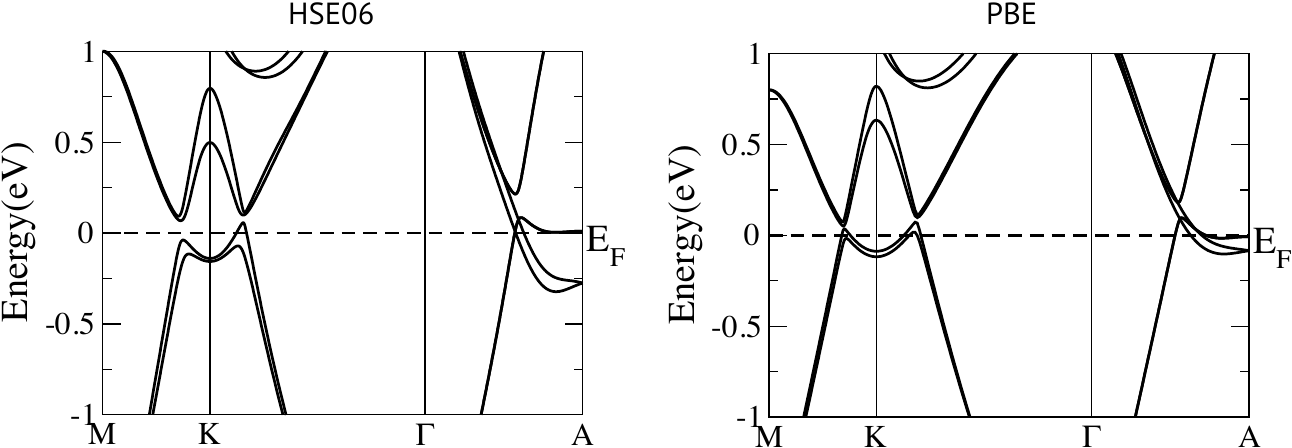} 
\caption{\textsf{\textbf{Band structure of ZrTe.} Band structure of ZrTe as computed from PBE+HSE06 (left) resp.~from PBE (right)  functional.}}\label{fig:zrte-pbe-hse06} 
\end{figure*}

\begin{figure*}[p!]
\centering
\includegraphics[width=0.69\linewidth]{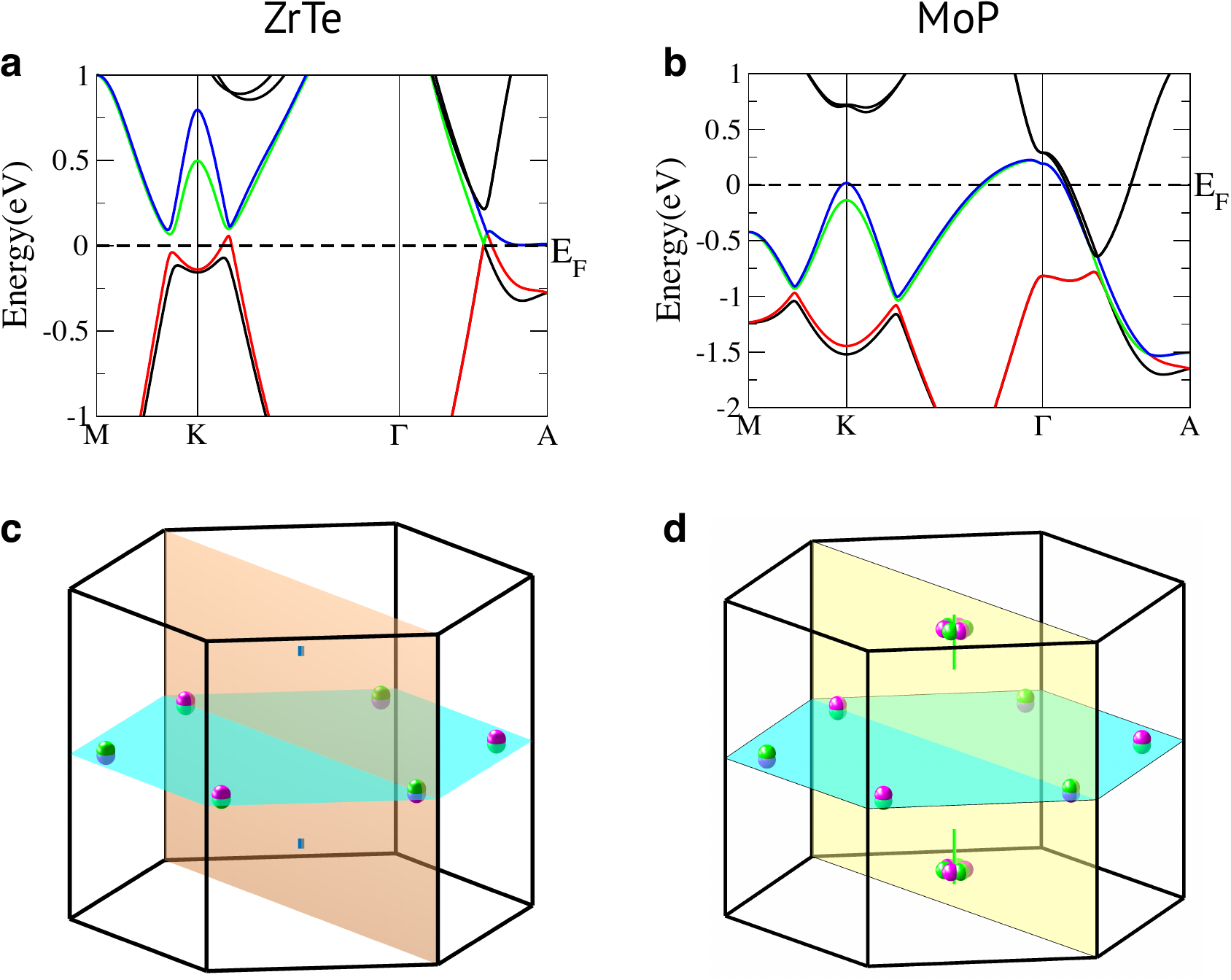} 
\caption{\textsf{\textbf{Comparison of principal band nodes of ZrTe and MoP.} Band structure and location of nodal points of ZrTe and MoP, obtained with PBE+HSE06 functional. The band nodes in panels {\textbf c} and {\textbf d} correspond to degeneracies of energy bands marked with green and red color in panels {\textbf a} and {\textbf b}.}}\label{fig:zrte-mop} 
\end{figure*}

\begin{figure*}[p!]
\centering
\includegraphics[width=0.99\linewidth]{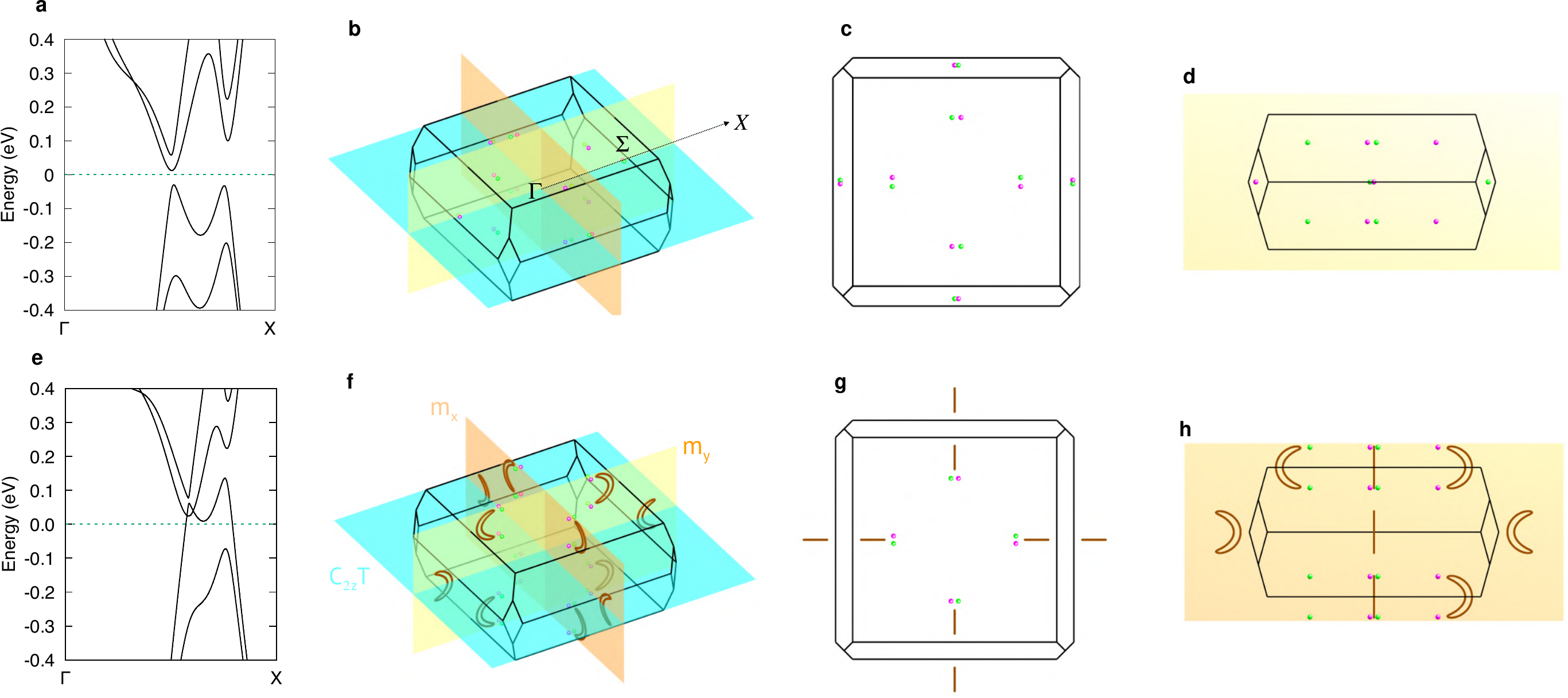} 
\caption{\textsf{\textbf{Band structure and principal band nodes of ambient TaAs and with strain.} Band structure and band node locations in the first Brillouin zone of TaAs {\textbf a}--{\textbf d} under ambient conditions, and {\textbf e}--{\textbf h} with 5\% [001] uniaxial strain obtained with PBE functional. The first column shows the band structure along the $\Gamma \textrm{X}$ line. The next three columns show the location of band nodes in 3-dimension view, top view, and front view, respectively. Weyl points with chirality +1 and -1 are respectively indicated with magenta and green spheres. Nodal lines of the strained TaAs are indicated as brown lines. The $C_2\mcT$-invariant plane is indicated with cyan color, while the vertical mirror planes are displayed in shades of yellow.}}\label{fig:taas-nodalconversion} 
\end{figure*}

\begin{figure*}[t!]
\centering
\includegraphics[width=0.45\linewidth]{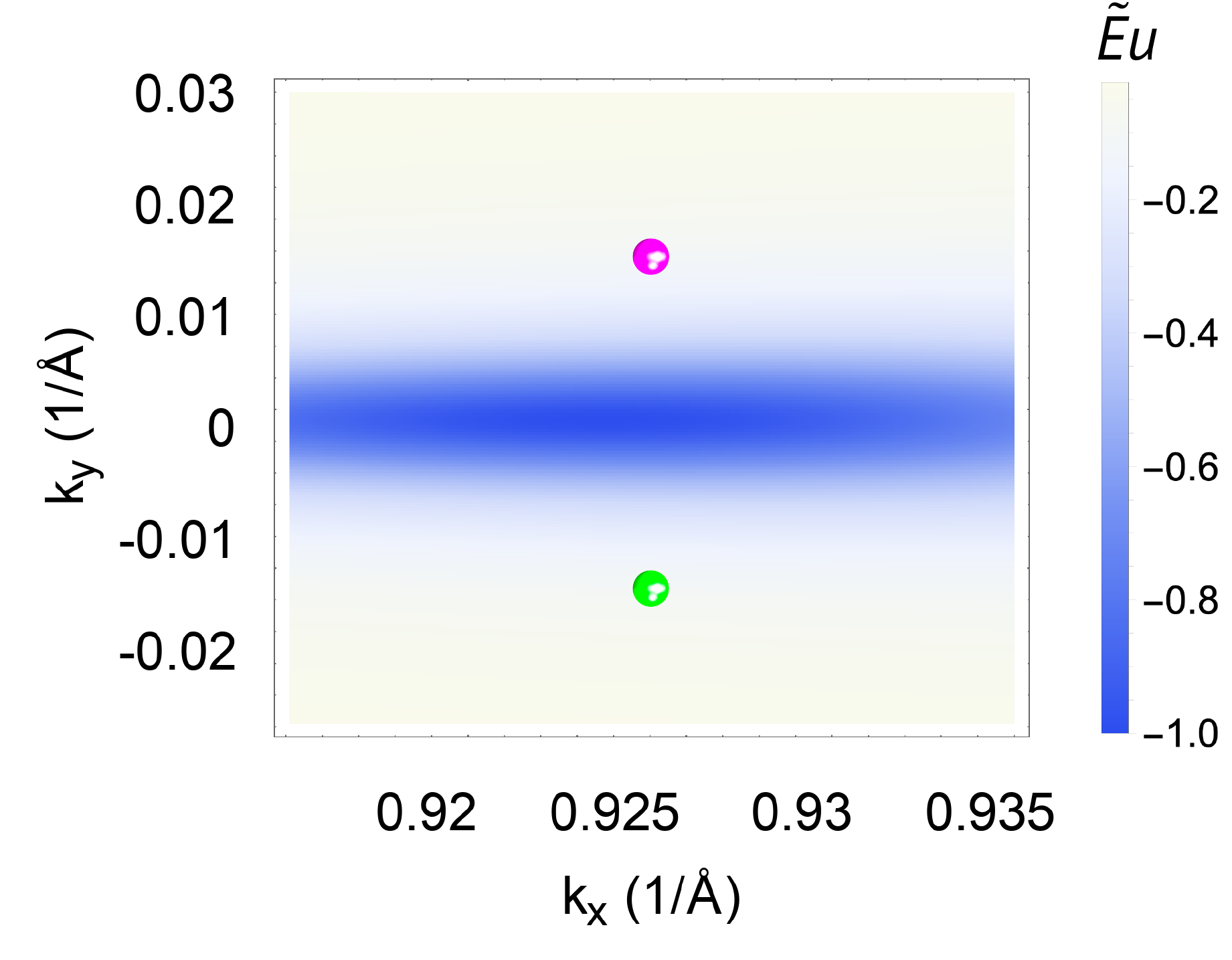} 
\caption{\textsf{\textbf{Euler curvature in TaAs.} Euler curvature of a pair of Weyl points inside the $C_2\mcT$-invariant plane of TaAs under ambient conditions. For a discussion, see Methods.}}~\label{fig:taas-eulercurvature} 
\end{figure*}

\begin{figure*}[p!]
\centering
\includegraphics[width=0.9\linewidth]{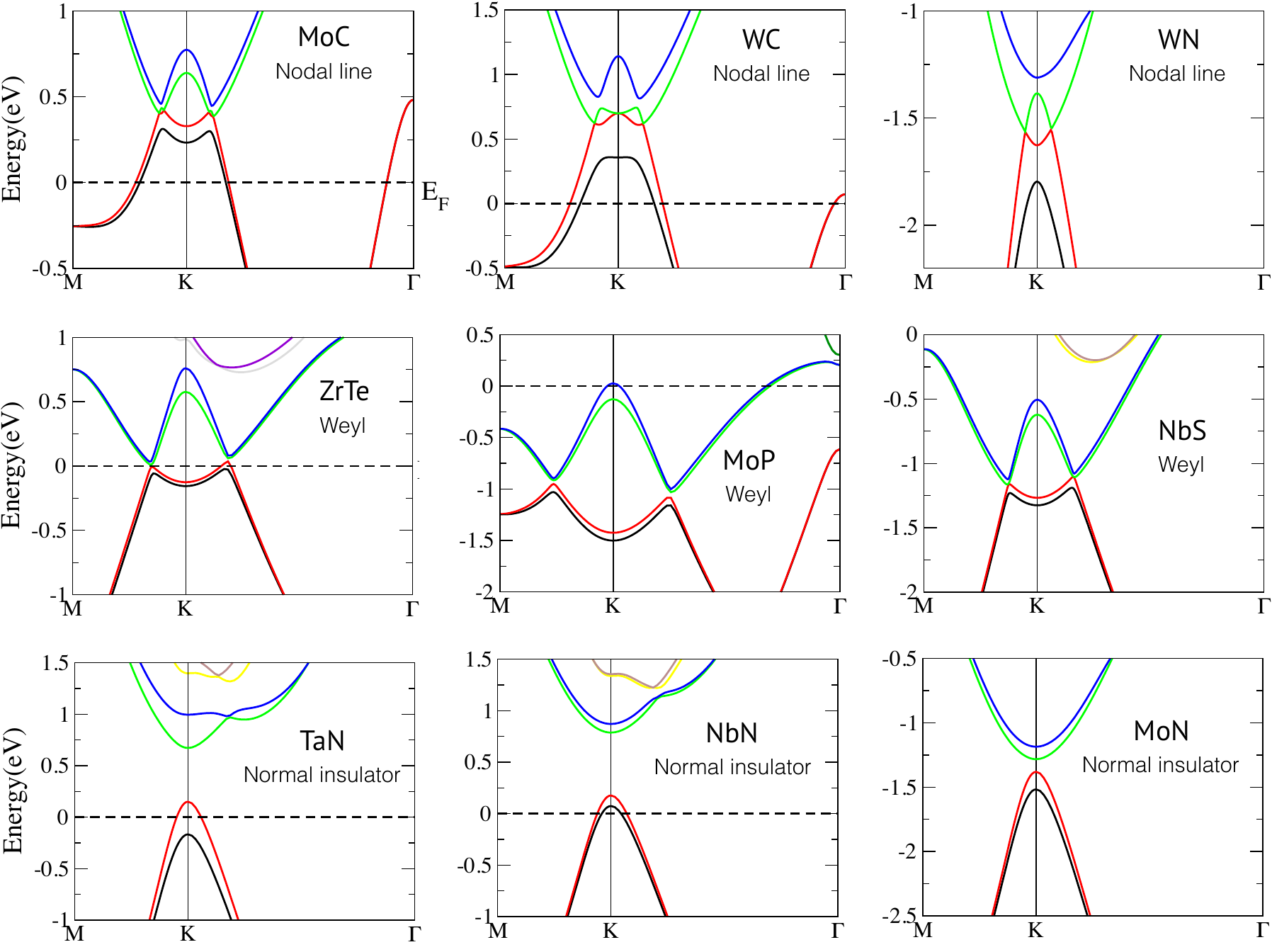} 
\caption{\textsf{\textbf{Band structure of nine WC-type materials obtained with PBE functional.} The corresponding nodal structure near the $\textrm{K}$-point under ambient conditions is indicated in each panel. }}\label{fig:9materials} 
\end{figure*}

\end{bibunit}

\FloatBarrier

\begin{bibunit}

\onecolumngrid
\pagebreak
\phantom{
\color{white}
\pagebreak}
\twocolumngrid

\newpage
\onecolumngrid

\setcounter{equation}{0}
\setcounter{page}{1}
\setcounter{figure}{0}
\renewcommand{\figurename}{\textsf{\textbf{Fig.}}}
\renewcommand{\thefigure}{\textsf{\textbf{{S\arabic{figure}}}}}\renewcommand{\theHfigure}{S\arabic{figure}}
\renewcommand{\theequation}{{S\arabic{equation}}}

\titleformat*{\section}{\normalsize\bfseries\sffamily\centering}
\titleformat*{\subsection}{\normalsize\bfseries\sffamily\centering}
\renewcommand\thesection{\Alph{section}}
\renewcommand\thesubsection{\Alph{section}\arabic{subsection}}
\makeatletter
\def\l@subsection#1#2{}
\def\l@subsubsection#1#2{}
\renewcommand{\p@subsection}{}
\makeatother


\title{\textsf{Supplementary Information for \texorpdfstring{\\ \vspace{0.2cm}}{} Non-Abelian reciprocal braiding of Weyl points and its manifestation in $\textrm{ZrTe}$}}

\author{Adrien Bouhon$^{1,2}$}\thanks{Contributed equally. Correspondence to \href{mailto:adrien.bouhon@su.se}{adrien.bouhon@su.se}, \href{mailto:quansheng.wu@epfl.ch}{quansheng.wu@epfl.ch}, and  \href{mailto:rjs269@cam.ac.uk}{rjs269@cam.ac.uk}.}
\author{QuanSheng Wu$^{3,4}$}\thanks{Contributed equally. Correspondence to \href{mailto:adrien.bouhon@su.se}{adrien.bouhon@su.se}, \href{mailto:quansheng.wu@epfl.ch}{quansheng.wu@epfl.ch}, and  \href{mailto:rjs269@cam.ac.uk}{rjs269@cam.ac.uk}.}
\author{Robert-Jan Slager$^{5,6}$}\thanks{Contributed equally. Correspondence to \href{mailto:adrien.bouhon@su.se}{adrien.bouhon@su.se}, \href{mailto:quansheng.wu@epfl.ch}{quansheng.wu@epfl.ch}, and  \href{mailto:rjs269@cam.ac.uk}{rjs269@cam.ac.uk}.}
\author{Hongming Weng$^{7,8}$}
\author{Oleg V. Yazyev$^{3,4}$}
\author{Tom\'{a}\v{s} Bzdu\v{s}ek$^{9,10,11}$}

\affiliation{\vspace*{0.2cm} $^{1}$Nordic Institute for Theoretical Physics (NORDITA), Stockholm, Sweden}
\affiliation{$^{2}$Department of Physics and Astronomy, Uppsala University, Box 516, SE-751 21 Uppsala, Sweden}
\affiliation{$^{3}$Institute of Physics, \'{E}cole Polytechnique F\'{e}d\'{e}rale de Lausanne, CH-1015 Lausanne, Switzerland}
\affiliation{$^{4}$National Centre for Computational Design and Discovery of Novel Materials MARVEL, Ecole Polytechnique F\'{e}d\'{e}rale de Lausanne (EPFL), CH-1015 Lausanne, Switzerland}
\affiliation{$^{5}$TCM Group, Cavendish Laboratory, University of Cambridge, J.~J.~Thomson Avenue, Cambridge CB3 0HE, United Kingdom}
\affiliation{$^{6}$Department of Physics, Harvard University, Cambridge, MA 02138}
\affiliation{$^{7}$Beijing National Laboratory for Condensed Matter Physics and Institute of Physics,
Chinese Academy of Sciences, Beijing 100190, China}
\affiliation{$^{8}$Songshan Lake Materials Laboratory, Guangdong 523808, China}
\address{$^{9}$Condensed Matter Theory Group, Paul Scherrer Institute, CH-5232 Villigen PSI, Switzerland}
\address{$^{10}$Department of Physics, University of Z\"{u}rich, Winterthurerstrasse 190, 8057 Z\"{u}rich, Switzerland}
\affiliation{$^{11}$Department of Physics, McCullough Building, Stanford University, Stanford, CA 94305, USA \vspace*{0.1cm}}

\date{\today}

\setcounter{page}{0}
\maketitle
\vfill
\vfill
\twocolumngrid
\newpage
{\small
\begin{center}
\onecolumngrid

\vfill
\vfill
\vspace{-0.8cm}

\begin{minipage}{0.8\textwidth}

\section*{LIST OF CONTENTS}
\renewcommand{\theenumi}{\Alph{enumi}}

\setstretch{1.1}

This document contains the Supplementary Information, which has been omitted from the main text to keep it brief and simple. We have organized the information into sections, as follows:\bigskip

\begin{enumerate}
\item \textsf{\textbf{Crystal structure of the minimal 2D model.}}\hfill \pageref{sec:TB_model} \\ 
We give the details of the crystal structure of the minimal 2D tight-binding model discussed in the main text and presented in Methods.
\item \textsf{\textbf{Alternative braiding model.}} \hfill\pageref{sec:k-dot-p} \\
We present a 3-band lattice model that exhibits a non-trivial braiding of band nodes without the nodes moving across the Brillouin zone boundary. 
\item \textsf{\textbf{Reality condition.}}
\hfill\pageref{sec:reality} \\
We show that the existence of a $\bs{k}$-local antiunitary symmetry squaring to $+1$ (such as $C_2\mcT)$ implies the existence of a basis in which the Bloch Hamiltonian is a real symmetric matrix.
\item \textsf{\textbf{Euler form in three-band models.}}\hfill \pageref{sec:three-band}\\
We prove the geometric interpretation of the Euler form in three-band models, which is presented in Fig.~4(b) of the main text. ~ 
\phantom{\color{white}\cite{Wen95_top, HasanKane10_RMP, Qi11_RMP, Majorana, Kitaev09_AIP, Ryu10_NJP, Fu11_PRL, Slager12_NatPhys, Bzduvsek16nodal, Fang:2016, Bzdusek:2017, Kruthoff17_PRX, Bouhon:global_top, Po17_NatCommun, Bradlyn17_Nat, Slager2019, Holler18_PRB, Zhang2019, topomat, Wan11, Weng:2015, Lv:2015, Xu:2015, Huang:2015,  Poenaru:1977, Volovik:1977, Madsen:2004, Alexander:2012, Prx2016, Wu:2018b, Tiwari:2019, Ahn:2019, Ahn:2018, Cao_2018, Ashvin_2019, Song2019, Bouhon2018_fragile, Francis:1994, Sjoeqvist_2004, Zhao:2017, Milnor:1975, Fukui:2005, Soluyanov:2015, Lv2017, Sun:2018, Zhu:2016, Weng:2016, He:2017, Ma:2018, vasp, wannier90, wanniertools, Guo:2017, Sie:2019, Li_2016, Wilczek:1984, Flaschner_2016, Alba:2011, Hauke:2014, Lu:2013, Lu:2015, Zilberberg_2018,Nakahara:2003,Hatcher:bundles,Chern:1945,Hatcher:2002,Bzdusek:Mathematica-Euler,PhysRevB.59.1758,PhysRevB.50.17953,PhysRevLett.77.3865,hse06,Orlygsson1999,Wu:conversion-data}}
\item \textsf{\textbf{Singularity of Euler form at principal nodes.}}\hfill \pageref{sec:singularity} \\
We study analytic properties of Euler form near principal nodes. If the calculation is performed in the eigenstate basis, the Euler form is integrable albeit non-differentiable at principal nodes. This is an obstruction for the na\"{i}ve application of Stokes' theorem.
\item \textsf{\textbf{Euler class for manifold with a boundary.}}\hfill \pageref{sec:boundary}\\
We generalize Euler class to manifolds with a boundary, and we show that it detects the capability of pairs of principal nodes to annihilate.
\item \textsf{\textbf{Non-Abelian frame-rotation charge.}}\hfill \pageref{sec:frame-rotation} \\
We review the homotopic derivation of the non-Abelian frame-rotation charge from Ref.~\cite{Wu:2018b}, and we prove its relation to the Euler class on manifolds with a boundary.
\item \textsf{\textbf{Numerical calculation of the Euler form.}}\hfill \pageref{sec:numerical}\\
We present a numerical algorithm that calculates the Euler class on a manifold with a boundary, and we present a way to regularize certain numerically induced divergences.
\vspace{0.2cm}
\item[] \textsf{\textbf{List of references}}\hfill \pageref{biblio}\\
\end{enumerate}
\end{minipage}

\end{center}
}

\newpage
\twocolumngrid
\setstretch{1.}

\section{Crystal structure of the minimal 2D model}\label{sec:TB_model}

\begin{figure}[h]
\centering
\includegraphics[width=0.9\linewidth]{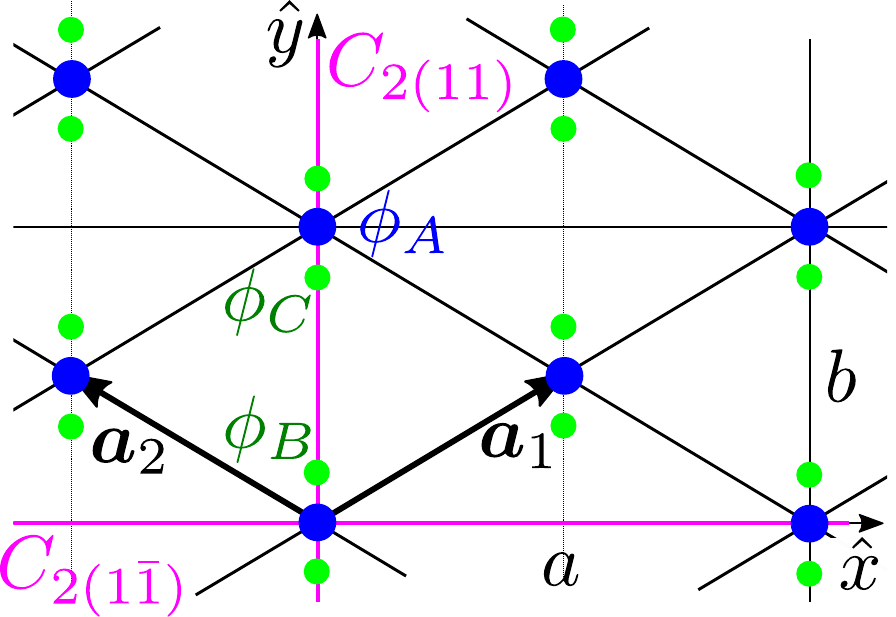}
\caption{\textsf{Crystal structure of the minimal 2D tight-binding model discussed in the main text and presented in Methods. The conventional unit cell (centered orthorhombic) is spanned by $a \hat{x}$ and $b\hat{y}$. There are three orbitals per each primitive cell spanned by the primitive vectors $\boldsymbol{a}_1$ and $\boldsymbol{a}_2$, one s-wave orbital labeled $\phi_A$ (blue), and two $p_z$-wave orbitals labeled $\phi_B$ and $\phi_C$ (green). The point group is $D_2 = \{C_{2z},C_{2(11)}\equiv C_{2y},C_{2(1\bar{1})}\equiv C_{2x}\}$ with $C_{2z}$ perpendicular to the basal plane.}}
\label{fig:2D_minimal_lattice} 
\end{figure}

The minimal 2D tight-binding model presented in Methods realizes the centered orthorhombic layer group $c222$ (LG$22$), with point group $D_2 = \{C_{2z},C_{2y},C_{2x}\}$~\cite{ITCE}. There are three orbitals per unit cell, one $s$-wave orbital at Wyckoff position $1a$ ($\phi_A$) and one (real) $p_z$-wave orbital at Wyckoff position $2g$ ($\phi_B,\phi_C$), see Fig.~\ref{fig:2D_minimal_lattice} where the primitive cell is spanned by the Bravais lattice vectors $\boldsymbol{a}_1 = (a \hat{x} + b \hat{y})/2$ and $\boldsymbol{a}_2 = (-a \hat{x} + b \hat{y})/2$.

The locations of the orbitals within the $n$-th unit cell are $\{\boldsymbol{R}_n+\boldsymbol{r}_{\alpha}\}_{\alpha=A,B,C}$, with $\boldsymbol{r}_A = (0,0)$ for $1a$, and $\boldsymbol{r}_B = (0,u)$ and $\boldsymbol{r}_C = (0,-u)$ for $2g$ (written in the Cartesian frame) with $u<b/2$. The tight-binding model is written in the Bloch state basis 
\begin{equation}
    \vert \phi_{\alpha} , \boldsymbol{k} \rangle = \dfrac{1}{\sqrt{N_{\alpha}}} \sum_{\boldsymbol{R}_n} e^{\imi \boldsymbol{k} \cdot \boldsymbol{R}_n }\vert w , \boldsymbol{R}_n + \boldsymbol{r}_{\alpha} \rangle \,, \;\alpha=A,B,C\,,
\end{equation}
that is the discrete Fourier transform of the Wannier states $\vert w , \boldsymbol{R}_n + \boldsymbol{r}_{\alpha} \rangle$ localized at $\boldsymbol{R}_n + \boldsymbol{r}_{\alpha}$, with $N_{\alpha}$ the number of lattice sites occupied by orbital $\alpha$. 

\section{Alternative braiding model}
\label{sec:k-dot-p}

The change of topological charge of band nodes upon exchange in momentum space, i.e.~the \emph{reciprocal braiding}, requires a minimum of three bands, corresponding to two distinct species of band nodes, An explicit tight-binding model demonstrating the reciprocal braiding is given in Eq.~(4) of Methods, and the corresponding braiding protocol is illustrated in Fig.~2 of the main text. That model, while being minimal in terms of hopping processes to be implemented, achieves the reciprocal braiding while moving the band nodes around the Brillouin zone (BZ) torus. However, the non-trivial braiding phenomenon is more general. To demonstrate this, we here provide an explicit tight-binding model that realizes the reciprocal braiding locally, i.e.~without either of the nodes passing though the BZ boundary.

\begin{figure}[t]
	\includegraphics[width=0.47\textwidth]{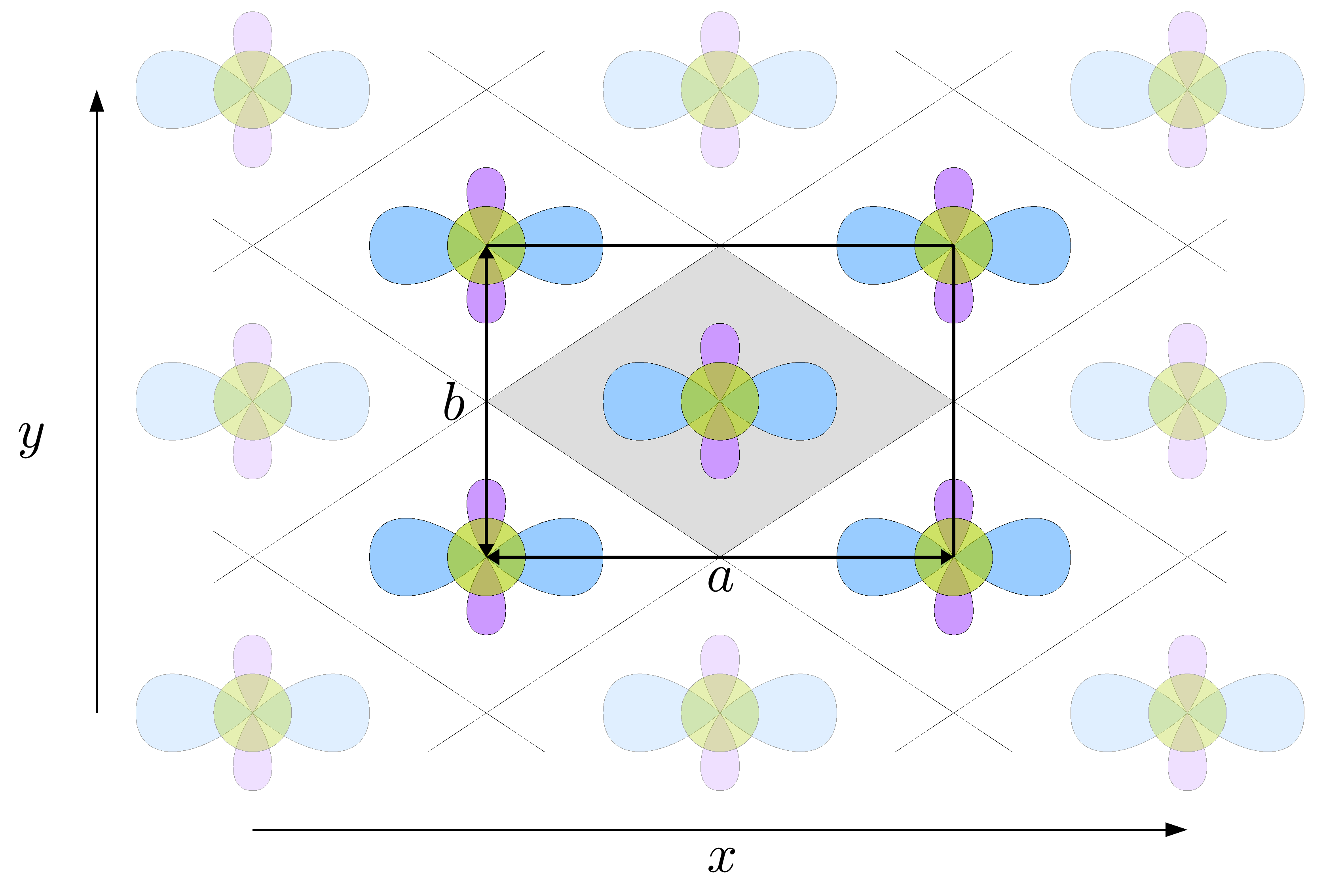}
	\caption{\textsf{\textbf{Model for the alternative braiding protocol.} Schematic illustration of the lattice model leading to the nearest-neighbor tight-binding Hamiltonian in Eq.~(\ref{eqn:model-2}). We consider $p_x$ (blue), $p_y$ (purple) and $s$ (green) orbital in the middle of a rhombic unit cell (gray). The conventional unit cell is a rectangle with dimensions $a$ and $b$, indicated by the black frame. We set $a=b=1$ for simplicity. We only consider hopping processes (both intra-orbital $T_{2,4,6}$ and inter-orbital $\pm T_{7,8,9}$) to the four nearest neighbor sites forming the corners of the conventional unit cell. On-site energies of the three orbitals are $T_{1,3,5}$.}}
	\label{fig:braid-model}
\end{figure} 

\begin{figure*}[t]
	\includegraphics[width=0.7\textwidth]{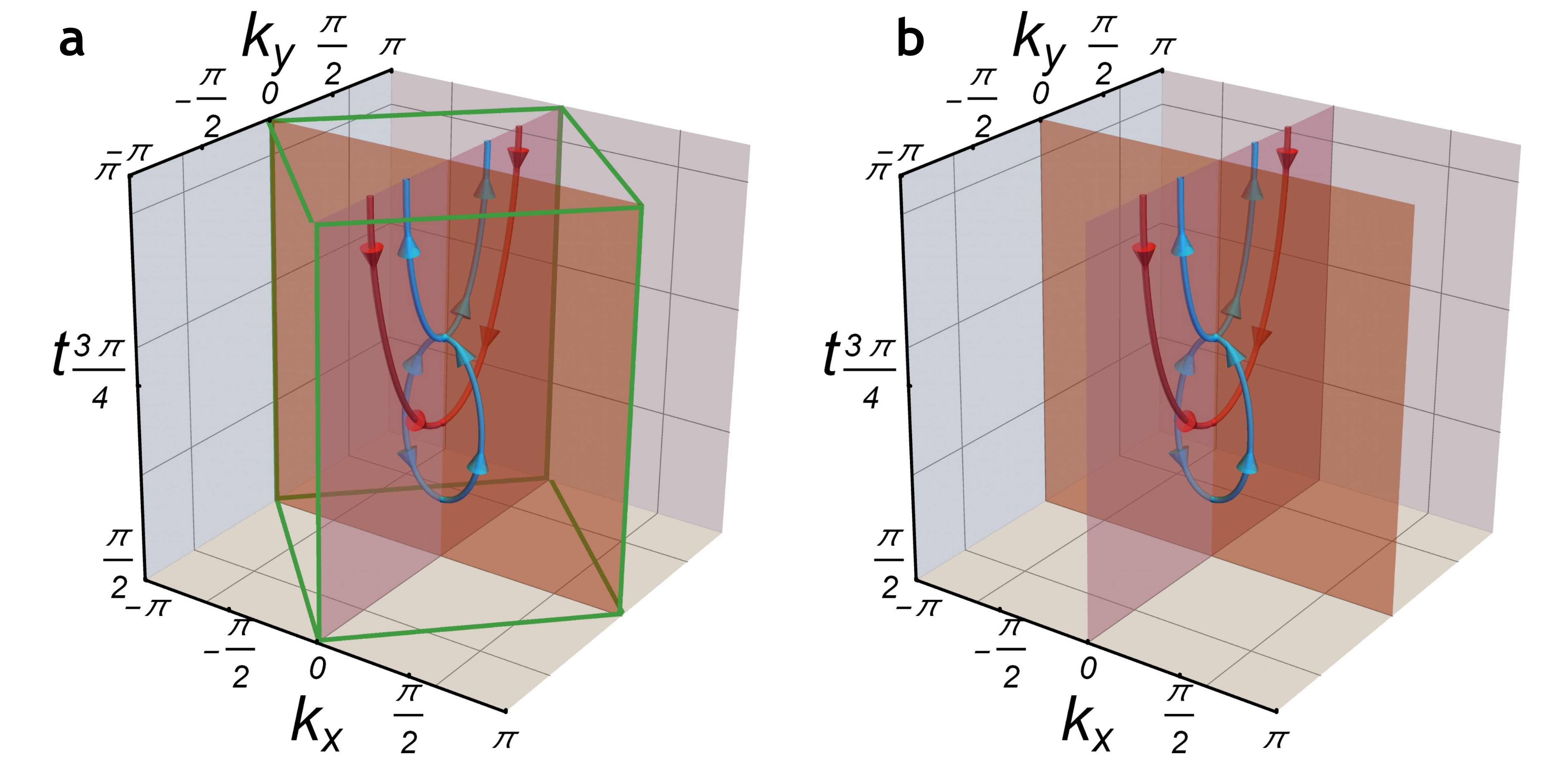}
	\caption{\textsf{\textbf{Node trajectories for the alternative braiding protocol.} \textbf{a.} Trajectory of band nodes of the 2D tight-binding model in Eq.~(\ref{eqn:model-2}) in the $(k_x,k_y)$-plane as a function of adiabatic parameter $t\in[\tfrac{\pi}{2},\pi]$. The red (blue) lines indicate trajectories of band nodes in the upper (lower) band gaps, and the indicated orientations follow the convention from Refs.~\cite{Wu:2018b} and~\cite{Tiwari:2019}. The green frame indicates the Brillouin zone of the tight-binding model, and the two diagonal planes indicate the mirror-invariant lines of the Brillouin zone. \textbf{b.} Analogous plot for the $\bs{k}\cdot\bs{p}$ model in Eq.~(\ref{eqn:model-3}). The node trajectories are almost identical (they are very slightly closer to $k_x = k_y = 0$). However, the periodic momentum space of the original tight-binding model is now replaced by the Euclidean $\reals^2$.}}
	\label{fig:braid}
\end{figure*} 

The dependence of the topological charge on the exchange of band nodes in momentum space implies robustness of the phenomenon, especially in contrast to certain other proposed schemes that flip the topological charge of a band node by relying on the periodicity of the momentum space. In particular, it has been reported by the insightful Ref.~\cite{Montambaux:2018} that in certain lattices the topological charge of a band node can be reversed simply by circumnavigating the BZ torus. Assuming the convention in which the Bloch Hamiltonian $H(\bs{k})$ is periodic in reciprocal lattice vectors, such phenomenon may arise when the $C_2\mcT$ operator explicitly depends on momentum ($\bs{k}$). Such a dependence often occurs for non-symmorphic lattices or for orbitals located on the unit cell boundary. Notably, this phenomenon is pertinent also to two-band models~\cite{Montambaux:2018}. However, as it explicitly relies on the periodicity of the momentum space, such a charge reversal cannot be reproduced by a $\bs{k}\cdot\bs{p}$ model. In contrast, the reciprocal braiding considered by our work does not rely on the periodicity of the momentum space but solely on the braided node trajectories, implying that it can be reproduced with a 3-band $\bs{k}\cdot\bs{p}$ model.

To obtain an explicit tight-binding model for such a ``local'' braiding, we consider a 2D lattice belonging to the $cmm$ wallpaper group, and we assume $p_x$, $p_y$ and $s$ orbitals located in the center of each unit cell. The conventional unit cell is a rectangle with dimensions $a$ and $b$, and the Brillouin zone is a rhombus with horizontal diagonal of length $2\pi/a$, and vertical diagonal of length $2\pi/b$. For simplicity, we set $a=1=b$ throughout the discussion. The lattice is symmetric under mirror $m_x:x\mapsto -x$ represented by $\hat{m}_x=\diag{(-1,+1,+1)}$, mirror $m_y:y\mapsto -y$ represented by $\hat{m}_y=\diag{(+1,-1,+1)}$, their composition $C_2:(x,y)\mapsto (-x,-y)$ represented by $\hat{C}_2=\diag{(-1,-1,+1)}$, as well as under time-reversal $\mcT$ represented by complex conjugation $\mcK$. The $C_2\mcT$ operator is rotated to the canonical form $\mcK$ [see Sec.~\ref{sec:reality} below] through a unitary rotation of the Hilbert space with $V = \diag{(+\imi,+\imi,+1)}$. The mirror symmetries are not essential for the node braiding, but they pin the band nodes to high-symmetry lines, simplifying the analysis. Mirror-breaking perturbations compatible with the $C_2\mcT$ symmetry preserve the non-trivial braiding characteristic of the model.

Setting the energy of the three orbitals to $T_{1,3,5}$, the intra-orbital hopping amplitudes between the four nearest-neighbor (NN) sites to $T_{2.4.6}$, and the inter-orbital hopping amplitudes for NN sites to $\pm T_{7,8,9}$, we obtain Bloch Hamiltonian
\begin{widetext}
\begin{equation}
H_\textrm{TB}(\bs{k}) = \left(\begin{array}{ccc}
T_1 + 4T_2 \cos k_x \cos k_y &
4T_7 \sin k_x \sin k_y       &
4\imi T_8 \sin k_x \cos k_y  \\
4T_7 \sin k_x \sin k_y       &
T_3 + 4T_4 \cos k_x \cos k_y &
-4\imi T_9 \cos k_x \sin k_y \\
-4\imi T_8 \sin k_x \cos k_y &
4\imi T_9 \cos k_x \sin k_y  &
T_5 + 4T_6 \cos k_x \cos k_y
\end{array}\right).\label{eqn:model-2}
\end{equation}
To realize the reciprocal braiding, we assume the following dependence of the parameters on time $t \in [\tfrac{\pi}{2},\pi]$,
\begin{eqnarray}
T_1 = 3.5 + 3 \cos t\qquad \qquad 
&T_2 = -0.1 + 0.2\cos t&\qquad\qquad 
T_3 = 0.25 - 0.25 \cos t \nonumber \\
T_4 = 0.1 \cos t \qquad\qquad \qquad\!\! 
&T_5 = -1.5 - \cos t\;\;\;\;& \qquad\qquad 
T_6 = 0.25 - 0.5 \cos t \\
T_7 = + 0.15 \qquad\qquad \qquad \;\,
&T_8 = - 0.5\cos t\;\;\;\;\;\;& \qquad\qquad 
T_9 = - 0.2.  \nonumber
\end{eqnarray}
The node trajectories as a function of time are plotted in  Fig.~\ref{fig:braid}{\bf a}. We observe that the pair of ``blue nodes'', which correspond to the lower/principal band gap and are pairwise created at time $t\approx \tfrac{16}{28}\pi$, fail to annihilate when they meet at a later time $t\approx \tfrac{22}{28}\pi$, after being braided with a ``red node'' (corresponding to the upper/adjacent band gap). The orientations of the trajectories displayed in Fig.~\ref{fig:braid} follow the convention explained in Refs.~\cite{Wu:2018b} and~\cite{Tiwari:2019}. A more detailed illustration of the spectrum along the two high-symmetry lines as a function of parameter $t$ is shown in Fig.~\ref{fig:braid-2}.

\begin{figure*}[t]
	\includegraphics[width=0.99\textwidth]{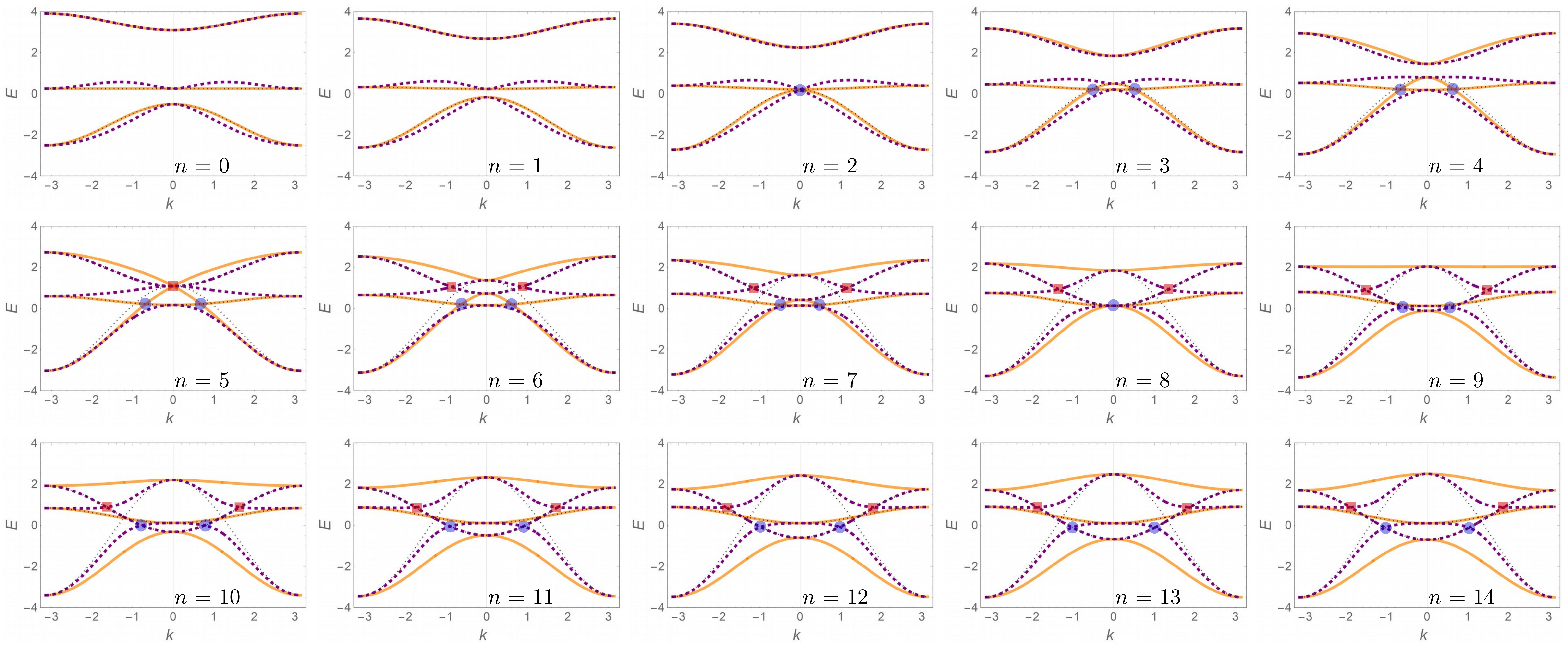}
	\caption{\textsf{\textbf{Spectrum of the alternative braiding Hamiltonian along the Brillouin zone diagonals.} Spectrum of the tight-binding Hamiltonian in Eq.~(\ref{eqn:model-2}) for times $t = \tfrac{\pi}{2}+n\tfrac{\pi}{28}$, plotted along the $k_x$-axis (solid orange lines), and along the $k_y$-axis (dashed purple lines). The blue circles (red squares) indicate band nodes formed inside the lower/principal (upper/adjacent) band gap. The gray dotted lines indicate the spectrum along either momentum axis when the orbital mixing terms $T_{7,8,9}$ are turned off.}}
	\label{fig:braid-2}
\end{figure*} 

Although the Hamiltonian in Eq.~(\ref{eqn:model-2}) is periodic in reciprocal lattice vectors, the change of the topological charge of the nodes after reciprocal braiding does \emph{not} depend on the periodicity of momentum space. One can confirm this by explicitly expanding the Hamiltonian to the second order in $k_{x,y}$, which implicitly replaces the BZ torus by $\reals^2$. The resulting $\bs{k}\cdot\bs{p}$ Hamiltonian
\begin{equation}
H_{\bs{k}\cdot\bs{p}}(\bs{k}) = \left(\begin{array}{ccc}
T_1 + 4T_2 \left(1-\tfrac{k_x^2 + k_y^2}{2}\right) &
4T_7 k_x k_y       &
4\imi T_8 k_x\\
4T_7 k_x k_y       &
T_3 + 4T_4 \left(1 - \tfrac{k_x^2 + k_y^2}{2} \right)&
-4\imi T_9 k_y \\
-4\imi T_8 k_x &
4\imi T_9 k_y &
T_5 + 4T_6 \left(1- \tfrac{k_x^2 + k_y^2}{2}\right))
\end{array}\right).\label{eqn:model-3}
\end{equation}
\end{widetext}
leads to node trajectories shown in Fig.~\ref{fig:braid}{\bf b}, which are just imperceptibly narrower than those of the tight-binding model in Eq.~(\ref{eqn:model-2}).

\section{Reality condition}\label{sec:reality}

In the main text we assumed that $C_2\mcT$ symmetry implies reality of the Bloch Hamiltonian. While this is not true in general, the conclusions presented in the main text still apply. More precisely, the presence of antiunitary operator $\mathpzc{A}$ that obeys (\emph{i}) $\mathpzc{A}^2 = +1$ and (\emph{ii}) $\forall\bs{k}:\mathpzc{A}\mcH(\bs{k})\mathpzc{A}^{-1} = \mcH(\bs{k})$, implies the existence of a Hilbert-space basis, in which the Hamiltonian $\mcH(\bs{k})$ is real. In this section we justify this claim by two different methods. We remark that the antiunitary $\mathpzc{A}$ that fulfills the two conditions can be realized as $C_2\mcT$ in two-dimensional spinful or spinless systems~\cite{Ahn:2018}, or as $\mcP\mcT$ in spinless systems of arbitrary dimension~\cite{Wu:2018b}. Therefore, for such symmetry settings, both the frame-rotation charge resp.~the Euler class can be defined if the right Hilbert-space basis has been adopted.

We first prove this statement formally, before providing a physical insight in the next paragraph. Every antiunitary operator $\mathpzc{A}$ can be represented as some unitary operator $\mathpzc{U}$ composed with the complex conjugation $\mathpzc{K}$~\cite{Wigner:1960}, i.e.~$\mathpzc{A}=\mathpzc{U}\mathpzc{K}$. Unitarity means that $\mathpzc{U}\mathpzc{U}^\dagger = \unit$, while $\mathpzc{A}^2 = +\unit$ implies that $\mathpzc{U}\mathpzc{U}^* = \unit$. It follows that $\mathpzc{U}=\mathpzc{U}^\top$. The Autonne-Takagi factorization~\cite{Horn:2012} then guarantees that $\mathpzc{U} = \mathpzc{V}\mathpzc{D}\mathpzc{V}^\top$ for some unitary $\mathpzc{V}$ and a diagonal matrix $\mathpzc{D} = \diag\{\e{\imi\varphi_j}\}_{j=1}^n$. Constructive and finite algorithms exist that find the Autonne-Takagi decomposition of a symmetric unitary matrix~\cite{mathSE:2026299}. Rotation of the Hilbert-space by unitary matrix $\sqrt{\mathpzc{D}^*} \mathpzc{V}^\dagger \equiv \mathpzc{W}$ then transform the antiunitary operator to $\mathpzc{W}\mathpzc{A}\mathpzc{W}^{\dagger} = \mathcal{K}$, i.e.~to the form assumed in the main text.

From a physical point of view, it is well known that an antiunitary symmetry squaring to $+1$ (rather than $-1$) 
does \emph{not} imply Kramers degeneracies~\cite{Kramers:1930}. In other words, eigenstates of $\mathpzc{A}$ form one-dimensional irreducible representations. By taking these eigenstates to form the basis of the Hilbert space, the antiunitary operator is represented by $\mathpzc{D}'\mathpzc{K}$, where $\mathpzc{D}' = \diag\{\e{\imi\varphi'_j}\}_{j=1}^n$ is a diagonal matrix of phase factors. Rotating the Hilbert space basis by $\mathpzc{W}' = \sqrt{\mathpzc{D}'^{*}}$ then transforms the antiunitary operator to the complex conjugation $\mathpzc{K}$. While the absence of Kramers doubling for such a symmetry is well accepted in the solid-state community, the formal proof of this statement actually follows from the Autonne-Takagi factorization, as discussed in the previous paragraph.  

\section{Euler form in three-band models.} \label{sec:three-band}

In this section, we elaborate on the analogy between the Euler class and the first Chern number by briefly focusing on the minimal models. More specifically, we show that both the Berry curvature $\mathsf{F}(\bs{k})$ of one band obtained from a two-band complex Hamiltonian, as well as the Euler form $\textrm{Eu}(\bs{k})$ of two bands obtained from a three-band real Hamiltonian, can be understood by considering geometry on a 2-sphere ($S^2$). The discussion below thus proves the geometric interpretation of Euler form presented in Fig.~4(b) of the main text. 

Let us first recall the mathematics behind the first Chern number of a complex Hamiltonian, defined in Eq.~(21) of Methods, for the case of two bands. Hermitian two-band Hamiltonians can be decomposed using the Pauli matrices and the unit matrix as
\begin{equation}
H(\bs{k}) = \bs{h}(\bs{k})\cdot\bs{\sigma} + h_0(\bs{k})\unit
\end{equation}
where $h_{0,x,y,z}(\bs{k})$ are real functions of momentum. Spectral flattening brings eigenvalues of the Hamiltonian to $\pm 1$ without changing the band topology~\cite{Ryu10_NJP}, and is associated with replacing $h_0\mapsto 0$ and $\bs{h} \mapsto \bs{h}/||{\bs{h}}|| \equiv \bs{n}$. The band topology of the two-band complex Hamiltonian is thus completely captured by the three-component unit vector $\bs{n}(\bs{k})\in S^2$. It is known~\cite{HasanKane10_RMP,Qi11_RMP} that the Berry curvature of one of the two bands can be expressed as
\begin{equation}
\mathsf{F}_{ij} = \tfrac{1}{2}\bs{n}\cdot( \partial_{k_i} \bs{n} \,\times\, \partial_{k_j} \bs{n})\label{eqn:original-Berry-curv},
\end{equation}
such that $\mathsf{F}_{ij} \, d k_i \, d k_j$ corresponds to one half of the (oriented) solid angle spanned by $\bs{n}$ on the $S^2$ as the momentum argument is varied over a rectangle of size $\de k_i \times \de k_
j$. For a \emph{closed} two-dimensional base manifold, the vector $\bs{n}$ has to wrap around the unit sphere an integer number of times, hence the integral of $\mathsf{F}_{ij} \, d k_i \, d k_j$, i.e.~the total (oriented) solid angle spanned by $\bs{n}$, must be quantized to integer multiples of $2\pi$. Therefore, $c_1(E)$ defined in Eq.~(21) of Methods is an integer in the case of two-band models. This simple argument does not generalize to models with more than two bands, in which case one has to follow the proof outlined in the paragraph \emph{Quantization of Euler class} in Methods. 

We find that a very similar geometric interpretation also applies to Euler form of two bands obtained from a three-band real Hamiltonian. In this case, spectral flattening brings the Hamiltonian with two occupied bands and one unoccupied band to
\begin{equation}
H(\bs{k}) = 2\, \bs{n}(\bs{k})\cdot  \bs{n}(\bs{k})^\top - \unit,\label{eqn:3-band-flattened}
\end{equation} 
where $\bs{n}(\bs{k}) = \bs{u}^1(\bs{k}) \times \bs{u}^2(\bs{k}) \in S^2$ is the (normalized) column vector representing the unoccupied state, which can be represented as cross product of the (normalized) occupied states $\bs{u}^1(\bs{k})$ and $\bs{u}^2(\bs{k})$. Note that, because of the reality condition, the left (``bra'') and the right (``ket'') eigenstates are componentwise equal to each other. The quadratic dependence of the Hamiltonian on the unit vector, manifest in Eq.~(\ref{eqn:3-band-flattened}), implies that vectors $\pm\bs{n}$ represent the same Hamiltonian. Therefore the space of \emph{unique} spectrally flattened 3-band Hamiltonians is $S^2/\ztwo \equiv \reals P^2$~\cite{Wu:2018b}. However, if the vector bundle defined by $H(\bs{k})$ is orientable (which is a necessary condition to define Euler form), then there are no closed paths $\gamma\subset B$ in the base manifold which would be mapped by the Hamiltonian to the non-contractible path in $\reals P^2$. Therefore, Euler form of an \emph{orientable} rank-2 bundle obtained from a three-band real Hamiltonian, is related to geometry on $S^2$. In fact, we show below that
\begin{equation}
\textrm{Eu}_{ij} = \bs{n} \cdot \left(\partial_{k_i} \bs{n} \times \partial_{k_j} \bs{n} \right), \label{eqn:Euler-3-band}
\end{equation}
which [besides the altered interpretation of $\bs{n}(\bs{k})$] qualitatively differs from Eq.~(\ref{eqn:original-Berry-curv}) only by the absence of the prefactor $\tfrac{1}{2}$. Following the same arguments as for the first Chern number, we find that for three-band models the Euler class $\chi(E)$ defined in Eq.~(20) of Methods must be an \emph{even} integer. This agrees with the known fact, that odd values of the Euler class (corresponding to a non-trivial second Stiefel-Whitney class) require models with at least two occupied and with at least two unoccupied bands~\cite{Ahn:2018}. We also remark that the cross-product definition of $\bs{n}(\bs{k})$ in terms of the two occupied states makes the expression in Eq.~(\ref{eqn:Euler-3-band}) invariant only under the \emph{proper} $\mathsf{SO}(2)$ gauge transformations of the occupied states, reminding us of the importance of orientability of the vector bundle.

The remainder of this section contains a proof of Eq.~(\ref{eqn:Euler-3-band}). While the logic of the proof is straightforward, some of the expressions are rather lengthy. We employ the Einstein summation convention, and we write
\begin{equation}
n_a = \epsilon_{abc} u_b^1 u_c^2
\end{equation}
where $\epsilon$ is the Levi-Civita symbol. The right-hand side of Eq.~(\ref{eqn:Euler-3-band}) can be expressed as
\begin{equation}
\textrm{(\ref{eqn:Euler-3-band})}= \epsilon_{abc} u_b^1 u_c^2 \epsilon_{ade} \partial_{k_i} (\epsilon_{dfg} u^1_f u^2_g)\partial_{k_j}(\epsilon_{ehi} u^1_h u^2_i).
\end{equation}
Using the identity $\epsilon_{abc}\epsilon_{ade} = \delta_{bd}\delta_{ce} - \delta_{be}\delta_{cd}$, and performing the summation over indices $b$ and $c$, we obtain
\begin{equation}
\textrm{(\ref{eqn:Euler-3-band})} = \epsilon_{dfg} \epsilon_{ehi}(u_d^1 u_e^2 - u_e^1 u_d^2)\partial_{k_i} ( u^1_f u^2_g)\partial_{k_j}( u^1_h u^2_i).\label{eqn:Euler-proof-2}
\end{equation}
To get rid off the remaining Levi-Civita symbols, we use
\begin{widetext}
\begin{equation}
\epsilon_{dfg} \epsilon_{ehi} = \delta_{de}(\delta_{fh}\delta_{gi}-\delta_{fi}\delta_{gh}) - \delta_{dh}(\delta_{fe}\delta_{gi}-\delta_{fi}\delta_{ge}) + \delta_{di}(\delta_{fe}\delta_{gh}-\delta_{fh}\delta_{ge}).\label{eqn:Euler-proof-3}
\end{equation}
This long identity has to be substituted into Eq.~(\ref{eqn:Euler-proof-2}). Note that the combinations of Kronecker symbols containing $\delta_{de}$ trivially lead to zero after the substitution, because $\delta_{de} (u_d^1 u_e^2 - u_e^1 u_d^2) = 0$. The remaining terms in Eq.~(\ref{eqn:Euler-proof-3}), after summing over indices $e$, $h$, and $i$, lead to
\begin{eqnarray}
\textrm{(\ref{eqn:Euler-3-band})} &=& -(u_d^1 u_f^2 - u_f^1 u_d^2)\partial_{k_i}(u_f^1 u_g^2)\partial_{k_j}(u_d^1 u_g^2) + (u_d^1 u_g^2 - u_g^1 u_d^2)\partial_{k_i}(u_f^1 u_g^2)\partial_{k_j}(u_d^1 u_f^2) \nonumber \\
&\phantom{=}& + (u_d^1 u_f^2 - u_f^1 u_d^2)\partial_{k_i}(u_f^1 u_g^2)\partial_{k_j} (u_g^1 u_d^2) - (u_d^1 u_g^2 - u_g^1 u_d^2)\partial_{k_i}(u_f^1 u_g^2)\partial_{k_j}(u_f^1 u_g^2) \label{eqn:Euler-proof-4}
\end{eqnarray}
Performing the derivatives by parts, Eq.~(\ref{eqn:Euler-proof-4}) expands into 32 individual terms
\begin{eqnarray}
\textrm{(\ref{eqn:Euler-3-band})} &=&
-{\color{blue}u_d^1} u_f^2 (\partial_{k_i} u_f^1) u_g^2 {\color{blue}(\partial_{k_j} u_d^1)} u_g^2
-u_d^1 u_f^2 (\partial_{k_i} u_f^1) {\color{blue}u_g^2} u_d^1 {\color{blue}(\partial_{k_j} u_g^2)}
-u_d^1 {\color{red}u_f^2 u_f^1} (\partial_{k_i} u_g^2) (\partial_{k_j} u_d^1) u_g^2
-u_d^1 {\color{red}u_f^2 u_f^1} (\partial_{k_i} u_g^2) u_d^1 (\partial_{k_j} u_g^2) \quad \nonumber \\
&\phantom{=}&
+{\color{blue}u_f^1} u_d^2 {\color{blue}(\partial_{k_i} u_f^1)} u_g^2 (\partial_{k_j} u_d^1) u_g^2
+{\color{blue}u_f^1} u_d^2 {\color{blue}(\partial_{k_i} u_f^1)} u_g^2 u_d^1 (\partial_{k_j} u_g^2)
+u_f^1 u_d^2 u_f^1 {\color{blue}(\partial_{k_i} u_g^2)} (\partial_{k_j} u_d^1) {\color{blue}u_g^2}
+u_f^1 {\color{red}u_d^2} u_f^1 (\partial_{k_i} u_g^2) {\color{red}u_d^1} (\partial_{k_j} u_g^2) \nonumber \\
&\phantom{=}& 
+ {\color{blue}u_d^1} u_g^2 (\partial_{k_i} u_f^1 ) u_g^2 {\color{blue}(\partial_{k_j} u_d^1)} u_f^2
+ {\color{green}u_d^1 u_g^2} (\partial_{k_i} u_f^1 ) {\color{green}u_g^2 u_d^1} (\partial_{k_j} u_f^2)
+ {\color{blue}u_d^1} u_g^2 u_f^1 (\partial_{k_i} u_g^2) {\color{blue}(\partial_{k_j} u_d^1)} u_f^2
+ u_d^1 {\color{blue}u_g^2} u_f^1 {\color{blue}(\partial_{k_i} u_g^2)}  u_d^1 (\partial_{k_j} u_f^2) \nonumber \\
&\phantom{=}& 
- {\color{red}u_g^1} u_d^2 (\partial_{k_i} u_f^1 ) {\color{red}u_g^2} (\partial_{k_j} u_d^1) u_f^2
- u_g^1 u_d^2 (\partial_{k_i} u_f^1 ) u_g^2 u_d^1 (\partial_{k_j} u_f^2)
- u_g^1 u_d^2 {\color{red}u_f^1} (\partial_{k_i} u_g^2) (\partial_{k_j} u_d^1) {\color{red}u_f^2}
- u_g^1 {\color{red}u_d^2} u_f^1 (\partial_{k_i} u_g^2)  {\color{red}u_d^1} (\partial_{k_j} u_f^2) \nonumber \\
&\phantom{=}& 
+ {\color{red}u_d^1} u_f^2 (\partial_{k_i} u_f^1) u_g^2 (\partial_{k_j} u_g^1) {\color{red}u_d^2}
+ u_d^1 u_f^2 (\partial_{k_i} u_f^1) {\color{red}u_g^2 u_g^1} (\partial_{k_j} u_d^2)
+ u_d^1 {\color{red}u_f^2 u_f^1} (\partial_{k_i} u_g^2) (\partial_{k_j} u_g^1) u_d^2
+ u_d^1 {\color{red}u_f^2 u_f^1}  (\partial_{k_i} u_g^2) u_g^1 (\partial_{k_j} u_d^2) \nonumber \\
&\phantom{=}& 
- {\color{blue}u_f^1} u_d^2 {\color{blue}(\partial_{k_i} u_f^1)} u_g^2 (\partial_{k_j} u_g^1) u_d^2
- {\color{blue}u_f^1} u_d^2 {\color{blue}(\partial_{k_i} u_f^1)} u_g^2 u_g^1 (\partial_{k_j} u_d^2)
- {\color{green}u_f^1 u_d^2 u_f^1} (\partial_{k_i} u_g^2) (\partial_{k_j} u_g^1) {\color{green}u_d^2}
- u_f^1 {\color{blue}u_d^2} u_f^1  (\partial_{k_i} u_g^2) u_g^1 {\color{blue}(\partial_{k_j} u_d^2)} \nonumber \\ 
&\phantom{=}& 
- {\color{red}u_d^1} u_g^2 (\partial_{k_i} u_f^1) u_g^2 (\partial_{k_j} u_f^1) {\color{red}u_d^2}
- u_d^1 u_g^2 {\color{blue}(\partial_{k_i} u_f^1)} u_g^2 {\color{blue}u_f^1} (\partial_{k_j} u_d^2 )
- u_d^1 u_g^2 {\color{blue}u_f^1}  (\partial_{k_i} u_g^2) {\color{blue}(\partial_{k_j} u_f^1)} u_d^2
- u_d^1 {\color{blue}u_g^2} u_f^1 {\color{blue}(\partial_{k_i} u_g^2)} u_f^1 (\partial_{k_j} u_d^2) \nonumber \\
&\phantom{=}& 
+ {\color{red}u_g^1} u_d^2 (\partial_{k_i} u_f^1) {\color{red}u_g^2} (\partial_{k_j} u_f^1) u_d^2
+ {\color{red}u_g^1} u_d^2 (\partial_{k_i} u_f^1) {\color{red}u_g^2} u_f^1 (\partial_{k_j} u_d^2 )
+ u_g^1 u_d^2 {\color{blue}u_f^1}  (\partial_{k_i} u_g^2) {\color{blue}(\partial_{k_j} u_f^1)} u_d^2
+ u_g^1 {\color{blue}u_d^2} u_f^1 (\partial_{k_i} u_g^2) u_f^1 {\color{blue}(\partial_{k_j} u_d^2)} \quad \nonumber
\end{eqnarray}
Most of these 32 terms vanish. To see this, we use orthonormality $u^1_a u^2_a = 0$ on the vector coordinates displayed in {\color{red}red}. Furthermore, the normalization $u^1_a u^1_a = 1 = u^2_a u^2_a$ implies that $u^1_a (\partial_{k_\beta} u^1_a) = 0 = u^2_a (\partial_{k_\beta} u^2_a)$, which we indicate in {\color{blue}blue}. Only two terms remain, in which we further use the normalization to $1$ on vector components displayed in {\color{green}{green}}. After renaming the repeated indices, we obtain
\begin{equation}
\textrm{Eu}_{ij} = (\partial_{k_i} u_f^1)(\partial_{k_j} u_f^2) - (\partial_{k_j} u_f^1)(\partial_{k_i} u_f^2) \equiv \braket{\partial_{k_i} u^1}{\partial_{k_j} u^2} - \braket{\partial_{k_j} u^1}{\partial_{k_i} u^2}.
\end{equation}
The last expression exactly corresponds to the components of the the Euler form over two bands as defined in Eq.~(19) of Methods. This completes the proof of Eq.~(\ref{eqn:Euler-3-band}).
\end{widetext}

\section{Analytic properties of Euler form at principal nodes} \label{sec:singularity} 

\subsection{General remarks}\label{sec:flag}

In the main text we consider the Euler form $\textrm{Eu}(\bs{k})$ defined by the two principal bands. Note that adjacent nodes pose problems for the mathematical construction. This is because circumnavigating an adjacent node reverses the orientation of one of the principal Bloch states (the one that participates in the formation of the adjacent node), but not the other one. Therefore, parallel transport around an adjacent node is associated with an \emph{improper} gauge transformation $X = \sigma_z \notin \mathsf{SO}(2)$. Since such vector bundle is not orientable, its Euler curvature cannot be defined. This is the reason why we only consider calculations over regions with no adjacent nodes.

In contrast, understanding the behavior of the Euler form near a \emph{principal} node is more subtle. First, circumnavigating a principal node reverses the sign of \emph{both} principal Bloch states, which corresponds to a \emph{proper} gauge transformation $X = -\unit \in \mathsf{SO}(2)$, such that Euler form of the bundle is well-defined on an annulus around the node. On the other hand, the eigenstates of the Hamiltonian are not continuous functions of $\bs{k}$ at the location of principal nodes, suggesting that the derivatives of the eigenstates are not well-defined at these points. Nevertheless, the rank-2 vector bundle spanned by the two principal bands is actually continuous and differentiable at principal nodes. The last two statements are not in contradiction! Indeed, a discontinuous basis of the bundle does not imply discontinuity of the bundle. There may be (and we show that there really is) a different basis which is perfectly continuous at principal nodes. 

However, one has to consider the relevance of the two bases for physical observables. Since the two principal bands are separated by a band gap away from the principal nodes, the discontinuous basis of the bundle corresponding to the principal eigenstates has a \emph{special physical significance}. Especially, we show in Sec.~\ref{sec:boundary} that this canonical (although discontinuous) basis encodes observable features, such as the path-dependent ability of band nodes to annihilate. This information is lost once we allow for mixing of the two principal eigenstates by a general $\mathsf{SO}(2)$ gauge transformation -- such as when going to the basis that reveals the continuity of the bundle.

The physical importance of the eigenstate basis can be naturally built into the mathematical description by introducing the notion of a \emph{flag bundle}~\cite{Fulton:1984,Bouhon:2020}, and by contrasting to the more familiar notion of a \emph{vector bundle}. On the one hand, the two principal bands span a two-dimensional vector space that varies continuously as a function of $\bs{k}$. In other words, the two principal bands constitute a rank-two \emph{vector bundle}. In this description, the two-dimensional vector space admits arbitrary $\mathsf{SO}(2)$ gauge transformations, and the physical eigenstate basis is not treated as being special. Therefore, in this description it is natural to adopt the continuous basis, where there are no singularities -- and thus also no topological information. This is compatible with the well-known fact that any smooth vector bundle is trivializable over a patch (disc). 

In contrast, a \emph{flag bundle} keeps information about the subdivision of the two-dimensional vector space into two one-dimensional vector spaces, and thus preserves information about the two physical eigenstates. In this language, the admissible (orientation-preserving) gauge transformations of the two dimensional vector space are just $\mathsf{S}[\mathsf{O}(1)\!\times\!\mathsf{O}(1)] \cong \ztwo$ (corresponding to $X=\pm \unit$, excluding $X = \pm\sigma_z$). The flag bundle exhibits a singularity at the principal node, which is unremovable by admissible gauge transformations, and which can be described using a topological invariant -- the Euler class. Note that the rank-2 flag bundle exhibits a singularity and topological invariant on a patch (disc), even though the corresponding rank-2 vector bundle is continuous and trivializable on the patch. This is because the vector bundle description carries less information, which is insufficient to describe the principal nodes.

In the remainder of this text, we avoid the notion of flag bundle. Instead, we speak of the \emph{computation in the eigenstate basis}, which (based on the remark in the previous paragraphs) we except to convey topological information. This is contrasted to the \emph{computation in the continuous} (or \emph{rotated}) basis, where no topological information is expected. The robustness of the invariant derived in the eigenstates basis cannot be rigorously justified within the vector bundle description, and only comes from the restricted $\mathsf{S}[\mathsf{O}(1)\!\times\!\mathsf{O}(1)] \cong \ztwo$ gauge transformation of the flag bundle formalism.

\subsection{Computation in the eigenstate basis}\label{sec:derivation-eigenbasis}

We begin our discussion by presenting the most general Hamiltonian near a principal node to the linear order in momentum. We treat the obtained Hamiltonian perturbatively, and we consider the \emph{eigenstate basis} to reveal the structure of the Euler form near the principal node. 

We first consider a two-band model that exhibits a node at $\bs{k} = \bs{0}$ at zero energy. To linear order, the Hamiltonian near the node must take the form
\begin{equation}
\mcH_\textrm{2-band}(\bs{k}) = \sum_{i=1}^2 \sum_{j=0}^3 k_i h_{ij} \sigma_j
\end{equation}
where $k_{1,2}$ are the two momentum coordinates, $h_{ij}$ are 8 real coefficients, and $\{\sigma_j\}_{j=0}^3$ is the unit matrix and the three Pauli matrices. It is known that by a suitable proper rotation and by a linear rescaling of the momentum coordinates to $\bs{\kappa}(\bs{k})$, we can bring the Hamiltonian to the form
\begin{equation}
\mcH_\textrm{2-band}({\bs{\kappa}}) = \left(\begin{array}{cc}
\kappa_1 + \varepsilon(\bs{\kappa})  &   \pm\kappa_2            \\
\pm\kappa_2   &   - \kappa_1 + \varepsilon(\bs{\kappa})         
\end{array}\right),
\end{equation}
where $\varepsilon(\bs{\kappa}) = v_1 \kappa_1 + v_2 \kappa_2$ describes the tilting of the band node~\cite{Soluyanov:2015}. Since the considered coordinate transformation is linear, the associated Jacobian $J_{ij} = \partial \kappa_i/\partial k_j$ is a  constant matrix. The $\pm$ sign corresponds to nodes with positive vs.~negative winding number $w \in \pi_1[\mathsf{SO}(2)] = \intg$, which we keep unspecified throughout the whole section. We assume $\det J > 0$, i.e.~the change of coordinates preserves the orientation of the bundle.

If there are additional bands, then the same rotation of the two basis vectors brings the corresponding $n$-band linear-order Hamiltonian to the form
\begin{equation}
\mcH_\textrm{$n$-band}({\bs{\kappa}}) \!=\! \left(\begin{array}{ccc}
\!\!\kappa_1 \!+\! \varepsilon(\bs{\kappa})\!\!  
&   \pm\kappa_2  
& f^\top(\bs{\kappa})             \\
\pm \kappa_2   
&   \!\!- \kappa_1 \!+\! \varepsilon(\bs{\kappa})\!\!  
& g^\top(\bs{\kappa})           \\
f(\bs{\kappa}) 
& g(\bs{\kappa})    
& \!\!\mathcal{E} \!+\! h(\bs{\kappa})\!\!
\end{array}\right)\!,\label{eqn:ugly-Hamiltonian}
\end{equation}
where $f(\bs{\kappa})$ and $g(\bs{\kappa})$ are $\bs{\kappa}$-linear $(n-2)$-component column vectors with real components $\{f_c(\bs{\kappa})\}_{c=3}^n$ and $\{g_c(\bs{\kappa})\}_{c=3}^n$, $\mathcal{E}$ is a non-degenerate diagonal matrix of $(n-2)$ non-zero band energies, and $h(\bs{\kappa})$ is $\bs{\kappa}$-linear Hermitian matrix of size $(n-2)\times(n-2)$. In Eq.~(\ref{eqn:ugly-Hamiltonian}) we explicitly assume that the additional $(n-2)$ basis vectors of the Hilbert space are given by the additional eigenstates of the Hamiltonian at the node, therefore $h(\bs{\kappa} \!=\! \bs{0}) = \triv$. Therefore, after adopting the properly rotated and rescaled momentum coordinates and the right Hilbert-space basis, the model in Eq.~(\ref{eqn:ugly-Hamiltonian}) represents the \emph{most general} $n$-band real Hamiltonian near a principal node to linear order in momentum.

To proceed, we split the Hamiltonian in Eq.~(\ref{eqn:ugly-Hamiltonian}) into $\mcH_0 = \diag(\epsilon(\bs{\kappa}),\epsilon(\bs{\kappa}),\mathcal{E})$, and a ``perturbation'' $\mcH'$ that contians all the terms linear in $\bs{\kappa}$, that is
\begin{equation}
\mcH'(\bs{\kappa}) = \left(\begin{array}{ccc}
\kappa_1        &   \pm\kappa_2        &   f^\top(\bs{\kappa}) \\
\pm\kappa_2        &   -\kappa_1       &   g^\top(\bs{\kappa}) \\
f(\bs{\kappa})  &   g(\bs{\kappa})  &   h(\bs{\kappa})
\end{array}\right).\label{eqn:perturbation}
\end{equation}
The first step of the pertubation theory requires us to find states that diagonalize the matrix $\mcH'_{ab} = \bra{a}\mcH'\ket{b}$ with $a,b\in\{1,2\}$ representating the degenerate states at the principal nodes. We take $\ket{1} = (1\; 0\; 0 \; \ldots)^\top$ and $\ket{2} = (0\; 1 \; 0 \; \ldots)^\top$, in which case this matrix corresponds simply to the upper-left $2\times 2$ block of Eq.~(\ref{eqn:perturbation}). If we further decompose $\bs{\kappa}$ using polar coordinates as $\kappa_1 = \kappa \cos \phi$ and $\kappa_2 = \kappa \sin \phi$, this block is diagonalized by
\begin{equation}
\!\!\!\ket{1^{(0)}\!} = \left(\begin{array}{c}
\!\pm\sin\tfrac{\phi}{2}\! \\ \!-\cos\tfrac{\phi}{2}\! \\ \bs{0}
\end{array}\right)\;
\textrm{and}
\; \ket{2^{(0)}\!} = \zeta\left(\begin{array}{c}
\!+\cos\tfrac{\phi}{2}\! \\ \!\pm\sin\tfrac{\phi}{2}\! \\ \bs{0}
\end{array}\right)\label{eqn:diag-perturb}
\end{equation}
where $\zeta = \pm 1$ corresponds to two different orientations of the bundle. Changing the relative sign between the two states corresponds to orientation-changing gauge transformation $X=\pm\sigma_z$. On the other hand, increasing $\phi \mapsto \phi + 2\pi$ flips the sign of \emph{both} bands, which corresponds to a proper gauge transformation $X=-\unit$. 

The first-order correction to the states in Eq.~(\ref{eqn:diag-perturb}) is given by 
\begin{equation}
|a^{(1)}\rangle = \sum_{c = 3}^n\frac{\bra{c}\mcH'\ket{a^{(0)}}}{\epsilon(\bs{\kappa}) - \mathcal{E}_c}\ket{c}
\end{equation}
where $\ket{c}$ is the $c^\textrm{th}$ element of the basis in which we expressed Eq.~(\ref{eqn:perturbation}). This prescription does not lead to a change in the first two compoments of the principal vectors, while for compoments with $c \geq 3$ we find
\begin{eqnarray}
\langle{c}|1^{(1)}\rangle &=& 
\tfrac{1}{\varepsilon(\bs{\kappa})-\mathcal{E}_c}\left[\pm f_c(\bs{\kappa})\sin\tfrac{\phi}{2}-g_c(\bs{\kappa})\cos\tfrac{\phi}{2}\right]  \\
\langle{c}|2^{(1)}\rangle &=& 
\tfrac{\zeta}{\varepsilon(\bs{\kappa})-\mathcal{E}_c}\left[+f_c(\bs{\kappa})\cos\tfrac{\phi}{2}\pm g_c(\bs{\kappa})\sin\tfrac{\phi}{2}\right].\label{eqn:pert-results}
\end{eqnarray}
Note that the expressions inside the square brackets are linear in $\bs{\kappa}$, while the prefactor can be approximated for $\bs{\kappa}$ close to $\bs{0}$ as
\begin{equation}
\frac{1}{\varepsilon(\bs{\kappa})-\mathcal{E}_c} \approx -\frac{1}{\mathcal{E}_c} + \frac{v_1 \kappa_1 + v_2 \kappa_2}{\mathcal{E}_c^2}.
\end{equation}
Therefore, if we are after terms of the lowest order in $\bs{\kappa}$, we can approximate the prefactor simply by $-1/\mathcal{E}_c$. Furthermore, notice that states $\ket{1^{(0+1)}}$ and $\ket{2^{(0+1)}}$, which we obtained by performing the first-order perturbation theory, are not properly normalized. However, since the lowest-order corrections $\ket{1^{(1)}}$ and $\ket{2^{(1)}}$ are \emph{linear} in $\bs{\kappa}$, the correction from the proper normalization would be \emph{quadratic} in $\bs{\kappa}$. More explicitly, normalizing the states would induce a prefactor of the form
\begin{equation}
\frac{1}{\sqrt{1+||\bs\kappa \mathcal{N}||^2}}\approx 1 - \frac{1}{2} ||\bs\kappa \mathcal{N}||^2
\end{equation}
where $||\bs{\kappa}\mathcal{N}||$ represents the normalization of the first-order correction. Since we are interested only in corrections to the principal states of the \emph{lowest order} in $\bs{\kappa}$, we safely ignore the normalization.

We have established the lowest-order (linear) corrections in $\bs{\kappa}$ to the principal Bloch states in Eq.~(\ref{eqn:diag-perturb}). We can use the obtained states to calculate the Euler connection and the Euler form \emph{in the eigenstate basis},
\begin{equation}
\!\!\textrm{a}_i \!=\! \braket{1}{\partial_{\kappa_i} 2}\;\;\,\textrm{and}\;\;\, \textrm{Eu} \!=\! \braket{\partial_{\kappa_1} 1}{\partial_{\kappa_2} 2} - \braket{\partial_{\kappa_2} 1}{\partial_{\kappa_1} 2}\label{eqn:calculate-euler}
\end{equation}
where we dropped the superscript ``$^{(0+1)}$'' for brevity. However, note that the symbolic computation of the derivatives in Eq.~(\ref{eqn:calculate-euler}) cannot be fully trusted for $k\to0$ because of the discontinuity of the states in Eq.~(\ref{eqn:diag-perturb}) at $k=0$. Nevertheless, for now we ignore this possible source of problem. To proceed, we use that
\begin{eqnarray}
\partial_{\kappa_1} &=& \frac{\partial \kappa}{\partial \kappa_1} \partial_\kappa + \frac{\partial \phi}{\partial \kappa_1} \partial_\phi = \cos\phi \partial_\kappa - \frac{\sin\phi}{\kappa}\partial_\phi \label{eqn:identities-1}\\
\partial_{\kappa_2} &=& \frac{\partial \kappa}{\partial \kappa_2} \partial_\kappa + \frac{\partial \phi}{\partial \kappa_2} \partial_\phi = \sin\phi \partial_\kappa + \frac{\cos\phi}{\kappa}\partial_\phi .
\end{eqnarray}
We also rewrite
\begin{eqnarray}
f_c(\bs{\kappa}) &=& \alpha_c \kappa \cos\phi + \beta_c \kappa \sin\phi \\
g_c(\bs{\kappa}) &=& \gamma_c \kappa \cos\phi + \delta_c \kappa \sin\phi.\label{eqn:identities-4}
\end{eqnarray}
With the help of \texttt{Mathematica}, we find the Euler connection to the leading order in $\kappa$ as
\begin{equation}
\mathbf{a} = \frac{\pm\zeta}{2\kappa}\left(\sin\phi,-\cos\phi\right) + \mathcal{O}(\kappa)\label{eqn:discont-gauge-a}
\end{equation}
which \emph{diverges} as we approach the node. In contrast, the Euler form to the leading order in $\kappa$ is
\begin{eqnarray}
\textrm{Eu} &=& -\,\zeta \sum_{c=3}^{n}\frac{1}{\mathcal{E}_c^2}\Big[ (\beta_c\gamma_c - \alpha_c\delta_c)    \qquad \label{eqn:Eu-general-result}\\
&\phantom{=}& \hspace{-0.4cm} \pm\tfrac{1}{2}(\alpha_c\cos\phi \!+\! \beta_c \sin\phi)^2\pm\tfrac{1}{2}(\gamma_c\cos\phi \!+\! \delta_c \sin\phi)^2\Big] \nonumber \\
&\phantom{=}& \textrm{(wrong! -- read end of Sec.~\ref{sec:derivation-eigenbasis})} \nonumber
\end{eqnarray}
which \emph{does not diverge}  at the node. All the $1/\kappa$ factors, present in some of the previous formulae, cancel out. Substituting back the original coordinates, $\bs{\kappa} \to \bs{k}$, corresponds to a double multiplication by the (constant) Jacobian matrix, which also does not induce a divergence. One observes (wrongly! as explained later) that $\textrm{Eu}$ is \emph{discontinuous} at principal nodes (note the dependence on $\phi$ for $\kappa\to 0$), which prevents the applicability of Stokes' theorem. Nevertheless, the singularity in Eq.~(\ref{eqn:Eu-general-result}) is integrable, which suggests that the definition of Euler class in Eq.~(1) of the main text is mathematically meaningful. By integrating over an infinitesimal disc $\mathcal{D}^\epsilon$ with radius $\epsilon$ around the principal node, we find $\lim_{\epsilon\to 0} \int_{\mathcal{D}^\epsilon} \textrm{Eu} = 0$ and $\lim_{\epsilon\to 0} \oint_{\partial\mathcal{D}^\epsilon} \textrm{a} = \mp \pi\zeta$, such that the Euler class on the disc [cf.~Eq.~(5) of the main text, and Sec.~\ref{sec:boundary} below]
\begin{equation}
\chi(\mathcal{D}^\epsilon) = \mp\tfrac{\zeta}{2}
\end{equation} 
is non-zero, and carries topological information.

However, as we warned before Eq.~(\ref{eqn:identities-1}), the \emph{presented computation cannot be fully trusted due to the discontinuity of states $\ket{1}$ and $\ket{2}$ at $\kappa=0$.} This means that both the Euler connection and Euler form are, strictly speaking, undefined at $\kappa=0$, and we have attempted to perform a series expansion around a singular point. It turns out that our result for Euler connection $\textbf{a}$ in Eq.~(\ref{eqn:discont-gauge-a}) is correct. In contrast, as we convincingly show in the next subsection, the result for the Euler connection $\textrm{Eu}$ in Eq.~(\ref{eqn:Eu-general-result}) is \emph{wrong!} -- namely the whole discontinuous contribution to Eu in \emph{the second line of Eq.~(\ref{eqn:Eu-general-result}) should not be there}. The Euler form turns out to be \emph{continuous} at the principal node, and what really prevents the applicability of the Stokes' theorem is the divergence of $\textbf{a}$ for $\kappa\to 0$ observed in Eq.~(\ref{eqn:discont-gauge-a}).

\subsection{Computation in the rotated (continuous) basis}

In the previous subsection, we attempted to derive the analytic properties of Euler connection and Euler curvature by performing the computation in the eigenstate basis. However, the discontinuous nature of the eigenstates at the principal node led to some uncontrolled steps, and the derived results cannot be fully trusted. Here, we approach the problem differently. We perform a gauge transformation to a \emph{rotated basis}, which reveals the continuity of the vector bundle spanned by the two principal bands and principal nodes. The continuity allows us to keep the computation of Euler form and Euler connection under control. At the end of the calculation, we perform a large gauge transformation back to eigenstate basis, and compare to the results obtained in Sec.~\ref{sec:derivation-eigenbasis}. While this is admittedly a somewhat indirect approach to derive the results, the advantage is that we only have to deal with the singularity at the very last step when we do the gauge transformation back to the eigenstate basis.

Recall that a rank-2 bundle is a collection of two-dimensional planes -- one plane per every point of the base space. The specific choice of these planes varies continuously between the points of the base space. Importantly, these planes need not in general be equipped with any intrinsic basis. The basis vectors that we use to span these planes are just an auxiliary tool. Performing an $\mathsf{SO}(2)$ gauge transformation on the two vectors spanning the rank-2 bundle does not correspond to a change of the bundle, just to a \emph{change of coordinates} that we use to describe the bundle. In particular, it is convenient to consider the ``mixed'' (i.e.~\emph{rotated}) states
\begin{eqnarray}
\ket{A(\bs{\kappa})} &=& \pm \sin\tfrac{\phi}{2} \ket{1^{(0+1)}} + \zeta\,\cos\tfrac{\phi}{2} \ket{2^{(0+1)}}  \label{eqn:rotated-basis-1} \\
\ket{B(\bs{\kappa})} &=& - \zeta\,\cos\tfrac{\phi}{2} \ket{1^{(0+1)}} \pm \sin\tfrac{\phi}{2} \ket{2^{(0+1)}} 
\end{eqnarray}
which are related to the eigenstates of the perturbed Hamiltonian by a proper gauge transformation
\begin{equation}
X(\bs{\kappa};\zeta) = \left(\begin{array}{cc}
\pm \sin\tfrac{\phi}{2}     &   +\zeta\cos \tfrac{\phi}{2}  \\
-\zeta\cos\tfrac{\phi}{2}   &   \pm \sin \tfrac{\phi}{2}
\end{array}\right) \label{eqn:singular-GT}
\end{equation}
Using trigonometric identities, we find that to linear order in $\bs{\kappa}$ these rotated vectors are
\begin{equation}
\!\!\!\!\ket{A(\bs{\kappa})} \!=\!\! \left(\begin{array}{c}
1 \\
0 \\
\!\!\{\tfrac{-f_c}{\mathcal{E}_c}\}_{c=3}^n \!\!
\end{array}\right) \!\quad\!
\ket{B(\bs{\kappa})} \!=\! \zeta\!\left(\begin{array}{c}
0 \\
1 \\
\!\!\{\tfrac{-g_c}{\mathcal{E}_c}\}_{c=3}^n\!\!
\end{array}\right).  \label{eqn:rotates-vectors-A}
\end{equation}
These are manifestly \emph{continuous and differentiable} at $\bs{\kappa}\to\bs{0}$, meaning that the vector bundle spanned by the two principal bands is continuous and differentiable at the principal node. We find that the Euler connection and the Euler form to the leading order in $\kappa$ \emph{in the rotated basis} are
\begin{equation}
\tilde{\mathbf{a}} \!=\! \braket{A}{\bs{\nabla}B} \!=\! \sum_{c=3}^n\frac{\zeta \left(\alpha_c \kappa_x +\beta_c \kappa_y\right) }{\mathcal{E}_c^2}(\gamma_c,\delta_c)\label{eqn:rotated-basis-conn}
\end{equation}
and
\begin{eqnarray}
\tilde{\textrm{Eu}} &=& \braket{\partial_1 A}{\partial_2 B} - \braket{\partial_2 A}{\partial_1 B} \nonumber \\ 
&=& -\zeta \sum_{c=3}^n\frac{1}{\mathcal{E}_c^2} \left( \beta_c\gamma_c - \alpha_c\delta_c\right)\label{eqn:fake-news}
\end{eqnarray}
which are both perfectly regular for $\bs{\kappa}\to\bs{0}$. In particular, the result in Eq.~(\ref{eqn:fake-news}) exactly reproduces the \emph{$\phi$-independent} contribution to Eq.~(\ref{eqn:Eu-general-result}). Moreover, one can show that Stokes' theorem applies, i.e. the surface integral of $\tilde{\textrm{Eu}}$ is exactly cancelled by the boundary integral of $\tilde{\mathbf{a}}$. Therefore, in the rotated basis the Euler class $\tilde{\chi}(\mathcal{D})$ cannot be meaningfully defined for a manifold $\mathcal{D}$ with a boundary [see Sec.~\ref{sec:boundary} for definition]. Euler class defined this way would always be zero.

We now translate the results in Eqs.~(\ref{eqn:rotated-basis-conn}) and~(\ref{eqn:fake-news}) to the eigenstates basis, which corresponds to a gauge transformation using $X^\top$ from Eq.~(\ref{eqn:singular-GT}). We must keep in mind that this is a large gauge transformation, i.e.~it introduces a singular winding at one point (at the principal node), but it is smooth elsewhere. First, it follows from Eq.~(18) of Methods that everywhere outside the principal node we must have $\textrm{Eu} = \tilde{\textrm{Eu}}$. Since the latter is continuous, the correct functional form of the Euler form in the eigenstate basis is
\begin{equation}
\textrm{Eu} = -\zeta \sum_{c=3}^n\frac{1}{\mathcal{E}_c^2} \left( \beta_c\gamma_c - \alpha_c\delta_c\right),
\end{equation}
which can be analytically continued to the point $\bs{\kappa} = \bs{0}$.

A more dramatic change occurs when applying Eq.~(13) of Methods to transform the connection $\tilde{\mathbf{a}}$ to the eigenstate basis. First, we derive
\begin{equation}
X^\top d X = \frac{d \phi}{2}\left(\begin{array}{cc}0 & \mp \zeta \\ \pm \zeta & 0\end{array}\right)\quad \textrm{(for $\kappa \neq 0$),}
\end{equation}
from where $\Pf(X^\top d X) = \mp\tfrac{1}{2}\zeta d\phi $ everywhere outside the principal node. From $\phi =\arctan(\frac{\kappa_y}{\kappa_x})$ we obtain 
\begin{equation}
d\phi = \frac{-\kappa_y d \kappa_x + \kappa_x d \kappa_y}{\kappa_x^2+\kappa_y^2} = \frac{-\sin\phi}{\kappa} d \kappa_x + \frac{\cos\phi}{\kappa}  d\kappa_y.
\end{equation}
Note that the term $\Pf(X^\top d X)$ diverges for $\kappa\to 0$ (consequence of $X$ being a \emph{large} gauge transformation around the principal node), and therefore it completely overshadows the expression in Eq.~(\ref{eqn:rotated-basis-conn}) which vanishes completely for $\kappa\to 0$. Therefore, according to Eq.~(13) of Methods, the Euler connection in the eigenstate basis to the leading order in $\kappa$ equals
\begin{equation}
{\mathbf{a}} = - X^\top d X = \frac{\pm \zeta}{2\kappa}(-\sin\phi,\cos\phi),
\end{equation}
which agrees with our result in Eq.~(\ref{eqn:discont-gauge-a}).

It is worth to emphasize that the finite quantized value of $\chi(\mathcal{D}^\epsilon)$ for the disc enclosing a principal node in the eigenstate basis comes entirely from the connection on the boundary $\partial\mathcal{D}^\epsilon$ and is inherited from the large gauge transformation from the continuous basis. This is compatible with our claim, mentioned in the beginning of Sec.~\ref{sec:singularity}, that it is expected that the eigenstate basis carries the topological information about the principal nodes on a patch, while the continuous basis does not have access to this information. Nevertheless, if one considers a \emph{closed} base manifold $B$ (e.g.~the whole BZ torus), then $\partial B = \varnothing$, and only the surface integral of the Euler form contributes to the Euler class $\chi(B)$. Since Euler form is the same both for the eigenstate basis and the continuous basis, the value of Euler class on a \emph{closed} manifold is quantized for both computations.

\section{Euler class for manifolds with a boundary}
\label{sec:boundary}

In this section, we generalize the notion of Euler class of a pair of principal bands in two important ways. First, we consider a base manifold \emph{with a boundary}. Throughout the discussion, we explicitly assume that the manifold is topologically a \emph{disc} (denoted $\mathcal{D}$), which has boundary homeomorphic to a circle ($\partial \mathcal{D} \simeq S^1$). Nevertheless, most of the presented results readily generalize to arbitrary manifolds with a boundary as long as the vector bundle spanned by the principal bunds remains orientable. Second, we admit the occurrence of principal nodes inside the base manifold. Crucially, to obtain useful information from such a generalization, it is essential to \emph{adopt the eigenstate basis}. We find that the value of the Euler class on a disc $\mathcal{D}$, denoted $\chi(\mathcal{D})$, indicates whether the collection of principal nodes can annihilate inside $\mathcal{D}$.

The first generalization is straightforward. If there are no band nodes in $\mathcal{D}$, then the eigenstate basis is continuous. Therefore, Stokes' theorem guarantees that
\begin{equation}
\chi(\mathcal{D}) \!=\! \frac{1}{2\pi}\! \left(\int_{\mathcal{D}}\!\!\textrm{Eu} - \oint_{\partial\mathcal{D}}\!\!\!\!\mathrm{a}\right) \!= 0 \!\!\quad\textrm{(if no nodes in $\mathcal{D}$)}\label{eqn:boring-invariant}
\end{equation}
is invariant. Let us further assume the presence of principal nodes in $\mathcal{D}$ (but no adjacent nodes as we want the bundle to remain orientable). We proved in the second part of Sec.~\ref{sec:singularity} that the vector bundle spanned by the two principal bands is continuous everywhere, including at the nodes. However one has to depart from the eigenstate basis to reveal this fact, and instead has to consider the continuous ``mixed'' basis, cf.~Eqs.~(\ref{eqn:rotated-basis-1}--\ref{eqn:fake-news}). In the continuous basis, Stokes' theorem applies, which implies the validity of Eq.~(\ref{eqn:boring-invariant}) even in the presence of principal nodes. One concedes that the vector bundle spanned by the two principal bands on a disc cannot support a non-trivial topological invariant. 

However, one should keep in mind the \emph{physical realization} of the vector bundle as a pair of energy bands that are non-degenerate away from the band nodes. This interpretation equips the bundle with a canonical basis, namely the \emph{eigenstate basis} discussed at length in Sec.~\ref{sec:singularity} (cf.~the comments on \emph{flag bundle} description in Sec.~\ref{sec:flag}). Therefore, one should only consider deformations of the vector bundle which preserve this additional structure. Indeed, we show below that such a constraint allows for a subtly modified definition of $\chi(\mathcal{D})$, which constitutes a \emph{half-integer} topological invariant. 

To develop the generalization of the Euler class, one should first recognize that each principal node is associated with a Dirac string~\cite{Ahn:2019}. Returning back to Eqs.~(\ref{eqn:diag-perturb}--\ref{eqn:pert-results}), we observe that the continuous real gauge for eigenstates near a principal node is \emph{double-valued}, namely the overall sign of both states is reversed if we increase $\phi \mapsto \phi + 2\pi$. Therefore, any \emph{single-valued} gauge must necessarily exhibit a discontinuity -- the Dirac string -- across which both principal states flip sign. Each principal node must constitute the end-point of a Dirac string. Furthermore, since away from band nodes the eigenstates basis is continuous, there are no other end-points for Dirac strings. Therefore, in the eigenstate basis there is a \emph{one-to-one correspondence} between the principal nodes and the end-points of Dirac strings. From this perspective, the gauge transformation to the continuous basis, analyzed in Eqs.~(\ref{eqn:diag-perturb}--\ref{eqn:pert-results}), can be understood as creation of a new Dirac string that \emph{exactly compensates} the ``physical'' Dirac string present in the eigenstate basis. This explains the difference between the Euler connection computed in the two different bases in Sec~\ref{sec:singularity}.

\begin{figure}[t]
	\includegraphics[width=0.33\textwidth]{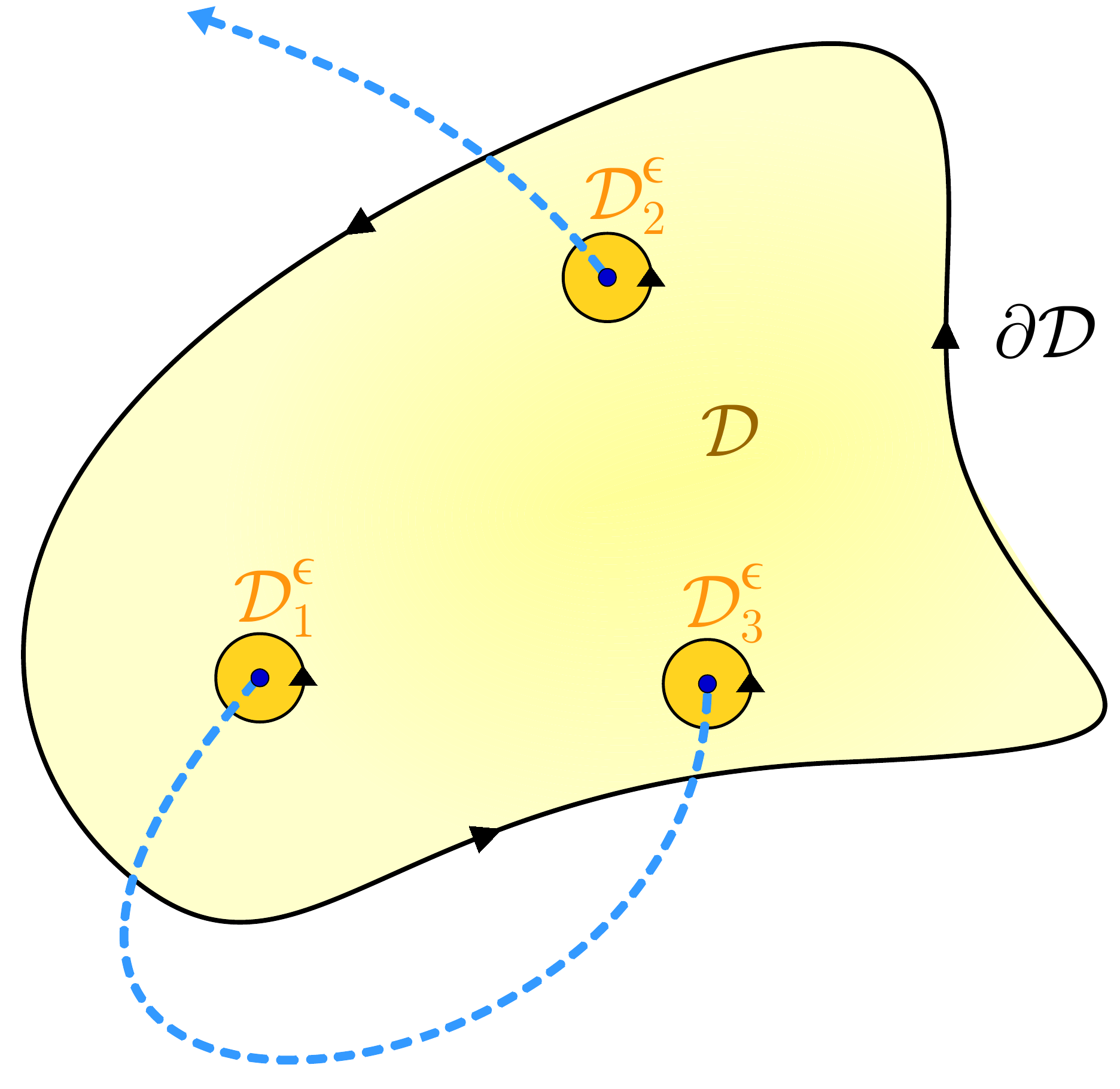}
	\caption{\textsf{\textbf{Relation between Euler class and Dirac strings.} Illustration of the use of Stokes' theorem as discussed in the text. Region $\mathcal{D}$ is the disc on which we study the real orientable vector bundle, and the blue dots are the principal nodes. We apply Stokes' theorem to $\mathcal{D}\backslash \cup_\alpha\! \mathcal{D}_\alpha^\epsilon$, i.e.~to the region with tiny discs around the nodes cut out. The blue dashed lines represent Dirac strings. Only the end-points of Dirac strings are physically meaningful. In the case of the eigenstate basis the Dirac string end-points coincide with principal nodes, whereas a gauge transformation that reguralizes the nodes creates a new Dirac string that exactly compensates the singular behavior.}}
	\label{fig:Stokes}
\end{figure} 

The exact position of the Dirac strings (i.e.~besides their fixed end-points) is arbitrary, and should not have any bearing on physical observables. Indeed, we find that both Euler connection and Euler form are \emph{continuous} at a Dirac string. This readily follows from the transformation rules presented in the paragraph on topology of real vector bundles in Methods, namely we see from Eqs.~(13) and~(18) of Methods that the orientation-preserving gauge transformation $X=-\unit$ ($\det X = +1$) leaves both $\textrm{Eu}(\bs{k})$ and $\textrm{a}(\bs{k})$ invariant. We conclude that the relation $\textrm{Eu}(\bs{k}) = d\,\mathrm{a}(\bs{k})$ remains valid along Dirac strings, meaning that they are no obstructions for the use of Stokes' thorem. Therefore, the \emph{only} obstructions for Stokes' theorem are the principal nodes themselves, since at their locations the derivatives of the principal eigenstates are not well defined and the Euler connection diverges, cf.~Sec.~\ref{sec:singularity}. We thus use Stokes' theorem to relate
\begin{eqnarray}
2\pi\chi(\mathcal{D}) 
&=& \int_{\mathcal{D}}\!\textrm{Eu} - \oint_{\partial\mathcal{D}}\!\!\!\textrm{a} \label{eqn:chi-d-def}\\
&=& \sum_\alpha \left(\int_{\mathcal{D_\alpha^\epsilon}}\!\!\!\textrm{Eu} -\oint_{\partial \mathcal{D}_\alpha^\epsilon}\!\!\!\!\!\mathrm{a}\,\right)\label{eqn:corrected-stokes}
\end{eqnarray}
where the summation index $\alpha$ indicates the principal nodes inside the region $\mathcal{D}$, and $\mathcal{D}_{\alpha}^\epsilon$ is a disc with radius $\epsilon$ centered at principal node $\alpha$. 

To simplify Eq.~(\ref{eqn:corrected-stokes}), we proceed as follows. First, we found in Sec.~\ref{sec:singularity} that Euler form is bounded near principal nodes, hence the integrals over $\mathcal{D}_\alpha^\epsilon$ converge to $0$ in the limit $\epsilon\to 0$. Furthermore, it follows from Eq.~(\ref{eqn:discont-gauge-a}) that $\mathbf{a}\cdot d\bs{\kappa} = \mp \tfrac{1}{2}\zeta d\phi$ for $\epsilon\to 0$, which integrates to $\mp \zeta \pi$ on $\partial\mathcal{D}_\alpha^\epsilon$. Plugging this result into Eq.~(\ref{eqn:corrected-stokes}), we find
\begin{equation}
\chi(\mathcal{D}) = \frac{\zeta}{2}\sum_\alpha w_\alpha \label{eqn:Eu-class-windings}
\end{equation}
where $w_\alpha = \pm 1$ describes the \emph{winding number} of the principal node, while $\zeta=\pm 1$ is the global orientation of the vector bundle. The result in Eq.~(\ref{eqn:Eu-class-windings}) proves that $\chi(\mathcal{D})$ of the orientable bundle spanned by the two principal states on region $\mathcal{D}$ with a boundary is a \emph{half-integer topological invariant} if one works in the eigenstate basis. 

Let us conclude the section with several remarks:
\begin{enumerate}
\item Note that Eq.~(\ref{eqn:Eu-class-windings}) expresses $\chi(\mathcal{D})$ as a sum of $\pm\tfrac{1}{2}$ quanta carried by the principal nodes. If the nodes were able to annihilate inside $\mathcal{D}$, then we would be left with a nodeless region, for which we proved in Eq.~(\ref{eqn:boring-invariant}) that $\chi(\mathcal{D}) = 0$. Therefore, non-trivial value $\chi(\mathcal{D})\neq 0$ is an obstruction for annihilating the principal nodes inside $\mathcal{D}$. 
\item The two admissible values of a winding number $w_\alpha = \pm 1$ are reminiscent of the frame-rotation angle associated with the node being either $+\pi$ or $-\pi$. We prove in Sec.~\ref{sec:frame-rotation} that this intuition is correct, i.e.~the two quantities are in an exact one-to-one correspondence.
\item One should take into account that Eq.~(\ref{eqn:Eu-class-windings}) is not as useful in pratice as it appears! To make sure that we take the same orientation of the vector bundle at all principal nodes, it is necessary to know the bundle along some trajectories connecting the principal nodes. To avoid this extra work, our numerical algorithm for computing $\chi(\mathcal{D})$, presented in Sec.~\ref{sec:numerical} is based on directly implementing Eq.~(\ref{eqn:chi-d-def}). 
\item In the presence of additional \emph{adjacent} nodes, the vector bundle ceases to be orientable, and the relative orientation of two principal nodes depends on the specific choice of trajectory connecting them. This foreshadows the non-Abelian conversion of band nodes which we discuss in the main text of the manuscript. This ``braiding'' phenomenon is more carefully exposed in the next section.
\end{enumerate}

\section{Non-Abelian frame-rotation charge}
\label{sec:frame-rotation}

In this section, we review the original derivation of the non-Abelian exchange of band nodes in $\bs{k}$-space, which was obtained by Ref.~\cite{Wu:2018b} using homotopy theory~\cite{Nakahara:2003}. We subsequently show that the same non-Abelian behavior is reproduced by considering the Euler class on manifolds with a boundary, as has been defined in Sec.~\ref{sec:boundary}. Similar observations on a less formal level were also made by Ref.~\cite{Ahn:2019}. The exact correspondence between the two approaches provides a proof that the two distinct mathematical descriptions of the braiding phenomena (homotopy theory vs.~cohomology classes) are two windows into the same underlying topological structure.

In the homotopic description~\cite{Mermin:1979}, one begins with identifying the space $M_N$ of $N$-band real symmetric matrices that do not exhibit level degeneracy. This corresponds to Bloch Hamiltonians at momenta lying away from band nodes. With this assumption, we can uniquely order \emph{all} eigenstates of $H(\bs{k})$ according to increasing energy into an $\mathsf{O}(N)$ matrix $\{|u^a({\bs{k})}\rangle\}_{a=1}^N \equiv F({\bs{k}})$, which can be interpreted as a an \emph{orthonormal $N$-frame}. We further adjust band energies $\{\eps^a({\bs{k})}\}_{a=1}^N$ to some standard values (\emph{e.g.}~$\eps^a = a$) while preserving their ordering. The space of such \emph{spectrally normalized} Hamiltonians is $M_N = \mathsf{O}(N)/\ztwo^N$, where the quotient corresponds to flipping the overall sign of any of the $N$ eigenvectors, which leaves the spectrally normalized Hamiltonian $H(\bs{k})=\sum_{a=1}^N \ket{u^a(\bs{k})} \eps^a \bra{u^a(\bs{k})}$ invariant. Band nodes correspond to obstructions for a unique ordering of bands by energy, and thus induce discontinuities of frame $F({\bs{k}})$.  

To describe the band nodes, we study the topology of space $M_N$. Since there are $2^N$ different frames (corresponding to the quotient $\ztwo^N$) which all represent the same Hamiltonian, the following scenario is possible: We start at some point $\bs{k}_0$, and we follow the continuously rotating frame $F({\bs{k}})$ that encodes the Hamiltonian $H(\bs{k})$ along $\bs{k}\in \gamma$, until we reach again $\bs{k}_0$ as the final point. Comparing the initial vs.~the final frame at $\bs{k}_0$, we may find that they are two \emph{different} of the $2^N$ frames representing the same Hamiltonian $\mcH(\bs{k}_0)$. We say that the Hamiltonian underwent a non-trivial \emph{frame rotation}, which represents a non-trivial closed path inside $M_N$. For example, we know from Secs.~\ref{sec:singularity} and~\ref{sec:boundary} that a band node leads to a $\pi$-rotation of the frame by $X = (1,\ldots,1,-1,-1,1,\ldots,1)$, where the two negative entries correspond to the two bands forming the node. More generally, since the handedness of the frame cannot change under the continuous evolution along $\gamma$, only $\tfrac{1}{2}\cdot 2^N  = 2^{N-1}$ of the elements with $\det X = +1$ can actually be reached. Since the frame rotation is quantized to a discrete set of elements, it constitutes a topological invariant of the Hamiltonian $H(\bs{k})$ along path $\gamma$, which cannot change under continuous deformations as long as the spectrum along $\gamma$ remains non-degenerate.

\begin{figure}[t]
	\includegraphics[width=0.3\textwidth]{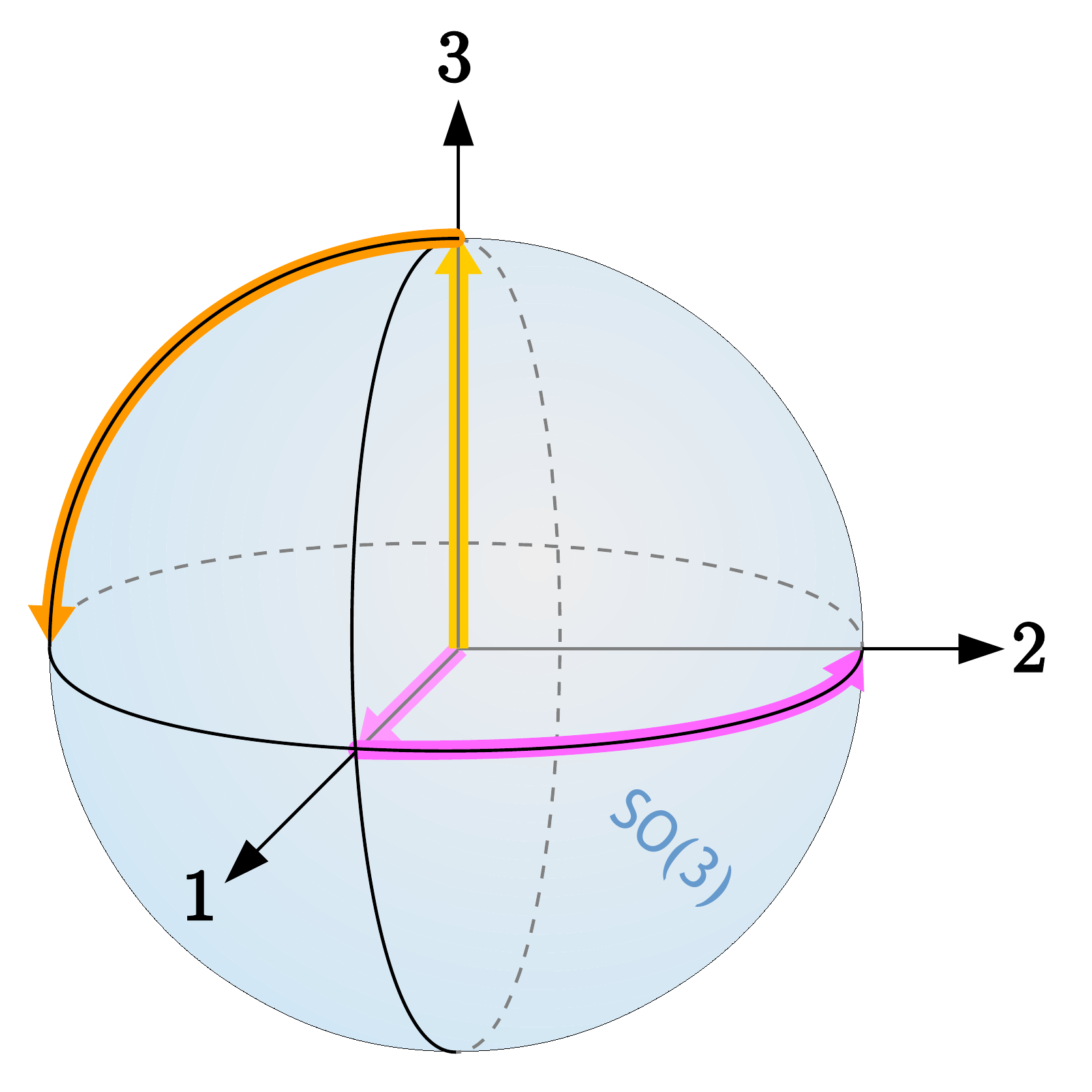}
	\caption{\textsf{\textbf{Commutativity of $\pi$ rotations in three dimensions.} If we represent clockwise rotation by angle $0 \leq \alpha \leq \tfrac{\pi}{2}$ around unit vector $\bs{n}$ by point with position $\bs{r} = \alpha \bs{n}$, then $\mathsf{SO}(3)$ looks like a ball with radius $\pi$ and with antipodal points on the surface identified. The orange vs.~the pink trajectory represent two ways of composing a $\pi$ rotation around axis $1$ with a $\pi$ rotation around axis $3$. The two trajectories cannot be continuously deformed into one another, i.e.~they are topologically distinct. From this observation we deduce that topological charges associated with band nodes in consecutive band gaps do not commute.}}
	\label{fig:so-3}
\end{figure} 

To explain the origin of the non-Abelian exchange of band nodes, let us briefly focus on models with $N=3$ bands. The same discussion also applies to \emph{any three consecutive bands} in models with $N\geq 3$ bands. A node formed by the lower (upper) two bands corresponds to a $\pi$-rotation in the first (last) two coordinates, i.e.~$X_{12}=\diag(-1,-1,+1)$ [$X_{23}=\diag(+1,-1,-1)$]. A path that encloses \emph{one of each} species of nodes is associated with total frame rotation $X_{13} = \diag(-1,+1,-1)$. However, while the geometric transformations 
\begin{equation}
X_{12}\cdot X_{23} = X_{23}\cdot X_{12} \label{eqn:pi-rot-commute}
\end{equation}
clearly commute, the continuous \emph{paths} in $\mathsf{SO}(3)$ that realize the left-hand vs.~the right-hand side of Eq.~(\ref{eqn:pi-rot-commute}) are topologically \emph{distinct}. To see this, recall that $\mathsf{SO}(3)$ can be visualized as a solid three-dimensional ball with radius $\pi$ and with antipodal points on the surface being pairwise identified. This relation is achieved by mapping $R(\alpha;\bs{n})$ (i.e.~a clockwise rotation by angle $0 \leq \alpha \leq \pi$ around axis given by unit vector $\bs{n}$), with a point inside the ball at position $\bs{r} = \alpha\bs{n}$. 
Then rotating first by $\pi$ around axis $1$ and then by $\pi$ around axis $3$ traces the pink path in Fig.~\ref{fig:so-3}, while performing the two rotations in reverse order produces the orange path in Fig.~\ref{fig:so-3}, which follows from
\begin{eqnarray}
\!\!\!\!\!\!R(\alpha;\hat{\bs{e}}_3)\cdot R(\pi;\hat{\bs{e}}_1) &=&  R(\pi;+\cos\tfrac{\alpha}{2}\hat{\bs{e}}_1+\sin\tfrac{\alpha}{2}\hat{\bs{e}}_2) \\
\!\!\!\!\!\!R(\alpha;\hat{\bs{e}}_1)\cdot R(\pi;\hat{\bs{e}}_3) &=&  R(\pi;+\cos\tfrac{\alpha}{2}\hat{\bs{e}}_3-\sin\tfrac{\alpha}{2}\hat{\bs{e}}_2)
\end{eqnarray}
where $\hat{\bs{e}}_i$ indicates the unit vector directed along axis $i\in\{1,2,3\}$. The two paths in Fig.~\ref{fig:so-3} both connect the center of the ball to the $\pi$-rotation around axis $2$. However, these paths cannot be continuously deformed into each other, i.e.~they are topologically distinct. This ultimately follows from the fact that $\mathsf{SO}(3)$ is not a simply connected space. As a consequence, the ordering of the group elements from the $\ztwo^N$ quotient matters, and the topological charges associated with a pair of band nodes inside consecutive band gaps \emph{do not commute}!

A careful analysis~\cite{Wu:2018b} reveals that the algebra of closed paths in space $M_N = \mathsf{O}(N)/\ztwo^N$ is governed by group $\mathsf{Q}_N$ (called ``Salingaros group''~\cite{Salingaros:1983}) which is uniquely characterized by the following four conditions~\cite{Tiwari:2019}. We use $+1$ to indicate the identity element.
\begin{itemize}
\item[(\emph{i})] There is a unique element $-1 \neq +1$ which has the property $(-1)^2 = +1$. 
\item[(\emph{ii})] For each band gap $1 \leq G \leq (N - 1)$ there is an associated element $g_G$ such $(g_G)^2 = -1$. 
\item[(\emph{iii})] $g_G \cdot g_{G'} = \epsilon g_{G'}\cdot g_G $, where $\epsilon = -1$ (anticommute) if $\abs{G - G'} = 1$  and $\epsilon = +1$ (commute) otherwise. 
\item[(\emph{iv})] All elements of $\mathsf{Q}_N$ can be expressed by composing elements $g_G$.
\end{itemize}
The element $-1$ corresponds to a $2\pi$ rotation (around any axis), and corresponds to the generator of the fundamental group $\pi_1[\mathsf{SO}(N)]=\ztwo$~\cite{Francis:1994}. Most interestingly, condition (\emph{iii}) states that band nodes in consecutive band gaps carry \emph{anticommuting charges}. This corresponds to the fact, visible in Fig.~\ref{fig:so-3}, that 
\begin{equation}
R(\pi,\hat{\bs{e}}_3)\cdot R(\pi,\hat{\bs{e}}_1) = \underbrace{R(2\pi,\hat{\bs{e}}_2)}_{``-1"} \cdot R(\pi,\hat{\bs{e}}_1) \cdot R(\pi,\hat{\bs{e}}_3) \label{eqn:anticommuting-patsh}
\end{equation}
if the rotations are interpreted as \emph{paths} in $\mathsf{SO}(3)$. The group $\mathsf{Q}_3$ coincides with the quaternion group $\{\pm 1,\pm\imi,\pm\imj,\pm\imk\}$, therefore $\mathsf{Q}_N$ for $N\geq 3$ has been dubbed ``generalized quaternions'' by Ref.~\cite{Wu:2018b}.

\begin{figure}[t]
	\includegraphics[width=0.44\textwidth]{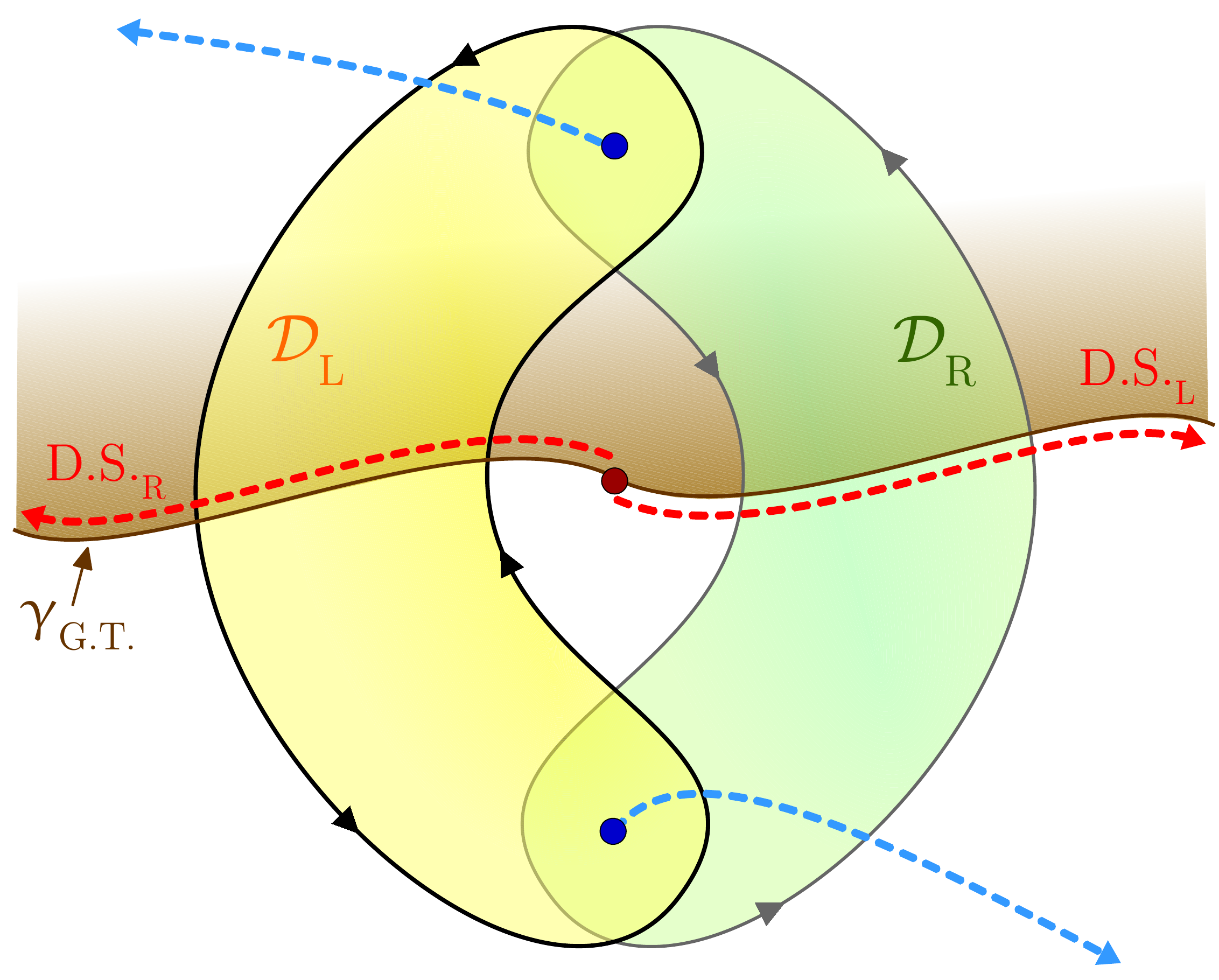}
	\caption{\textsf{\textbf{Relation of the non-commutative topological charge to Dirac strings.} We consider two principal nodes (blue dots) near an adjacent node (red dot). Principal nodes are end-points of Dirac strings associated with a proper gauge transformation $X=-\unit$ (dashed blue lines), which does not affect the continuity of Euler connection and Euler form of the vector bundle spanned by the two principal bands. On the other hand, the adjacent node is an end-point of a Dirac string associated with an improper transformation $X=\pm\sigma_z$ (dashed red line), which reverses the orientation of the bundle. The yellow resp.~the green disc $\mathcal{D}_{\textrm{L},\textrm{R}}$ correspond to two topologically different coverings of the principal nodes, for which the bundle can be made orientable by choosing an appropriate position of the red Dirac string (indicated by $\textrm{D.S.}_\textrm{L}$ resp.~$\textrm{D.S.}_\textrm{R}$). The two gauges are related by a gauge transformation which has an orientation-reversing discontinuity along $\gamma_\textrm{G.T.}$ (solid brown line). This gauge transformation reverses the orientation of the bundle near one of the nodes, thus flipping the relative contribution of the two nodes to the sum in Eq.~(\ref{eqn:Eu-class-windings}).}}
	\label{fig:non-commute}
\end{figure} 

We finally show that the same anticommuting behaviour follows by studying the Euler form on manifolds with a boundary as defined in Sec.~\ref{sec:boundary}. Our proof thus successfully bridges the homotopic description of Ref.~\cite{Wu:2018b} with the cohomological description proposed by Ref.~\cite{Ahn:2019} and further elaborated by the present work. To observe the non-trivial exchange, let us consider the situation, illustrated in Fig.~\ref{fig:non-commute}, with two principal nodes (blue dots) near an adjacent node (red dot). We know from Sec.~\ref{sec:boundary} that principal nodes are end-points of Dirac strings carrying a proper gauge transformation $X\!=\!-\unit$ on the two principal bands (dashed blue lines). We showed in the same section that such a gauge transformation is harmless for the calculation of the Euler class. On the other hand, adjacent nodes are end-points of Dirac strings carrying an \emph{improper} gauge transformation $X\!=\!\pm\sigma_z$, which flips the sign of \emph{just one} of the principal bands. Therefore, the bundle spanned by the two principal bands is non-orientable on regions containing such ``adjacent'' Dirac strings. Especially, an annulus enclosing the adjacent node is traversed by such a Dirac string for any single-valued gauge of the eigenstate basis, i.e.~the bundle spanned by the principal bands is non-orientable on such an annulus.

Nevertheless, the total Euler class of the two principal nodes can still be calculated, provided that one covers the nodes with a disc lying to the side of the adjacent node. We show in Fig.~\ref{fig:non-commute} two such discs, labelled $\mathcal{D}_{\textrm{L},\textrm{R}}$, which lie to the left (yellow) resp.~to the right (green) of the adjacent node. Both discs admit a gauge with a well-defined orientation of the bundle, which is achieved by appropriately positioning the adjacent Dirac string (dashed red lines $\textrm{D.S.}_\textrm{L}$ resp.~$\textrm{D.S.}_\textrm{R}$) such that it lies \emph{outside} of the corresponding disc. Importantly, these two gauges are related by a gauge transformation that has a discontinuity along path $\gamma_{\textrm{G.T.}}= \textrm{D.S.}_\textrm{L} \,\cup\, \textrm{D.S.}_\textrm{R}$ (solid brown path in Fig.~\ref{fig:non-commute}). This gauge transformation rotates $\textrm{D.S.}_\textrm{L}$ into $\textrm{D.S.}_\textrm{R}$ (and vice versa), and is simply equal to $X=\unit$ on one side and to $X=\pm\sigma_z$ on the other side of the path $\gamma_\textrm{G.T.}$. Such a gauge transformation necessarily reverses the orientation of the bundle at \emph{exactly one} of the two principal nodes. It follows that the relative contribution of the two principal nodes to the sum in Eq.~(\ref{eqn:Eu-class-windings}) is \emph{reversed} due to the reversed orientation $\zeta$ near one of the nodes. Therefore, if the contributions of the two nodes to the Euler class cancel on $\mathcal{D}_\textrm{L}$ [e.g.~$\chi(\mathcal{D}_\textrm{L}) \!=\! \tfrac{1}{2} - \tfrac{1}{2} \!=\! 0$], then the Euler class is automatically non-trivial on $\mathcal{D}_\textrm{R}$ [corresponding to $\chi(\mathcal{D}_\textrm{R}) \!=\! \pm \left(\tfrac{1}{2} + \tfrac{1}{2}\right) \!=\! \pm 1$]. Following the discussion at the end of Sec.~\ref{sec:boundary}, the two principal nodes annihilate if brought together along a trajectory inside $\mathcal{D}_\textrm{L}$, but are incapable to annihilate if brought together along a trajectory inside $\mathcal{D}_\textrm{R}$. We thus conclude that the anticommutation relation in Eq.~(\ref{eqn:anticommuting-patsh}) [resp.~in axiom (\emph{iii}) on the previous page] is exactly reproduced by the behaviour of Euler class on manifolds with a boundary.

\section{Numerical calculation of the Euler form.} \label{sec:numerical}

To test the presented theory numerically, we have implemented a \texttt{Mathematica} code that takes as input (1) an $N$-band real-symmetric Bloch Hamiltonian, (2) two (consecutive) band indices, and (3) a rectangular region containing no adjacent nodes. The program outputs the Euler class on the defined region (with a boundary) for the selected pair of consecutive bands, by implementing Eq.~(\ref{eqn:chi-d-def}) in the eigenstate basis. The code requires setting one hyper-parameter, namely the subdivision of the sides of rectangular region into a discrete set of points.
The code is briefly described below, and we have made it available online~\cite{Bzdusek:Mathematica-Euler}.

\begin{figure*}[t]
	\includegraphics[width=0.98\textwidth]{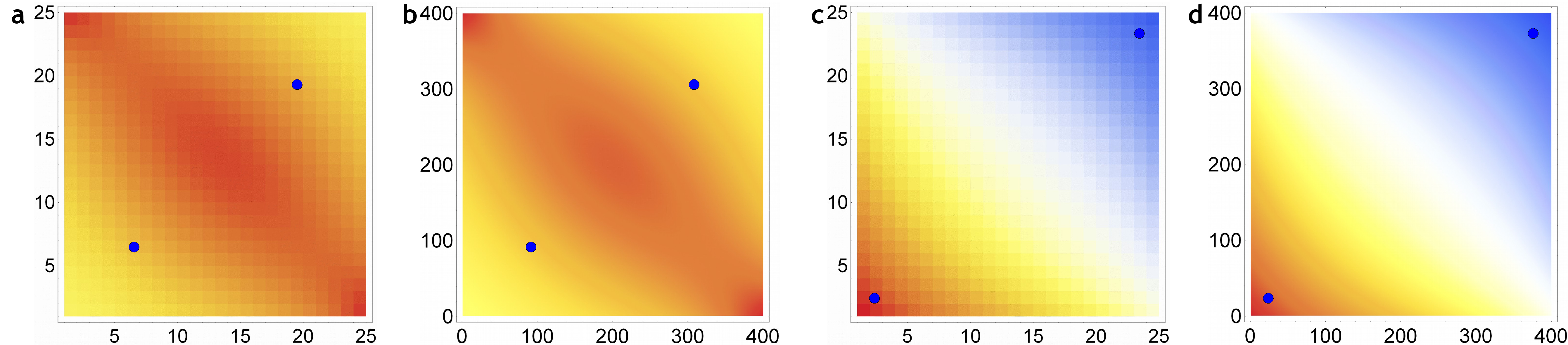}
	\caption{\textsf{\textbf{Euler form and Euler class for the braiding protocol from the main text.} {\textbf a} and {\textbf b}. Numerically computed Euler form for the model in Eq.~(4) of Methods for $t\!=\!-2.5$. The considered rectangular region $k_1\!\in\![-1,1]$ and $k_2\!\in\![-1,1]$ contains two principal nodes (blue dots) and no adjacent nodes. The two panels differ in the mesh size, $\texttt{pts}\!\in\!\{25,400\}$. The color scheme shows positive (negative) values in red (blue) tones, and white values in zero. The code adjusts the color scheme range to fully fit the computed range of $\abs{\textrm{Eu}(\bs{k})}$. The estimated value of the Euler class is -0.984220 in {\textbf a}, and -0.999934 in {\textbf b}. This is numerical approximation of $\chi(\mathcal{D}) = -1$, implying that the two nodes cannot annihilate inside the region. The result is negative, because the positive area integral of $\textrm{Eu}$ (note the red tones of the plots) is overcompensated by a larger negative boudnary integral of Euler connection {\bf a}. {\textbf c} and {\textbf d}. Analogous data for $t=-5.5$ and region $k_1\!\in\![\pi-1,\pi+1]$ and $k_2\!\in\![\pi-1,\pi+1]$. For both choices of mesh, the computed Euler class is zero within $\approx 10^{-7}$, which implies that the nodes can annihilate inside the region.}}
	\label{fig:regularized}
\end{figure*} 

The code sequentially implements the following steps. It begins by initializing the input parameters. We define a Bloch Hamiltonian \texttt{H[k1,k2]}, two (consecutive) band indices \texttt{LowerBand} and \texttt{UpperBand} that label the two principal bands, and a rectangular domain $k_1\in[\texttt{k1Min},\texttt{k1Max}]$, $k_2\in[\texttt{k2Min},\texttt{k2Max}]$. The code automatically extracts the total number of bands \texttt{TotalBands}. The labelling of the bands is such that the lowest-energy band is indexed by \texttt{1}, and the highest-energy band is indexed by the value \texttt{TotalBands}. We further set the value of hyper-parameter \texttt{pts} which defines the discretization of the sides of the rectangular region into $\texttt{pts}\times\texttt{pts}$ infinitesimal squares of size $\texttt{dk1}\times\texttt{dk2}$, where $\texttt{dk1} = (\texttt{k1Max}-\texttt{k1Min})/\texttt{pts}$ and similarly for $\texttt{dk2}$. 

In the next stage, we prepare the data for the numerical calculation of the Euler connection and Euler form. We save the two numerically obtained principal eigenstates of the Hamiltonian at the regular mesh of points into a $(\texttt{pts}+1)\times(\texttt{pts}+1)\times 2$ array called \texttt{AllStates}. Note that each entry of this array is itself an array of size $\texttt{TotalBands} \times 1$ (i.e.~a right eigenstate). However, numerical diagonalization of the Hamiltonian finds the principal bands with a {random} $+/-$ gauge, which has to be smoothed before computing the derivatives. Furthermore, we know from Sec.~\ref{sec:boundary} that each principal node is a source of a Dirac string associated with a $X=-\unit$ gauge transformation. This implies the absence of a single-valued continuous gauge on regions containing principal nodes. To deal with these two issues, the code proceeds as follows. First, it computes the Berry phase~\cite{Berry:1984} on each of the $\texttt{pts}\times\texttt{pts}$ infinitesimal squares of the mesh, and stores the information in an array \texttt{Fluxes}. The default value is $+1$, while squares containing a node are indicated by value $-1$. Positions of all the nodes are then extracted and saved as pairs of numbers in array \texttt{Nodes}. Knowing the position of all the principal nodes inside the region, the code follows a set of rules to fix the position of the Dirac strings. The chosen trajectories of the Dirac strings are saved in array $\texttt{Strings}$. The infinitesimal squares traversed by a Dirac string are characterized by \texttt{Strings[[i,j]]=-1}, else the default value is \texttt{+1}. Finally, the two principal states are gauged such that they vary smoothly away from the Dirac strings, while both of the states simultaneously flip sign across each Dirac string. This gauge is then used to update all the states stored in array \texttt{AllStates}.

In the final stage, the code takes the gauged eigenstates saved in \texttt{AllStates}, and uses them to compute Euler form inside the region and Euler connection on the boundary of the region. Following Eq.~(\ref{eqn:chi-d-def}), these two quantities are integrated to obtain the Euler class on the rectangular region. The code computes Euler connection along the boundary by implementing Newton's method of finite differences to approximate the derivative in the expression for $\mathrm{a}(\bs{k})$ in Eq.~(\ref{eqn:calculate-euler}). The numerical integral of the Euler connection is then saved as \texttt{EulerConnectionIntegral}.

To integrate the Euler form inside the rectangular region, we use the complexification discussed in the Methods section. The complexified principal bands are stored in an $(\texttt{pts}+1) \times (\texttt{pts}+1)$ array \texttt{ComplexifiedStates}. As stated by Eq.~(28) in Methods, Euler form of the two principal bands is exactly reproduced as the Berry curvature of the complexified band. However, following Ref.~\cite{Fukui:2005}, we directly compute the \emph{flux} of the Euler form through each infinitesimal square, which exactly obeys 
\begin{widetext}
\begin{eqnarray}
\textrm{Eu}(\bs{k}) \,dk_1 dk_2 
&=& \arg\tr{[P_{\bs{k}+dk_2}P_{\bs{k}+dk_1+dk_2}P_{\bs{k}+dk_1}P_{\bs{k}}]}\label{eqn:Euler-Berry-trick}\\
&=& \arg{[\braket{\psi_{\bs{k}}}{\psi_{\bs{k}+dk_2}}\braket{\psi_{\bs{k}+dk_2}}{\psi_{\bs{k}+dk_1+dk_2}}\braket{\psi_{\bs{k}+dk_1+dk_2}}{\psi_{\bs{k}+dk_1}}\braket{\psi_{\bs{k}+dk_1}}{\psi_{\bs{k}}}]}\nonumber
\end{eqnarray}
\end{widetext}
where $P_{\bs{k}} = \ket{\psi_{\bs{k}}}\bra{\psi_{\bs{k}}}$ is a projector onto the complexified state at momentum $\bs{k}$. With this trick, we elegantly avoid the accumulation of numerical errors in computing the derivatives of states in the right Eq.~(\ref{eqn:calculate-euler}). For infinitesimal squares not traversed by a Dirac string we readily compute the Euler form via Eq.~(\ref{eqn:Euler-Berry-trick}). For infinitesimal squares traversed by a Dirac string, we perform an extra gauge transformation to guarantee a continuous gauge on the four vertices of the square. Note also that the code is set to skip the computation of Euler form on infinitesimal squares containing the principal nodes (i.e.~those with $\texttt{Fluxes[[i,j]]}=-1$). We numerically integrate the Euler form by summing up the contributions from all infinitesimal squares, and the obtained approximation of the integral is saved as $\texttt{EulerFormIntegral}$. 
We finally output \texttt{EulerClass}, which is the difference of \texttt{EulerFormIntegral} and \texttt{EulerConnectionIntegral} divided by $2\pi$, cf.~the definition in Eq.~(\ref{eqn:chi-d-def}). 

To demonstrate the performance of the code, we consider the braiding protocol in Eq.~(4) of Methods for two values of $t$. First, for $t=-2.5$, the system exhibits two principal nodes near the \emph{center} of the Brillouin zone. It is observed that these two nodes fail to annihilate at a collision for $t=-2$. To analyze the situation, we consider a square region $k_1\!\in\![-1,1]$ and $k_2\!\in\![-1,1]$, which contains the two principal nodes and no adjacent nodes. The estimated value of Euler class is very close to $-1$, cf.~Fig.~\ref{fig:regularized}{\textbf{a}} and~{\textbf{b}}, consistent with the observed stability of the nodes. 

We contrast this to the situation with $t=-5.5$, when the region $k_1\!\in\![\pi-1,\pi+1]$ and $k_2\!\in\![\pi-1,\pi+1]$ contains no adjacent nodes but a pair of principal nodes, which have been pairwise created inside the region at an earlier time. In this case, the numerically computed Euler class is zero within $\approx 10^{-7}$, cf.~Fig.~\ref{fig:regularized}{\textbf{c}} and~{\textbf{d}}, consistent with the fact that the nodes were pairwise created inside the region. 

It is worth to emphasize that the two principal nodes studied for $t=-2.5$ resp.~for $t=-5.5$ are \emph{the same} pair of nodes, but we have computed their Euler class inside different regions of the Brillouin zone. The Euler class on the two regions reaches different value, confirming that the capability of the principal nodes to annihilate depends on the choice of path used to bring them together.

\onecolumngrid
\bigskip

\PRLsep

\section*{List of references}\label{biblio}

%

\end{bibunit}
\end{document}